\documentclass[11pt,urlcolor=blue, linkcolor=blue]{article}
\usepackage{cite}

\usepackage{amsmath, amsthm, amssymb}
\usepackage{ifpdf}
\ifpdf
  \usepackage[pdftex]{graphicx}
  \usepackage{epstopdf}
\else
  \usepackage[dvips]{graphicx}
\fi
\parskip=\baselineskip

\usepackage[usenames, dvipsnames]{color}

\allowdisplaybreaks[1]

\definecolor{mygray}{gray}{0.6}

\usepackage{upgreek}
\usepackage{bbm}

\newcommand{\GSD}{\text{GSD}}

\newenvironment{myfont}[2][]{\csname#2\endcsname[#1]}{}

\newcommand{\bea}{\begin{eqnarray}}
\newcommand{\eea}{\end{eqnarray}}
\def\be{\begin{equation}}
\def\ee{\end{equation}}

\usepackage{color}
\usepackage[colorlinks,citecolor=blue]{hyperref}
\definecolor{red}{rgb}{1,0,0}
\definecolor{blue}{rgb}{0,0,1}
\definecolor{dblue}{rgb}{0,0,0.4}
\definecolor{green}{rgb}{0,1,0}
\definecolor{black}{rgb}{0,0,0}
\definecolor{white}{rgb}{1,1,1}

\definecolor{brn}{rgb}{.8,.4,.0}
\definecolor{redo}{rgb}{1,.5,.0}
\definecolor{ddgrn}{rgb}{0,0.4,0}
\definecolor{dgrn}{rgb}{0,0.55,0}
\definecolor{dbl}{rgb}{0,0,0.5}

\usepackage[bbgreekl]{mathbbol}
\usepackage{amscd}

\newcommand{\Z}{\mathbb{Z}}

\newcommand{\R}{\mathbb{R}}

\renewcommand{\v}[1]{\boldsymbol{#1}}

\newcommand{\ii}{\hspace{1pt}\mathrm{i}\hspace{1pt}}

\newcommand{\eqn}[1]{Eq.~(\ref{#1})}

\newcommand{\Tr}{{\rm Tr}}

\newcommand{\bpm}{\begin{pmatrix}}
\newcommand{\epm}{\end{pmatrix}}
\newcommand{\bmm}{\begin{matrix}}
\newcommand{\emm}{\end{matrix}}

\newcommand{\cH}{ {\cal H} }

\newcommand{\cS}{ {\cal S} }
\newcommand{\cT}{ {\cal T} }

\newcommand{\al}{\alpha}

\def\CM{{\cal M}}

\def\Z{{\mathbb{Z}}}

\def\R{{\mathbb{R}}}

\def\Tr{{\mathrm{Tr}}}

\newcommand{\diag}{\mathop{\mathrm{diag}}}
\def\cW{{\cal W}}
\def\cN{{\cal N}}

\usepackage{slashed}
\usepackage[makeroom]{cancel}
\usepackage[normalem]{ulem}
\usepackage{soul}
\newcommand{\stkout}[1]{\ifmmode\text{\sout{\ensuremath{#1}}}\else\sout{#1}\fi}

\usepackage{tikz}
\usetikzlibrary{arrows}
\usetikzlibrary{calc}
\pgfmathsetseed{3}
\usetikzlibrary{shapes.arrows}

\def\ra{{\mathrm{a}}}
\def\rb{{\mathrm{b}}}
\def\rc{{\mathrm{c}}}
\def\rT{{\mathrm{T}}}
\def\ti{i}
\def\rt{{\mathrm{t}}}
\def\GL{{\mathrm{GL}}}
\def\SL{{\mathrm{SL}}}
\def\rn{{\mathrm{n}}}
\def\rm{{\mathrm{m}}}
\def\rU{{\mathrm{U}}}
\def\rSU{{\mathrm{SU}}}
\newcommand{\cblue}[1]{\textcolor{black}{#1}}

\numberwithin{equation}{section}

\begin{document}

\begin{titlepage}
\begin{flushright}
\end{flushright}
\vskip .5in

\begin{center}

{\bf\LARGE{Theory of \\[6.2mm]
Disordered $\nu = 5/2$ Quantum Thermal Hall State:\\[6.6mm]
Emergent Symmetry and Phase Diagram}}


\vskip 1.5cm
\Large{Biao Lian$^{1}$ and Juven Wang$^{2}$}

\vskip.5cm

 \vskip.2cm
{\small{\textit{$^1$
Princeton Center for Theoretical Science, Princeton University,
Princeton, NJ 08544, USA}\\}}
 \vskip.3cm
{\small{\textit{$^2$School of Natural Sciences, Institute for Advanced Study, Princeton, NJ 08540, USA}\\}}


\end{center}
\vskip.5cm
\baselineskip 16pt
\begin{abstract}
Fractional quantum Hall (FQH) system at Landau level filling fraction $\nu=5/2$ has long been suggested to be non-Abelian,
either Pfaffian (Pf) or antiPfaffian (APf) states by numerical studies, both with quantized Hall conductance $\sigma_{xy}=5e^2/2h$.
Thermal Hall conductances of the Pf and APf states are quantized at $\kappa_{xy}=7/2$ and $\kappa_{xy}=3/2$ respectively in a proper unit.
However, a recent experiment shows the thermal Hall conductance of $\nu=5/2$ FQH state is $\kappa_{xy}=5/2$.
It has been speculated that the system contains random Pf and APf domains driven by disorders, and the neutral chiral Majorana modes on the domain walls may undergo a percolation transition to a $\kappa_{xy}=5/2$ phase. In this work,
we do perturbative and non-perturbative analyses on the domain walls between Pf and APf.
We show the domain wall theory possesses an emergent SO(4) symmetry at energy scales below a threshold $\Lambda_1$, which is lowered to an emergent U(1)$\times$U(1) symmetry at energy scales between $\Lambda_1$ and a higher value $\Lambda_2$, and is finally lowered to the composite fermion parity symmetry $\mathbb{Z}_2^F$ above $\Lambda_2$. Based on the emergent symmetries, we propose a phase diagram of the disordered $\nu=5/2$ FQH system, and show that a $\kappa_{xy}=5/2$ phase arises at disorder energy scales $\Lambda>\Lambda_1$.
Furthermore, we show the gapped double-semion sector of $N_D$ compact domain walls
contributes non-local topological degeneracy $2^{N_D-1}$,
causing a low-temperature peak in the heat capacity.
We implement a non-perturbative method to bootstrap generic topological 1+1D domain walls (2-surface defects)
applicable to any 2+1D non-Abelian topological order.
We also identify potentially relevant spin topological quantum field theories (TQFTs) for various $\nu = 5/2$ FQH states in terms of fermionic version of U(1)$_{\pm 8}$ Chern-Simons theory$\times \mathbb{Z}_8$-class TQFTs.

\end{abstract}
\end{titlepage}

\tableofcontents

\section{Introduction}\label{Sec1}
The filling fraction $\nu=5/2$ fractional quantum Hall (FQH) state in $2+1$ dimensional ($2+1$D) spacetime is one of the few non-Abelian state candidates which show experimental evidences \cite{willett1987}. Exact diagonalization (ED) and density matrix renormalization group (DMRG) studies in the past \cite{morf1998,rezayi2000,peterson2008,feiguin2009,wangh2009,storni2010,rezayi2011,papic2012,zaletel2015,pakrouski2015} have shown the $\nu=5/2$ ground state favors either the Moore-Read Pfaffian (Pf) state \cite{moore1991,read2000} or its particle-hole (PH) conjugate, the anti-Pfaffian (APf) state \cite{levin2007,lee2007}, both of which are non-Abelian. (See also an early theoretical work on non-Abelian states\cite{1991Wen}.)
More precisely, the ground state is found to be either the Pf state or the APf state at half filling in the spin polarized first Landau level, together with two fully occupied spin up and down zeroth Landau levels. While both states exhibit a quantized Hall conductance $\sigma_{xy}=5/2$ in units of $e^2/h$ where $e$ is the electron charge and $h$ is the Planck constant, the thermal Hall conductance $\kappa_{xy}$ of the Pf state and the APf state are quantized differently at $\kappa_{xy}=7/2$ and $\kappa_{xy}=3/2$ in units of $\pi^2k_B^2T/3h$, respectively, where $k_B$ is the Boltzmann constant and $T$ is the temperature. Theoretically, $\kappa_{xy}$ is the total chiral central charge of the $1+1$D edge conformal field theory (CFT) of a bulk-gapped $2+1$D topological state \cite{kane1997} (Appendix \ref{app:data}). For Pf and APf states, the half-integer $\kappa_{xy}$ is due to the existence of odd number of neutral chiral Majorana-Weyl fermions on the edge \cite{WenPhysRevLett.70.355}
in addition to charged chiral bosons (or complex fermions).

\begin{figure}[htbp]
  \centering
  \includegraphics[width=4.4in]{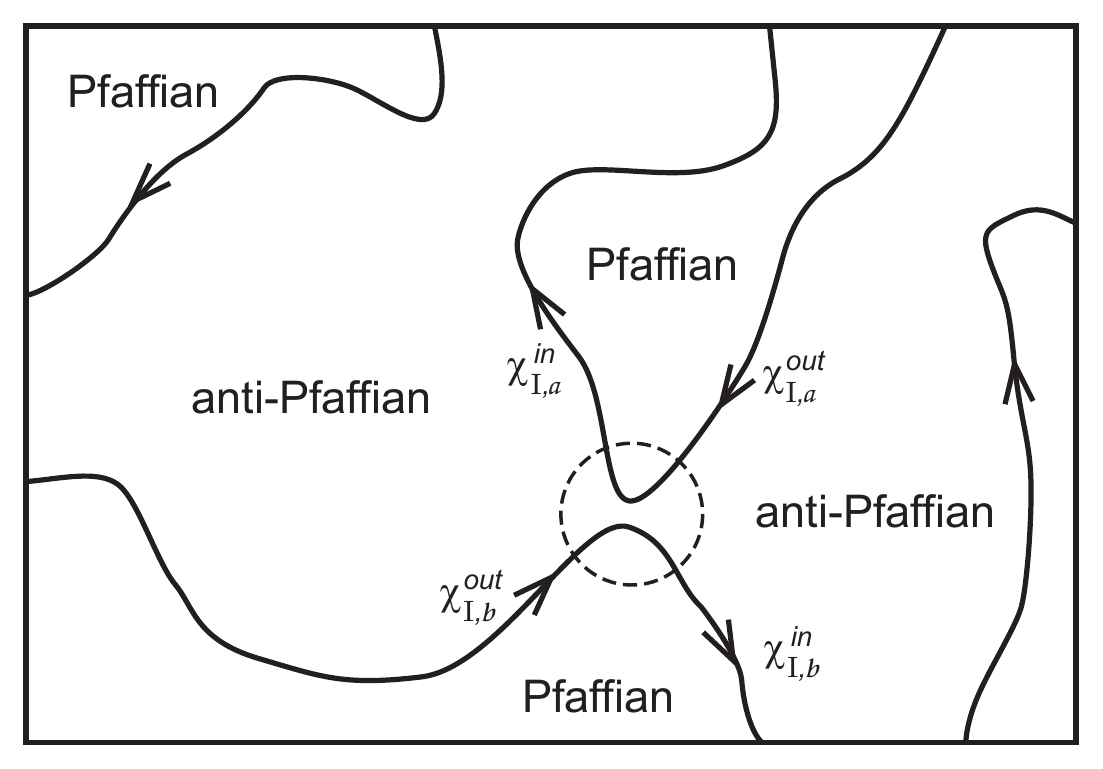}
  \caption{Random domains of Pfaffian and anti-Pfaffian states with percolating domain walls, where each domain wall possesses four chiral Majorana fermion modes along the direction of the arrow. When two domain walls get close to each other as shown by the dashed circle, the chiral Majorana fermions on them may tunnel between them.}\label{RD}
\end{figure}

Recently, a measurement by Banerjee \emph{et al.} \cite{banerjee2017} observed that the thermal Hall conductance of the $\nu=5/2$ FQH state is $\kappa_{xy}=5/2$, which is in contradiction to both the Pf state and the APf state.
Instead, this experimental result \cite{banerjee2017} agrees with a different non-Abelian state candidate known as the particle-hole Pfaffian (PH-Pf) state \cite{zucker2016,son2015,chen2014} (Appendix \ref{app:data}). However, it is generically believed the PH-Pf state is not energetically favored compared to the Pf or APf state, with evidences from numerical calculations (see discussion in \cite{wangc2017}). Most recently, it is suggested that the presence of disorders may stabilize the PH-Pf state, and various other possible phases under disorders are discussed \cite{wangc2017,mross2017}. The idea is that the disorders may drive the $2+1$D system into random domains of Pf and APf states (Fig.~\ref{RD}),
where each domain wall carries chiral central charge $c_- \equiv c_L-c_R=2$ and hosts four chiral Majorana edge states.
The percolation of these chiral Majorana edge states in the bulk of the system may then yield different phases in the infinite size thermodynamic limit, including a possible $\kappa_{xy}=5/2$ phase which is identified with the PH-Pf state. However, several problems still remain unsettled, which include a detailed analysis of the edge theory on the domain walls, emergent symmetries at low energies, the disorder strength for the $\kappa_{xy}=5/2$ phase to be stabilized, and the energy cost of domain walls in the system, etc.

\begin{figure}[htbp]
  \centering
  \includegraphics[width=4.9in]{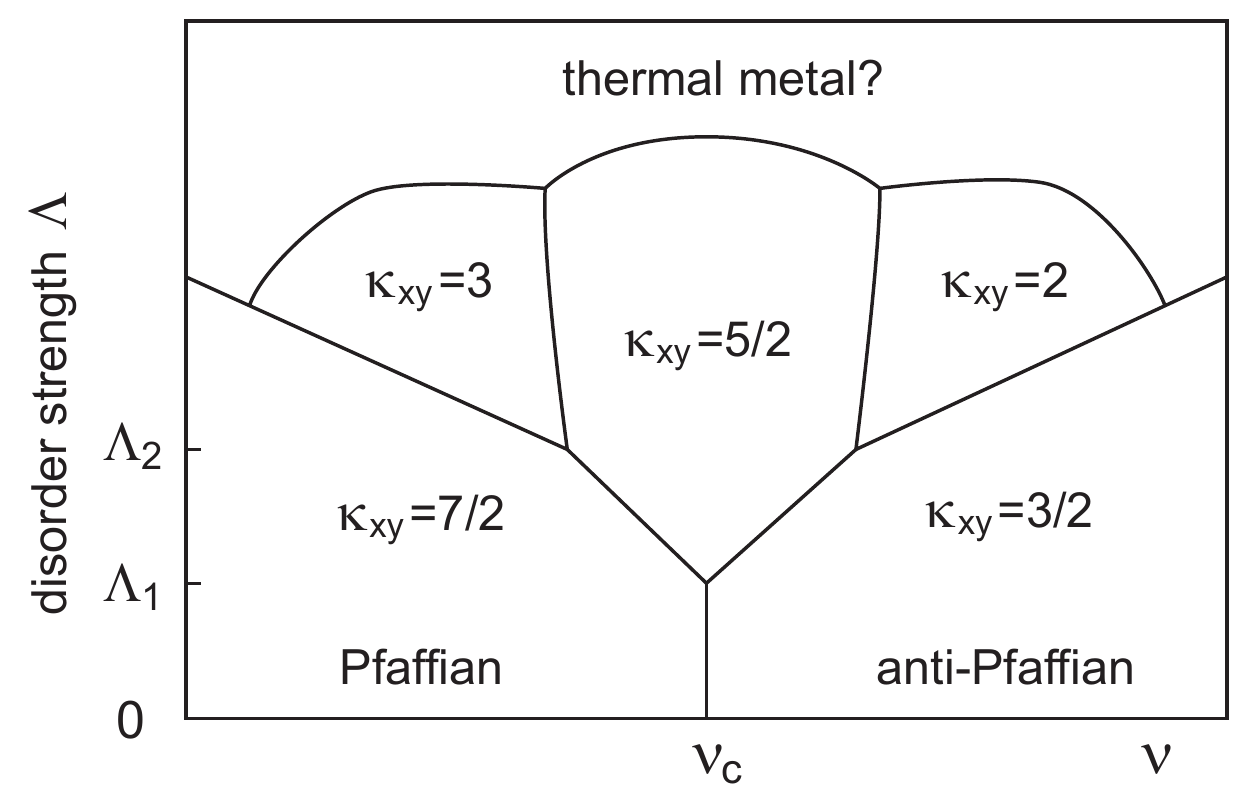}
  \caption{We propose the specific phase diagram of disordered $\nu=5/2$ state, where $\nu$ is the mean filling fraction, $\Lambda$ is the disorder energy scale inversely proportional to Pf or APf domain size, and the phases are labeled by thermal Hall conductance $\kappa_{xy}$. The system exhibits an SO(4) emergent symmetry for $\Lambda<\Lambda_1$, and has a U(1)$\times$U(1) emergent symmetry for $\Lambda_1<\Lambda<\Lambda_2$.}\label{PD}
\end{figure}

In this work, we study the edge theory and possible emergent symmetries on the domain wall between Pf and APf states, based on which,
we propose a possible yet more specific phase diagram of the disordered system. We show the effective theory of the domain wall has an emergent SO(4) symmetry at low energies below an energy scale $\Lambda_1$, which breaks down to a U(1)$\times$U(1) emergent symmetry at intermediate energy scales between $\Lambda_1$ and $\Lambda_2$, and finally to
 a fermion parity $\mathbb{Z}_2^F$ symmetry at energy scales above $\Lambda_2$. This leads to our phase diagram in the vicinity of $\nu=5/2$ as shown in
 Fig.~\ref{PD}, where $\nu$ is the filling fraction, and $\Lambda$ is the disorder strength. In the absence of disorders, as shown by previous numerical studies \cite{morf1998,rezayi2000,peterson2008,feiguin2009,wangh2009,storni2010,rezayi2011,papic2012,zaletel2015,pakrouski2015},
the ground state is the Pf state (with quasiholes) for $\nu<\nu_c$, and is the APf state (with quasielectrons) for $\nu>\nu_c$, where $\nu_c\approx5/2$ is the critical filling fraction. When the energy scale of disorders $\Lambda<\Lambda_1$ is weak, there is just a single transition from Pf to APf phase with respect to $\nu$ as ensured by the emergent SO(4) symmetry. At intermediate disorder energy scales $\Lambda_1<\Lambda<\Lambda_2$, the emergent symmetry is lowered to U(1)$\times$U(1), and the single transition with respect to $\nu$ splits into two transitions, with a new gapped phase of $\kappa_{xy}=5/2$ arises between the Pf and APf phases. Here the disorder energy scale $\Lambda$ is roughly inversely proportional to the size of a Pf or APf domain.
For higher disorder energy scales $\Lambda>\Lambda_2$ where only $\mathbb{Z}_2^F$ symmetry remains, the system may undergo four phase transitions with respect to $\nu$, each of which changes $\kappa_{xy}$ by $1/2$, or the system may enter a thermal metal phase where the bulk becomes gapless \cite{senthil2000,chalker2001,fulga2012}.
We note that since emergent symmetries are not ``true" exact symmetries, it is possible the above picture is only approximately true, namely, the single phase transition at $\Lambda<\Lambda_1$, and two phase transitions at $\Lambda_1<\Lambda<\Lambda_2$ with respect to $\nu$, may be broadened and are not sharp transitions. Such broadenings are, however, expected to be at least exponentially suppressed by factors $e^{-\Lambda_1^2/\Lambda^2}$ and $e^{-\Lambda_2^2/\Lambda^2}$ \cite{wangc2017,imry1975,binder1983}, respectively, and are probably beyond the resolution of the experiments. Nevertheless, it is still possible that there are no broadenings at all due to dynamical fluctuations, and all phase transitions in the phase diagram Fig.~\ref{PD} are sharp, which calls for a future study.
Finally, we show that the charged sector of the domain walls, although being gapped out, also has a nontrivial contribution to the non-local topological ground state degeneracy \cite{1212.4863WW, levin2013}, which affects the heat capacity and longitudinal thermal conductance of the system.


We first briefly review the topological properties of the Pf and APf states. In the original work by Moore and Read \cite{moore1991}, the Pf state is a filling fraction $\nu=1/2$ wave function in the zeroth Landau level
\begin{equation}
\Psi_{\text{Pf}}=\prod_{1=i<j}^{N}(z_i-z_j)^2\mbox{Pf}\left(\frac{1}{z_i-z_j}\right) \prod_{i=1}^{N}e^{-|z_i|^2/4\ell_B^2}\ ,
\end{equation}
where $N$ is the number of electrons which is even, $\ell_B=\sqrt{\hbar c/eB}$ is the magnetic length of magnetic field $B$, $z_i=x_i+iy_i$ is the complex coordinate of the $i$-th electron, $\mbox{Pf}$ gives the Pfaffian of the antisymmetric matrix $M_{ij}=1/(z_i-z_j)$, and all the electrons are spin polarized. In the context here, both the spin up and down
in the zeroth Landau levels are fully occupied, and the Pf state is formed in the spin polarized first Landau level, so the total filling fraction is around $\nu=5/2$. 
The gapped bulk of the Pf state allows the existence of both charge $\pm e/2$ semions which are Abelian, and charge $\pm e/4$ quasiparticles which obey non-Abelian statistics \cite{read1996}. The gapless edge of the Pf state contains a left-moving charge $e/2$ chiral boson mode and a left-moving neutral chiral Majorana fermion mode \cite{milo1996}. Together with two left-moving charge $e$ chiral complex fermion modes from the spin up and down zeroth Landau levels, they contribute a Hall conductance $\sigma_{xy}=5/2$ and a thermal Hall conductance $\kappa_{xy}=7/2$.

The APf state is obtained by applying a particle-hole transformation \cite{girvin1984a} to the Pf state in the first Landau level. The edge theory of the APf state consists of a left-moving charge $e$ chiral complex fermion, a right-moving charge $e/2$ chiral boson and a right-moving neutral chiral Majorana fermion. Under disorders, such an edge theory renormalizes into a charge $e/2$ left moving chiral boson and an SO(3) symmetric triplet of right-moving neutral chiral Majorana fermions, therefore develops an SO(3) emergent symmetry \cite{lee2007,levin2007}. At $\nu=5/2$, this yields a Hall conductance $\sigma_{xy}=5/2$ and a thermal conductance $\kappa_{xy}=3/2$. Meanwhile, the bulk of APf state also hosts charge $\pm e/2$ semions and charge $\pm e/4$ non-Abelian quasiparticles, but slightly different from those in the bulk of Pf state.
Use the above information, we identify 
the detailed bulk topological quantum field theories (TQFTs) and gapless edge theories (conformal field theories coupling to external background fields)
of Pf, APf, PH-Pf states and other related $\nu=5/2$-states, which can be found in  Appendix \ref{app:data}.

A useful perspective often adopted in literature is to view the Pf and APf states as superconductors of composite fermions with different pairing symmetries \cite{read2000,halperin1993,son2015}. A composite fermion is defined as an electron bound with two statistical gauge field fluxes, which cancel the external magnetic field on average for 
a half-filled Landau level
\cite{jain1989,halperin1993,son2015}. It is believed to be a good starting point to assume the composite fermions in the first Landau level form a fermi liquid with a single fermi surface \cite{son2015,mross2015}. Recent studies suggest the fermi liquid may be a Dirac fermi liquid, and the fermi surface has an intrinsic $\pi$ Berry phase \cite{son2015, mross2015, geraedts2016,wangc2016,potter2016,geraedts2017}. The fermi surface can then be gapped out by forming Cooper pairs. If we denote the composite fermion at momentum $\mathbf{k}$ as $f_\mathbf{k}$, the possible pairing amplitude $\Delta(\mathbf{k})f_\mathbf{k}f_{-\mathbf{k}}$ near the fermi surface must satisfy $\Delta(\mathbf{k})=-\Delta(-\mathbf{k})$ because of the anticommutation relation of $f_{\mathbf{k}}$ and $f_{-\mathbf{k}}$. Namely, the pairing must have an odd parity. In particular, a $p+ip$ pairing amplitude $\Delta(\mathbf{k})=\Delta_{\text{Pf}}e^{i\theta_\mathbf{k}}$ near the fermi surface leads to the Pf state, while a $f-if$ pairing amplitude $\Delta(\mathbf{k})=\Delta_{\text{APf}}e^{-3i\theta_\mathbf{k}}$ corresponds to the APf state, where
$\theta_{\mathbf{k}}=\mbox{arg}(k_x+ik_y)$ is the polar angle of the
momentum \cite{son2015,read2000}.\footnote{In the literature of
 Dirac fermi liquid,
the fermi surface $\pi$ Berry phase is embedded in the definition of $f_\mathbf{k}$, namely, $f_\mathbf{k}\rightarrow e^{i\theta_\mathbf{k}/2}f_\mathbf{k}$,
so the
$p+ip$-pairing Pf state, $p-ip$-pairing PH-Pf state and
$f-if$-pairing APf state are denoted as $d+id$, $s$ and $d-id$ pairing, respectively. This is just a different definition of basis
and does not change any physics. To be precise, by $s$, $p\pm ip$, $d\pm id$ and $f\pm if$ pairings we mean $\Delta(\mathbf{k})\propto1$, $k_x\pm ik_y$, $(k_x\pm ik_y)^2$ and $(k_x\pm ik_y)^3$, respectively. 
See also discussions in Appendix \ref{app:data}.
\label{foot:composite-pair}}
This picture correctly reproduces the neutral sector of edge theories of the Pf and APf states, i.e., a left-moving chiral Majorana mode on the Pf state edge and three right-moving chiral Majorana modes on the APf state edge. Besides, the PH-Pf state corresponds to a $p-ip$ pairing $\Delta(\mathbf{k})=\Delta_{\text{PH}}e^{-i\theta_\mathbf{k}}$ of the fermi surface \cite{son2015,zucker2016}, although this state may be energetically unfavored. We note that, however, the above pairing picture cannot reproduce the charged sector, namely, the $e/2$ left-moving chiral boson mode on the edges of Pf, APf and PH-Pf states.


In the presence of chemical potential disorders, the local filling fraction $\nu(\mathbf{r})$ may vary spatially above and below $\nu_c$, and the system may be driven into random domains of Pf and APf states as shown in Fig. \ref{RD}. The domain wall between Pf and APf states carries a chiral central charge $c_-=2$, and it is easy to see from the above pairing picture that there are four neutral chiral Majorana fermion modes of the same chirality on the domain wall.
The system is then a random percolation system of chiral Majorana fermions.
When two domain walls are close to each other as shown by the dashed circle in Fig. \ref{RD}, the chiral Majorana fermions $\chi_I$ ($1\le I\le4$) on them may tunnel into each other. Such a tunneling can be generically expressed as $\chi_{I,\tau}^{out}=\left(O^{out,\tau}_{IJ}S^{J}_{\tau\tau'}O^{in,\tau'}_{JI'}\right)\chi^{in}_{I',\tau'}$, where $\chi_{I,\tau}^{in/out}$ ($\tau=a,b$ is the domain wall label) are in/out chiral Majorana modes around the dashed circle as shown in Fig. \ref{RD}, $O^{in,\tau}$ and $O^{out,\tau}$ are random SO(4) rotation matrices mixing the four Majorana modes of a domain wall and obeying a certain distribution (e.g., Gaussian), and $S^{I}_{\tau\tau'}$ are the $2\times2$ scattering matrices defined by
\begin{equation}\label{tunnel}
S^I =\left(\begin{array}{cc}\cos\alpha_I&\sin\alpha_I\\ -\sin\alpha_I&\cos\alpha_I\end{array}\right)\ ,
\end{equation}
where $\alpha_I$ ($1\le I\le4$) are usually called scattering angles.
In addition, due to disorder,
the four chiral Majorana modes on the same domain wall may be mixed and thus propagate into each other. In addition, the propagation of chiral Majorana fermions $\chi_I$ on domain wall $\tau$ may also involve a flavor mixing due to disorders, namely, $\chi_{I,\tau}(x')=O_{IJ}^{\tau}(x',x)\chi_{J,\tau}(x)$, where the propagation matrix $O^\tau(x,x')$ is a random SO(4) matrix obeying certain distributions, while $x$ and $x'$ denote the 1D coordinate of the domain wall.
Such a percolation system can be studied numerically using random network models \cite{chalker1988,wangc2017,mross2017,kramer2005}.
In particular, if none of the angles $\alpha_I$ are equal, the system belongs to the D symmetry class
(with only a fermion parity $\mathbb{Z}_2^F$ symmetry) and is the least symmetric \cite{altland1997,kramer2005}.

\begin{figure}[htbp]
  \centering
  \includegraphics[width=5.5in]{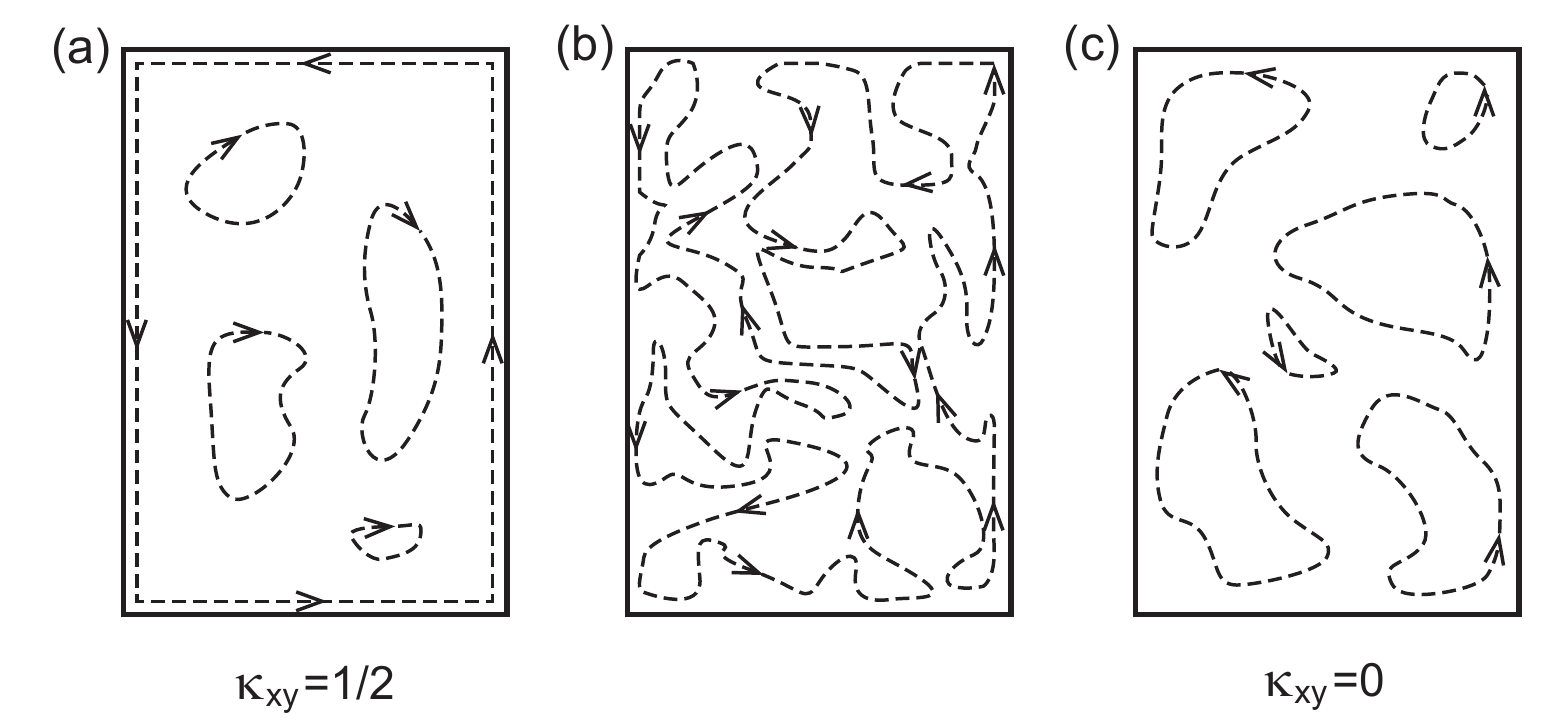}
  \caption{Illustration of the percolation transition of a single chiral Majorana fermion mode, which is represented by the dashed line with arrows indicating the chirality. (a) Before percolation transition, where all the bulk chiral Majorana states are localized, and there is a chiral Majorana edge state on the edge contributing $\kappa_{xy}=1/2$. (b) At percolation transition point, the bulk chiral Majorana mode is delocalized and connects with the edge state randomly. (c) After the percolation transition, the original edge state breaks up into localized states, and the thermal Hall conductance becomes $\kappa_{xy}=0$.
{
To make an analogy, in Fig.~\ref{RD}, the Pf-APf forms an archipelago-sea relation. When the sea level (filling-fraction $\nu$) decreases ($\nu < \nu_c$ in Fig.~\ref{PD}), the land (Pf) delocalizes and the sea (APf) localizes. When the sea level increases ($\nu > \nu_c$ in Fig.~\ref{PD}), the land (Pf) localizes and the sea (APf) delocalizes. At the transition ($\nu \simeq \nu_c$ in Fig.~\ref{PD}), both the land (Pf) and sea (APf) delocalize, which means the domain walls also delocalize, percolating to the boundary (of the experimental sample).
} 
}\label{Perc}
\end{figure}

The spatial mean values $\langle\alpha_I\rangle$ of $\alpha_I$ ($1\le I\le4$) are monotonic functions of the average filling fraction $\nu$. When $\nu$ is far above (below) $\nu_c$, the system is in the APf (Pf) state, and all the $\langle\alpha_I\rangle$ tend to $0$ ($\pi/2$). When one increases $\nu$ from far below $\nu_c$, whenever an $\langle\alpha_I\rangle$ becomes equal to $\pi/4$, a chiral Majorana fermion mode will be delocalized in the bulk and extend to the edge between the system and the vacuum, which is the percolation transition point. After $\langle\alpha_I\rangle$ has passed by $\pi/4$, a chiral Majorana edge state on the edge of the system will be eliminated or created, and the chiral central charge on the edge of the system will change by $1/2$. Fig.~\ref{Perc} shows an example how a chiral Majorana edge state is eliminated (created) after the percolation transition of a chiral Majorana mode in the bulk. Therefore, if the system is in the D symmetry class, one expects four phase transitions with respect to $\nu$ for disorders not too strong (which potentially causes thermal metal), during which $\kappa_{xy}$ undergoes four transitions from $7/2\rightarrow 3\rightarrow 5/2\rightarrow2\rightarrow3/2$ \cite{wangc2017,mross2017}. For strong disorders especially those coming from random $\pi$ flux vortices, the system may enter a gapless thermal metal phase where $\kappa_{xy}$ is no longer quantized \cite{senthil2000,chalker2001,fulga2012}.

There is, however, a possibility that the domain walls between Pf and APf states possess certain emergent symmetries, which enforce two or more $\alpha_I$ angles to be equal everywhere. In this case, the system will undergo less phase transitions with respect to $\nu$. For instance, if the four chiral Majorana fermions on the domain wall possess an SO(4) rotational symmetry during scattering, all the four scattering angles $\alpha_I$ will be equal, and one would expect a single phase transition directly from the Pf state to the APf state. Such emergent symmetries may generically arise from interactions and disorders \cite{kane1994,haldane1995}.
For instance, an SO(3) symmetry emerges on the edge of APf state due to disorder \cite{lee2007,levin2007}.
The following sections are devoted to explore the possibility of emergent symmetries, and how the phase diagram will be modified.

\section{Example of Disordered Superconductor with Emergent U(1) Symmetry}\label{Sec2}
Before we proceed to the $\nu=5/2$ FQH system, it is useful to study a simpler but similar $2+1$D system of random $p+ip$ and $p-ip$ superconducting domains, which we shall show has an emergent U(1) symmetry. Importantly, we assume the system before superconducting has a single fermi surface, described by Hamiltonian
\begin{equation}
H_0=\sum_\mathbf{k} \left(\frac{k^2}{2m_0}-\mu\right) c_\mathbf{k}^\dag c_\mathbf{k}\ ,
\end{equation}
where $c_\mathbf{k}$, $c_\mathbf{k}^\dag$ are the electron annihilation and creation operators with $\mathbf{k}$ being momentum, $\mu>0$ is the chemical potential, $m_0>0$ is the effective electron mass, and we have set the Plank constant $\hbar=1$. The single fermi surface requires the parity of the pairing amplitude $\Delta(\mathbf{k})$ to be odd. Here we assume the pairing depends on an interaction parameter $\lambda$, so that the system prefers a $p+ip$ pairing $\Delta(\mathbf{k})=\Delta_+e^{i\theta_\mathbf{k}}$ for $\lambda>0$, and prefers a $p-ip$ pairing $\Delta(\mathbf{k})=\Delta_-e^{-i\theta_\mathbf{k}}$ for $\lambda<0$, with $\theta_{\mathbf{k}}=\mbox{arg}(k_x+ik_y)$. Accordingly, the clean system is a chiral topological superconductor with a left-moving (right-moving) chiral Majorana fermion on the edge when $\lambda>0$ ($\lambda<0$) \cite{read2000}, and has a thermal Hall conductance $\kappa_{xy}=1/2$ ($\kappa_{xy}=-1/2$). As long as the chemical potential $\mu>0$, there will be no trivial superconductor phase in between, and only a single phase transition exists between the $p\pm ip$ superconductor phases with respect to $\lambda$.

In the presence of disorders, $\lambda$ may have a spatial fluctuation, and random domains of $p+ip$ and $p-ip$ superconductivity will occur when the spatial mean value $\langle\lambda\rangle$ is near $0$. Each domain wall has two chiral Majorana fermions of the same chirality. If there is no additional symmetry, the system is in the D symmetry class, and the system should exhibit two delocalization phase transitions with respect to $\langle \lambda\rangle$, with $\kappa_{xy}$ changing from $1/2\rightarrow0\rightarrow-1/2$. However, one expects the $\kappa_{xy}=0$ phase to vanish for sufficiently weak disorders (when the fermi surface picture is still valid), since such a gapped phase does not correspond to any pairing of a single fermi surface. It is therefore more natural to expect the $\kappa_{xy}=0$ phase does not occur until the disorder strength reaches a certain threshold.

\begin{figure}[htbp]
  \centering
  \includegraphics[width=3.5in]{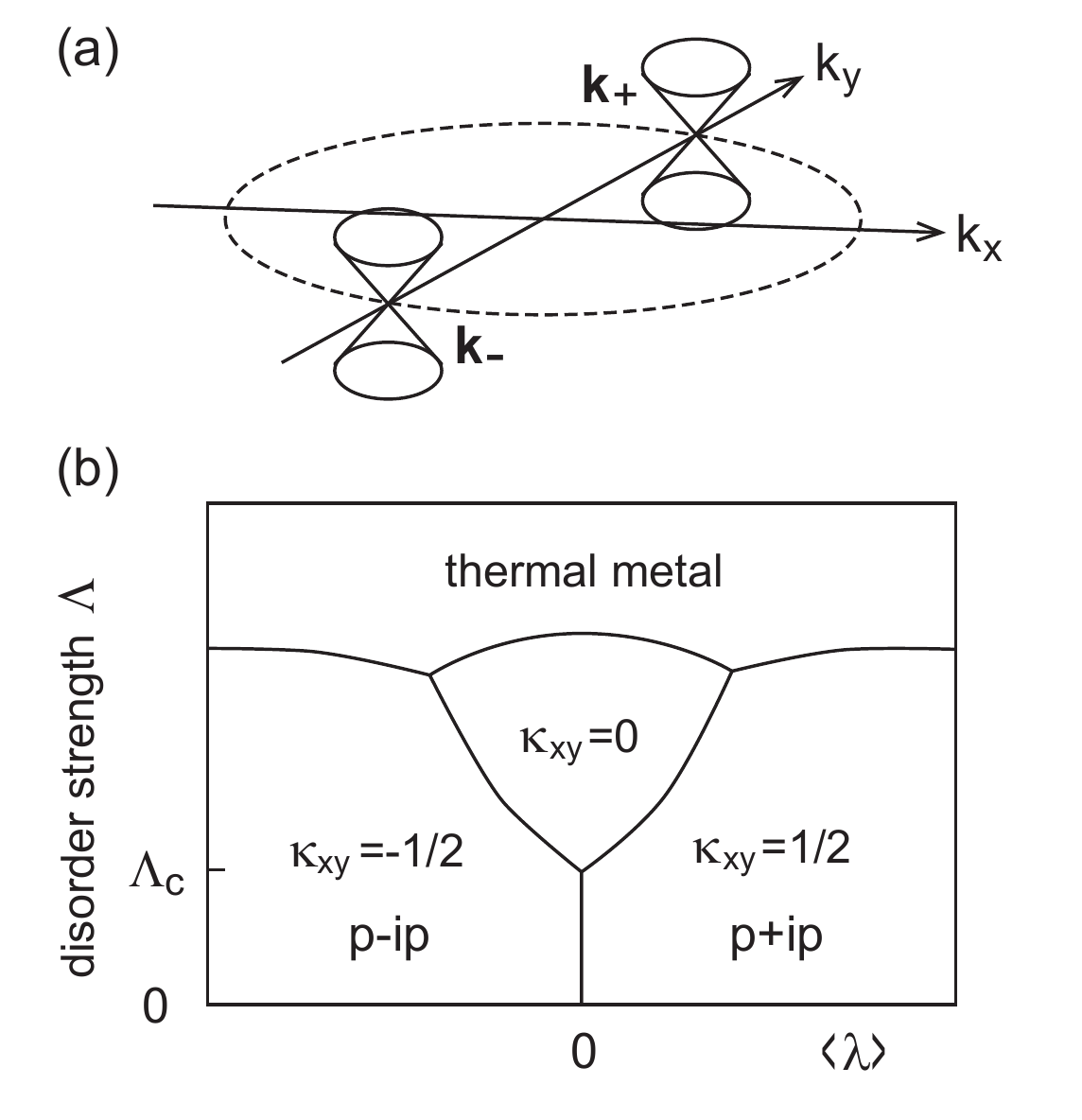}
  \caption{(a) The low energy BdG bands on the domain wall between $p+ip$ and $p-ip$ superconductivity are two Dirac cones at $\mathbf{k}_+$ and $\mathbf{k}_-$, respectively. (b) Expected phase diagram for the disordered $p\pm ip$ superconducting system, where $\Lambda$ is the disorder strength.}\label{SCDW}
\end{figure}

We shall show this is ensured by an emergent U(1) symmetry on the domain walls at low energies. The pairing amplitude near a domain wall can be generically written as
\begin{equation}
\Delta(\mathbf{k})=\Delta_+(\mathbf{r})e^{i\theta_\mathbf{k}}+\Delta_-(\mathbf{r})e^{-i\theta_\mathbf{k}}\ ,
\end{equation}
where $\Delta_+(\mathbf{r})=\Delta_+$ and $\Delta_-(\mathbf{r})=0$ on the $p+ip$ side away from the domain wall, while $\Delta_+(\mathbf{r})=0$ and $\Delta_-(\mathbf{r})=\Delta_-$ on the $p-ip$ side. The domain wall is then located where $|\Delta_+(\mathbf{r})|=|\Delta_-(\mathbf{r})|$. For the moment, assume both $\Delta_+(\mathbf{r})$ and $\Delta_-(\mathbf{r})$ are real and positive. The Bogoliubov-de Gennes (BdG) Hamiltonian of the superconductor then becomes gapless at momentum $\mathbf{k}_\pm=(0,\pm k_F)$ on the domain wall, as shown in Fig. \ref{SCDW}(a). In the vicinity of the domain wall, the low energy BdG Hamiltonian at momentum $\mathbf{k}$ near $\mathbf{k}_+$ is
\begin{equation}\label{BdG}
H_{\text BdG}(\mathbf{k})=\left(\begin{array}{cc}v_F(k_y-k_F)&v_\Delta k_x+im(\mathbf{r})\\v_\Delta k_x-im(\mathbf{r})&-v_F(k_y-k_F)\end{array}\right)\ ,
\end{equation}
where the basis is the Nambu basis $(c_{\mathbf{k}},c_{-\mathbf{k}}^\dag)^\rT$, $k_F=\sqrt{2m_0\mu}$ and $v_F=k_F/m_0$ are the electron Fermi momentum and Fermi velocity, $v_\Delta=[\Delta_+(\mathbf{r})+\Delta_-(\mathbf{r})]/k_F$, and $m(\mathbf{r})=\Delta_+(\mathbf{r})-\Delta_-(\mathbf{r})$. The BdG Hamiltonian near $\mathbf{k}_-=-\mathbf{k}_+$ is simply the particle-hole transformation of Eq. (\ref{BdG}), and describes exactly the same degrees of freedom.

The edge states on a single domain wall can be easily solved from the BdG Hamiltonian (\ref{BdG}). For a domain wall perpendicular to the direction $\mathbf{n}=(\cos\varphi,\sin\varphi)$, the mass term $m(\mathbf{r})$ can be approximated as a function of $x'=\mathbf{n}\cdot\mathbf{r}$ which is positive (negative) for $x'>0$ ($x'<0$), where we assume the domain wall is located at $x'=0$. If we approximate $v_\Delta$ as a constant, the edge state at momentum $k$ along the domain wall can be solved to be a chiral complex fermion
\begin{equation}
\psi_k=\left(\begin{array}{c}\cos\zeta\\ -\sin\zeta\end{array}\right)e^{iky'+ip(k,\varphi)x'-\kappa(\varphi)\int_0^{x'} m(x'')dx''}\ ,
\end{equation}
where $y'=-x\sin\varphi+y\cos\varphi$ is the coordinate along the domain wall, $\kappa(\varphi)=(v_F^2\sin^2\varphi+v_\Delta^2\cos^2\varphi)^{-1/2}$, the angle $\zeta=\arctan\left[\frac{\kappa(\varphi)v_F\sin\varphi}{1+\kappa(\varphi)v_F\cos\varphi}\right]$, and $p(k,\varphi)$ is a real function of $k$ and $\varphi$ which is not important here. The energy of the edge mode is
\begin{equation}\label{dispersion}
\epsilon(k)=v(\varphi)(k-k_F\cos\varphi)\ ,
\end{equation}
where $v(\varphi)=|v_F^{-1}\cos2\zeta\cos\varphi+v_\Delta^{-1}\sin2\zeta\sin\varphi|^{-1}$ is the edge state velocity, which oscillates between $v_F$ and $v_\Delta$ as a function of $\varphi$ with an oscillation period $\pi$. Therefore, the velocity $v(\varphi)$ depends on the direction of the domain wall. One can rewrite the complex fermion $\psi_k$ as two chiral Majorana fermions $\chi_1+i\chi_2$, then both $\chi_1$ and $\chi_2$ will propagate at the same velocity $v(\varphi)$.
We note that if the pairing amplitudes $\Delta_+(\mathbf{r})$ and $\Delta_-(\mathbf{r})$ on the two sides of the domain wall have a phase difference $\varphi_\Delta\neq0$, the edge state velocity will be shifted to $v(\varphi-\varphi_\Delta)$, and accordingly, $\mathbf{k}_\pm$ will be rotated to $\mathbf{k}_\pm=\pm(-k_F\sin\varphi_\Delta, k_F\cos\varphi_\Delta)$.

Regardless of the Nambu basis, the form of the Hamiltonian in Eq. (\ref{BdG}) is no different from that of a charge conserved $2+1$D Dirac fermion with a spatial dependent mass $m(\mathbf{r})$. In fact, if the mass term varies slow enough so that $|\nabla m(\mathbf{r})/m(\mathbf{r})|< k_F$, we can define an emergent U(1) symmetry for the BdG Hamiltonian at low energies as
\begin{equation}\label{EMU1}
c_\mathbf{k}\rightarrow e^{i\phi}c_\mathbf{k}\ , \quad c_{-\mathbf{k}}^\dag\rightarrow e^{i\phi}c_{-\mathbf{k}}^\dag\ ,
\end{equation}
where $\mathbf{k}$ and $-\mathbf{k}$ are momentums near $\mathbf{k}_+$ and $\mathbf{k}_-$, respectively. Such a U(1) symmetry is analogous to the chiral U(1) symmetry defined in Weyl semimetals \cite{wan2011,balents2011}, where electrons in different neighborhoods of the momentum space are associated with different phase rotations. As shown in Fig. \ref{SCDW}(a), the two Dirac cones at $\mathbf{k}_\pm$ do not overlap with each other until the energy scale reaches the bulk gap $\Delta_\pm$ of the $p\pm ip$ superconductors. Therefore, as long as the energy scale under consideration is below a certain value $\Lambda_c$ of the order of $\Delta_\pm$, the momentums $\mathbf{k}$ and $-\mathbf{k}$ in Eq. (\ref{EMU1}) are well separated, and the emergent U(1) symmetry is well-defined.

The configuration of random domains of $p\pm ip$ superconductivity is generically determined by a certain mass function $m(\mathbf{r})$, and the percolation of chiral Majorana fermions on the domain walls is entirely governed by the BdG Hamiltonian (\ref{BdG}). We can define an disorder energy scale for the system as
\begin{equation}\label{Escale}
\Lambda=\overline{v}/\ell_0\ ,
\end{equation}
where $\overline{v}$ is the mean value of the edge state velocity $v(\varphi-\varphi_\Delta)$, and $\ell_0$ is the length scale of a single domain (i.e., the length of a link in the network model language \cite{chalker1988}). When $\Lambda$ is below $\Lambda_c$, our argument above shows the system has the emergent U(1) symmetry defined in Eq. (\ref{EMU1}), so the system is in the A symmetry class (which is U(1) symmetric) instead of the D symmetry class \cite{altland1997,kramer2005}. In other words, the two chiral Majorana fermions $\chi_{1}$ and $\chi_{2}$ on each domain wall behave as a single complex fermion with a conserved U(1) charge. Accordingly, there will be only a single phase transition as a function of $\langle\lambda\rangle$ which changes $\kappa_{xy}$ by $1$, as shown in Fig. \ref{SCDW}(b). Such a phase transition is similar to the Hall conductance plateau transition of integer quantum Hall (IQH) effect \cite{chalker1988,pruisken1988,wangjing2014}.

When $\Lambda>\Lambda_c$, the two Dirac cones at $\mathbf{k}_\pm$ begin to mix with each other, and the emergent U(1) symmetry is lost. What remains is the $\mathbb{Z}_2^F$ symmetry of superconductors which takes $c_\mathbf{k}\rightarrow -c_\mathbf{k}$ and $c_{-\mathbf{k}}^\dag\rightarrow -c_{-\mathbf{k}}^\dag$ for all $\mathbf{k}$, and the system will be in the D symmetry class. In this case, the two chiral Majorana fermions will undergo two separate percolation transitions as a function of $\langle\lambda\rangle$, and an intermediate $\kappa_{xy}=0$ phase arises. Furthermore, the D symmetry class allows the existence of a thermal metal phase for strong enough ($\pi$ flux vortex) disorders, which has a divergent longitudinal thermal conductance $\kappa_{xx}$ and a non-quantized $\kappa_{xy}$. The expected phase diagram is shown in Fig. \ref{SCDW}(b).

We note that in this example, the absence of gapped trivial superconductor with $\kappa_{xy}=0$ at zero disorder (due to single fermi surface) is important for the emergent U(1) symmetry to arise. Roughly speaking, this binds tightly the two chiral Majorana fermions between $p+ip$ and $p-ip$ regions within a single domain wall, so that they undergo percolation transitions together. This is different from the models where trivial superconductor is allowed in the absence of disorder, in which case the two chiral Majorana fermions between $p+ip$ and $p-ip$ regions will be spatially separated (by trivial regions) and percolate differently \cite{he2017,lian2017}.


\section{Pairing Picture of Emergent Symmetries on Pfaffian-antiPfaffian Domain Wall}\label{Sec3}


The disordered $\nu=5/2$ FQH system is similar to the example in Sec.~\ref{Sec2}, in the sense that the ground state at zero disorder is believed to be restricted to either the Pf state or the APf state, which can be understood as $p+ip$ and $f-if$ superconductors of composite fermions, respectively. In this section, we shall illustrate the emergent symmetries of the domain wall between Pf and APf states in the composite fermion pairing picture. In the below, we shall show the emergent symmetries on the domain wall between Pf and APf under different energy scales based on the pairing picture.

\subsection{Emergent U(1)$\times$U(1) symmetry at energy scales $\Lambda_1<\Lambda<\Lambda_2$}

The Hamiltonian near the single fermi surface of composite fermions in the first Landau level before pairing can be written as
\begin{equation}
H_0^{cf}=\sum_\mathbf{k} v_F(|\mathbf{k}|-k_F) f_\mathbf{k}^\dag f_\mathbf{k}\ ,
\end{equation}
where $v_F$ and $k_F$ are the Fermi velocity and Fermi momentum, respectively. At filling fraction $\nu=5/2$, one has $k_F=1/\ell_B$ by the Luttinger's theorem \cite{luttinger1960,balram2015}. Depending on whether the filling fraction $\nu$ is either below or above $\nu_c$, the Fermi surface will form either a $p+ip$ pairing (the Pf state) or a $f-if$ pairing (the APf state), as we have explained in Sec.~\ref{Sec1}.

\begin{figure}[htbp]
  \centering
  \includegraphics[width=3.5in]{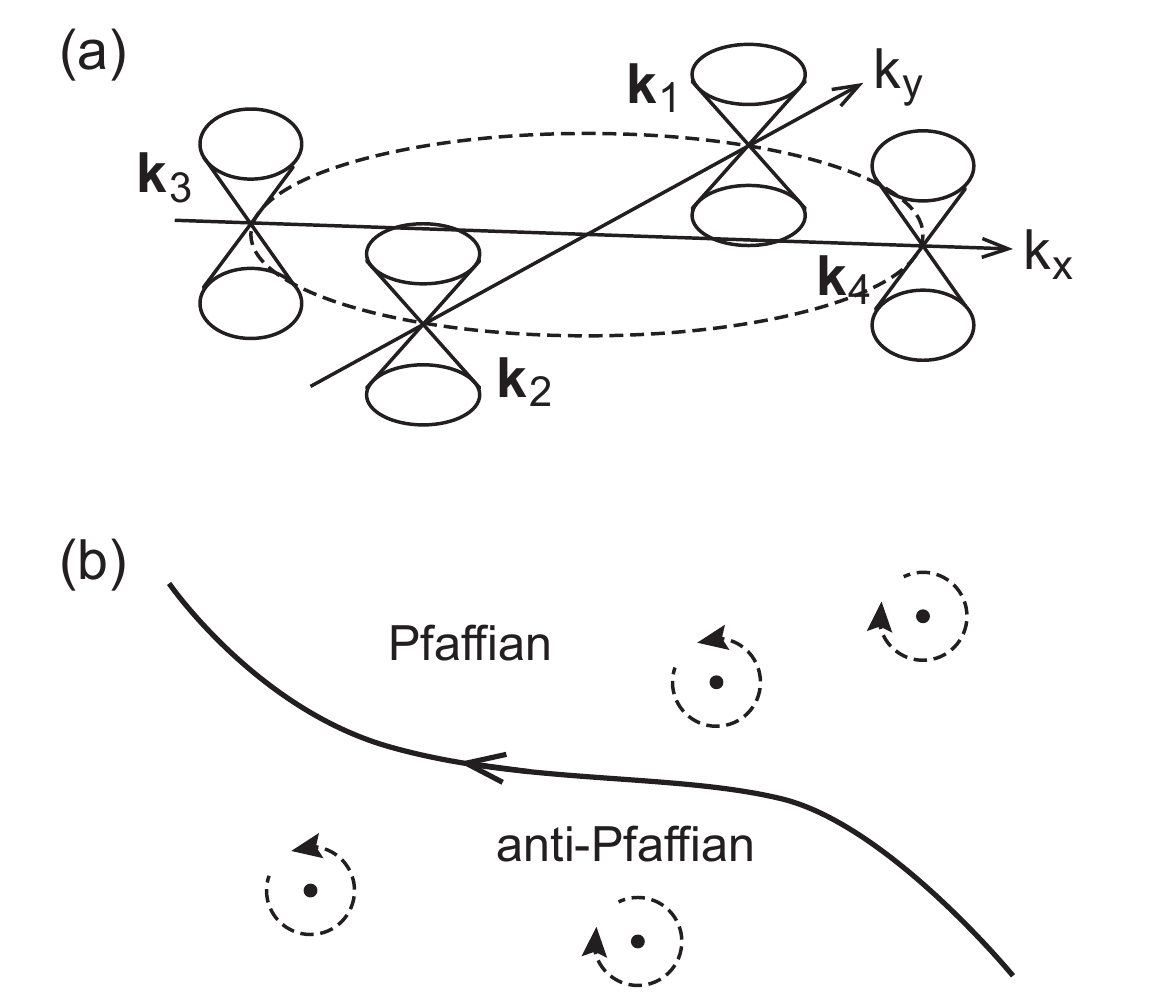}
  \caption{(a) The low energy BdG bands on the domain wall between Pf and APf states consist of four Dirac cones at $\mathbf{k}_i$ ($1\le i\le4$). (b)Top view of a domain wall. The pairing phase difference $\varphi_\Delta$ on the domain wall acquires fluctuations from random $\pm\pi$ flux vortices on the Pf or APf side and other small fluctuations.}\label{PfDW}
\end{figure}

In the disordered $\nu=5/2$ system consisting of Pf and APf domains, the pairing amplitude of composite fermions near a domain wall takes the form
\begin{equation}
\Delta(k)=\Delta_{\text{Pf}}(\mathbf{r})e^{i\theta_k}-\Delta_{\text{APf}}(\mathbf{r})e^{-3i\theta_k}\ ,
\end{equation}
where $\Delta_{\text{Pf}}(\mathbf{r})=\Delta_{\text{Pf}}$, $\Delta_{\text{APf}}(\mathbf{r})=0$ far away on the Pf side of the domain wall, and $\Delta_{\text{Pf}}(\mathbf{r})=0$, $\Delta_{\text{APf}}(\mathbf{r})=\Delta_{\text{APf}}$ far away on the APf side. The location of the domain wall is determined by $|\Delta_{\text{Pf}}(\mathbf{r})|=|\Delta_{\text{APf}}(\mathbf{r})|$. It is easy to show the BdG Hamiltonian of composite fermions is gapless at four momentums $\mathbf{k}_1=-\mathbf{k}_2=(-k_F\sin\varphi_\Delta,k_F\cos\varphi_\Delta)$ and $\mathbf{k}_3=-\mathbf{k}_4=(k_F\cos\varphi_\Delta,k_F\sin\varphi_\Delta)$ on the domain wall
(Fig.~\ref{PfDW}(a)), where $\varphi_\Delta$ is the phase difference between $\Delta_{\text{Pf}}(\mathbf{r})$ and $\Delta_{\text{APf}}(\mathbf{r})$. By defining $v_\Delta=(|\Delta_{\text{Pf}}(\mathbf{r})|+|\Delta_{\text{APf}}(\mathbf{r})|)/k_F$ and the mass term $m(\mathbf{r})=|\Delta_{\text{Pf}}(\mathbf{r})|-|\Delta_{\text{APf}}(\mathbf{r})|$, the low energy BdG Hamiltonian in the vicinity of $\mathbf{k}_i$ ($1\le i\le4$) takes the form of Dirac fermions similar to
Eq.~(\ref{BdG}). Following the argument in
Sec.~\ref{Sec2}, when the energy scale is below a certain value $\Lambda_2$ around the order of $\Delta_{\text{APf}}$ or $\Delta_{\text{Pf}}$, the four Dirac cones of the BdG band are well separated in momentum space, and one can define two emergent U(1) symmetries
\begin{equation}
f_\mathbf{k}\rightarrow e^{i\phi}f_\mathbf{k}\ , \quad f_{-\mathbf{k}}^\dag\rightarrow e^{i\phi}f_{-\mathbf{k}}^\dag
\end{equation}
for $\mathbf{k}$ near $\mathbf{k}_1=-\mathbf{k}_2$ and $\mathbf{k}$ near $\mathbf{k}_3=-\mathbf{k}_4$, respectively. The two emergent U(1) symmetries are independent of each other, so the total emergent symmetry is U(1)$\times$U(1).\footnote{
The U(1)$\times$U(1) symmetry is effectively the same as SO(2)$\times$SO(2).
The SO(2) symmetry is the rotational symmetry that bonds two chiral Majorana-Weyl fermions together.
We also denote U(1) symmetry to make connection to the A class of Cartan notations in electronic disordered systems.
} 
For a domain wall perpendicular to $\mathbf{n}=(\cos\alpha,\sin\alpha)$, the Dirac cone at $\mathbf{k}_1=-\mathbf{k}_2$ yields a complex chiral fermion $\psi_{k}^a$ with velocity $v_a=v(\varphi-\varphi_\Delta)$, while the Dirac cone at $\mathbf{k}_3=-\mathbf{k}_4$ yields a complex chiral fermion $\psi_{k}^a$ with velocity $v_b=v(\varphi-\varphi_\Delta+\pi/2)$, where $v(\varphi)$ is defined below
Eq.~(\ref{dispersion}). By rewriting the two complex chiral fermions into four chiral Majorana fermions as $\psi^a=\chi_1+i\chi_2$ and $\psi^b=\chi_3+i\chi_4$, we can write down a action of the domain wall:
\begin{equation}
S=\int dtdx \sum_{I=1}^4\chi_I(i\partial_t-iv_I\partial_x)\chi_I\ ,
\end{equation}
where $v_I=(v_a,v_a,v_b,v_b)$, $x$ is the spatial coordinate along the domain wall, and the origin of momentum $k$ is properly redefined. The emergent U(1)$\times$U(1) symmetry then corresponds to the SO(2) rotation between $\chi_1$ and $\chi_2$ and the SO(2) rotation between $\chi_3$ and $\chi_4$. Since the only energy scale within the first Landau level is the Coulomb interaction $e^2/\epsilon\ell_B$ where $\epsilon$ is the dielectric constant, we expect both $v_F$ and $v_\Delta$, and thus $v_a$ and $v_b$, to be of order $e^2/\epsilon\ell_B\hbar k_F\sim e^2/\epsilon\hbar$. For the reason, we expect the energy scale for the emergent U(1)$\times$U(1) symmetry to break down is
\bea
\Lambda_2\lesssim e^2/\epsilon\ell_B.
\eea

\subsection{Emergent SO(4) symmetry at energy scales $\Lambda<\Lambda_1$}\label{subsec32}


The above U(1)$\times$U(1) emergent symmetry is further enhanced in the presence of spatial or temporal fluctuations of the pairing phase difference $\varphi_\Delta=\mbox{arg}(\Delta_{\text{Pf}}/\Delta_{\text{APf}})$, which controls the velocities $v_a$ and $v_b$ of chiral Majorana fermions. Such fluctuations are generically present due to interactions, disorders and finite temperatures. To see the enlarging of the emergent symmetry, we can rewrite the action as
\begin{equation}\label{LEDGE}
S=\int dtdx \sum_{I=1}^4\left[\chi_I(i\partial_t-i\bar{v}\partial_x)\chi_I-i\delta v_I(x,t)\chi_I\partial_x\chi_I\right]\ ,
\end{equation}
where $\bar{v}=\frac{1}{2\pi}\int_0^{2\pi}v(\varphi) d\varphi$ is the mean edge state velocity, while $\delta v_I(x,t)\approx \frac{v_F-v_\Delta}{2}\cos[\varphi_\Delta(x,t)-\varphi_{0I}]$ is the anisotropic velocity of $\chi_I$, where $\varphi_{0I}$ is some constant. The correlation function of $\delta v_I$ is thus closely related to the correlation function of $\varphi_\Delta$, namely, $\langle\delta v_I(x)\delta v_I(x')\rangle\propto\langle \cos[\varphi_\Delta(x)-\varphi_\Delta(x')]\rangle$, while the mean value $\langle\delta v_I(x)\rangle=0$.

At zero disorder, the Berezinskii-Kosterlitz-Thouless (BKT) mechanism \cite{berezinski1971,kosterlitz1973} will yield either a power law or an exponential correlation $\langle e^{i\varphi_\Delta(x)-i\varphi_\Delta(x')}\rangle$ for temperature $T$ below or above the BKT transition temperature $T_{\text{BKT}}$, therefore
\[
\langle\delta v_I(x)\delta v_I(x')\rangle=
\begin{cases}
W_v|x-x'|^{-2\eta}\quad\ \ (T<T_{\text{BKT}}),\\
W_v\xi_v^{-1}e^{-|x-x'|/\xi_v}\  (T>T_{\text{BKT}}),
\end{cases}
\]
where $W_v$ characterises the strength of the correlation, $\eta>0$, and $\xi_v$ depends on interactions.
In the presence of disorders, random $\pm\pi$ flux vortices generically arise in both Pf and APf domains and are pinned by disorders. Due to these vortices, $\varphi_\Delta(\mathbf{r})$ at $\mathbf{r}$ on the domain wall becomes $\varphi_\Delta(\mathbf{r})=\varphi_\Delta^{0}+\sum_j\eta_j\theta(\mathbf{r}-\mathbf{r}_j)/2$, where $\varphi_\Delta^{0}$ are small fluctuations, $\mathbf{r}_j$ is the 2D spatial coordinate of the $j$-th vortex, $\theta(\mathbf{r})$ stands for the polar angle of $\mathbf{r}$ in polar coordinates, and $\eta_j=\pm1$ for $\pm\pi$ ($\mp\pi$) vortices on the Pf (APf) side of the domain wall. Given $\mathbf{r}_j$ and $\eta_j$ fully random, one would have
\[
\langle\delta v_I(x)\delta v_I(x')\rangle=W_v\delta(x-x')\ ,
\]
where $x$, $x'$ denote the 1D coordinate along the domain wall. In any case, one would find the scaling dimension of $W_v$ to be negative, which is either $d_{v}=-2\eta$ or $d_{v}=-1$. The renormalization group (RG) equation for $W_v$ is given by
\begin{equation}
\frac{dW_v}{d\log\Lambda}=-d_vW_v,
\end{equation}
where $\Lambda$ is the energy scale of the physical process under consideration. Therefore, $W_v$ is irrelevant at low energies, and the second term in
Eq.~(\ref{LEDGE}) can be dropped at low energies. The remaining action then has an enlarged emergent symmetry SO(4), i.e., rotations in the four dimensional space spanned by $\chi_I$ ($1\le I\le4$). This conclusion can also be drawn by examining correlations of $\delta v_I(x,t)$ in the time direction. The energy scale for the emergent SO(4) symmetry to break down is roughly
\bea
\Lambda_1\sim \bar{v}(\bar{v}^2/W_v^*)^{-1/d_v},
\eea
where $W_v^*$ is the value of $W_v$ at the ultraviolet (UV) cutoff. In general, we expect $\Lambda_1<\Lambda_2$, since the fluctuations of $\varphi_\Delta$ usually appear at longer distances compared to $\ell_B$.

When the $\nu=5/2$ FQH system forms random domain of Pf and APf states, an energy scale of disorder strength can be defined as $\Lambda=\bar{v}/\ell_0$ similar to Eq.~(\ref{Escale}), where $\ell_0$ is the size of a single Pf or APf domain, or the length of a link when formulated in network models \cite{chalker1988}. In particular, the inter-domain-wall tunneling area indicated by the dashed circle in Fig. \ref{RD} roughly has a length scale around $\ell_0$, therefore all the scattering angles $\alpha_I$ ($1\le I\le4$) are determined at the energy scale $\Lambda=\bar{v}/\ell_0$. The expected phase diagram of the system is then as shown in Fig. \ref{PD}. When $\Lambda<\Lambda_1$, the emergent SO(4) symmetry enforces all the four chiral Majorana fermions $\chi_I$ ($1\le I\le4$) to have the same scattering angles $\alpha_I$, so there is only a single phase transition from Pf state to APf state with respect to $\nu$. When $\Lambda_1<\Lambda<\Lambda_2$, the emergent symmetry is lowered to U(1)$\times$U(1), which only ensures $\chi_1$ is identical to $\chi_2$, and $\chi_3$ is identical to $\chi_4$. Therefore, there are two phase transitions with respect to $\nu$, and an intermediate $\kappa_{xy}=5/2$ phase arises. For strong disorder strengths $\Lambda>\Lambda_2$, only the $\mathbf{Z}_2$ symmetry $f_{\mathbf{k}}\rightarrow-f_\mathbf{k}$ remains, and the system is in the D symmetry class. The system may undergo four phase transitions with $\kappa_{xy}$ changing by $1/2$ each time, or it may enter the thermal metal phase. However, if the energy scale $\Lambda_2$ is comparable to the bulk gap of Pf and APf states, the analysis for $\Lambda>\Lambda_2$ may become invalid, since the concept of domains of Pf and APf states is no longer well-defined. As a result, the phases of $\kappa_{xy}=2$ and $3$ may not exist.

We note the phase diagram we predicted in Fig.~\ref{PD} differs from that predicted in Refs. \cite{wangc2017,mross2017} in some aspects. This is due to the low energy emergent symmetries we identified in the above, which are not fully considered in the two earlier papers. In particular, we expect the thermal metal phase to arise only beyond disorder energy scale $\Lambda_2$, since the system is effectively not in D class for $\Lambda<\Lambda_2$ where there is a U(1)$\times$U(1) or SO(4) emergent symmetry. Besides, in Refs. \cite{wangc2017,mross2017} the authors suggest the identification of the $\kappa_{xy}=2$, $5/2$ and $3$ phases with known FQH candidates, the $K=8$, PH-Pf and 113 states \cite{yang2013}. Here we shall not attempt to make such identifications in this work, as the three phases in the phase diagram Fig.~\ref{PD} do not exist at zero disorder, therefore may have bulk theories different from the above known FQH states at zero disorders. Instead, we leave this problem of phase identification for future studies.

Although we do not attempt to identify all phases in phase diagrams, we can still identify the spin TQFTs for these Pf/PH-Pf/APf states (in Appendix \ref{app:data}), which provides essential data to confirm the phases and their underlying topological orders in the future.

\section{Edge Theory of Disordered Pfaffian-antiPfaffian Domain Wall}\label{Sec4}

In this section, we give an understanding of emergent symmetries on the domain wall between Pf and APf states from the usual FQH edge theory formalism, where the physical meaning of disorders becomes clearer. We shall also show how the charged modes are gapped out, leaving a domain wall with four charge neutral chiral Majorana fermions. 
Previous analysis of a domain wall between Pf and APf states limited to Abelian sector can be found in Refs.~\cite{barkeshli2015,wan2016}, while here we carry out the analysis under the most possible physical assumptions in consistency with the experiments, also suitable for non-Abelian topological orders (See Appendix \ref{app:quantize-cboson}-\ref{app:Bootstrap}). 
Furthermore, we reveal the existence of a nonlocal topological degeneracy associated with multiple domain walls in Sec.~\ref{Sec5}.


\begin{figure}[htbp]
  \centering
  \includegraphics[width=4.5in]{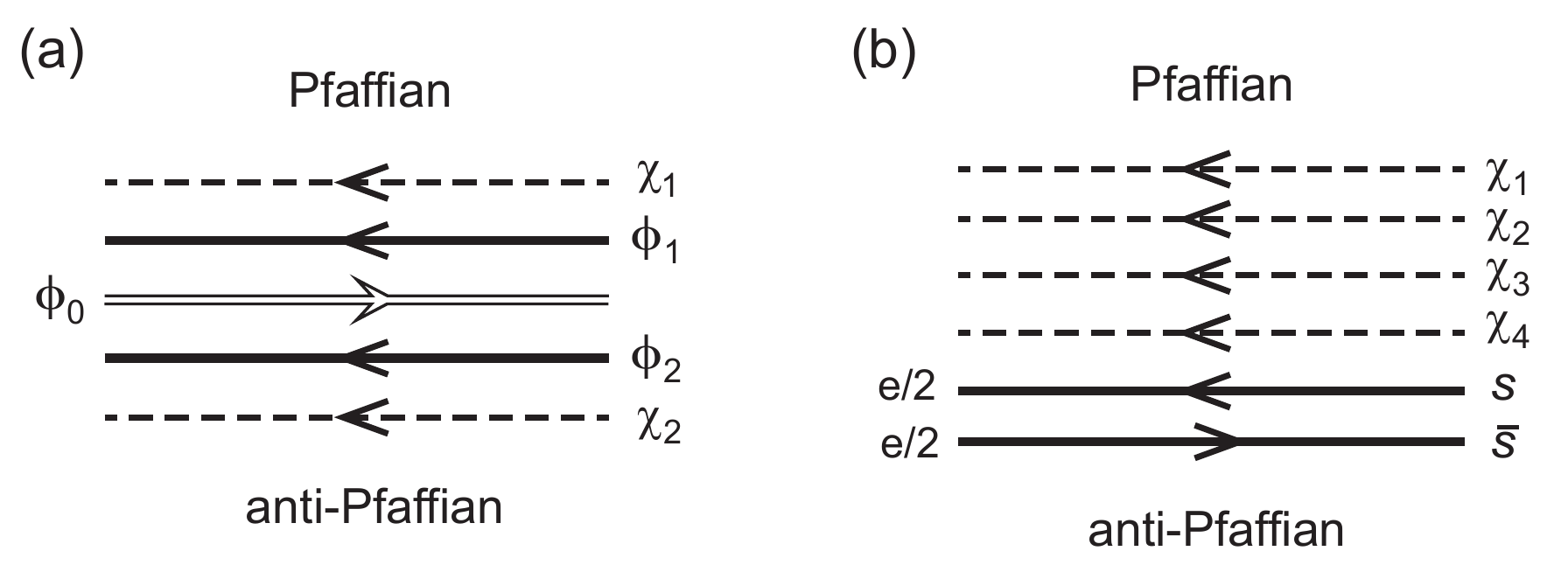}
  \caption{(a) Edge modes on the domain wall between Pf state and APf state before charge sector is gapped out, which include two left-moving chiral Majorana modes $\chi_1$ and $\chi_2$, two left-moving charge $e/2$ level-$2$ chiral bosons, and a right-moving charge $e$ level-$1$ chiral boson. (b) After a disorder induced recombination, the modes between Pf and APf states become four neutral chiral Majorana modes $\chi_I$ ($1\le I\le4$), and two opposite chiral boson modes of charge $e/2$ (semions) which gap themselves out due to interactions.}\label{Modes}
\end{figure}

The $1+1$D edge theory of the domain wall between Pf and APf states can be derived from their bulk theories (Appendices A and B) \cite{milo1996}, which has an effective action
\begin{equation}
\begin{split}
&S=S_\phi+S_\chi+S_{\text{int}}+S_d\ ,\\
&S_\phi=\int \frac{dtdx}{4\pi}\sum_{I,J=0}^2(K_{IJ}\partial_t\phi_I\partial_x\phi_J-V_{IJ}\partial_x\phi_I\partial_x\phi_J),\\
&S_{\chi}=\int dtdx \sum_{I=1}^2\chi_I(i\partial_t-iv_I\partial_x)\chi_I\ ,
\end{split}
\end{equation}
where the $K$ matrix and the chiral boson basis are given by
\begin{equation}
K=\left(\begin{array}{ccc}
2&0&0\\
0&-1&0\\
0&0&2\\
\end{array}\right),\
\left(\begin{array}{c}
\phi_1\\
\phi_0\\
\phi_2\\
\end{array}\right),
\end{equation}
$V_{IJ}$ is the velocity matrix which is positive definite and symmetric, and $S_{\text{int}}$ and $S_d$ denote the contributions of other interactions and disorders, respectively.
As shown in Fig.~\ref{Modes}(a), $\phi_1$ and $\phi_2$ are two left-moving charge $e/2$ level $2$ chiral bosons, $\phi_0$ is a right-moving charge $e$ level $1$ chiral boson which is identical to a complex fermion (electron), while $\chi_1$ and $\chi_2$ are two left-moving neutral chiral Majorana fermions with velocities $v_1>0$ and $v_2>0$, respectively. The edge modes $\phi_1$ and $\chi_1$ come from the Pf state, while $\phi_0$, $\phi_2$ and $\chi_2$ are from the APf state \cite{milo1996,lee2007,levin2007}. Note that Fig.~\ref{Modes}(a) is only illustrative and does not represent the physical positions of the five modes on the domain wall. The total electron charge density of the domain wall is given by $\rho=(e/2\pi)q_I\partial_x\phi_I$ (hereafter we assume repeated indices $I$ are automatically summed), where $q_I=(1,1,1)^\rT$ is the charge vector. Accordingly, the vertex operator $e^{i\phi_I}$ ($e^{-i\phi_I}$) creates (annihilates) a fractionalized quasiparticle of charge $\mathrm{t}_I=(K^{-1})_{IJ}q_J$ in units of $e$. Therefore, $V_{IJ}$ can be viewed as interactions between charge densities $\partial_x\phi_I/2\pi$ and $\partial_x\phi_J/2\pi$.

\subsection{Emergent U(1)$\times$U(1) symmetry at energy scales $\Lambda_1<\Lambda<\Lambda_2$}

For convenience of analysis, we first make an SL($3,\mathbb{Z}$) transformation $K'=U^{-1T}KU^{-1}$ to the $K$ matrix to a new basis $\phi_s=\phi_1$, $\phi_{\bar{s}}=\phi_2+\phi_0$ and $\phi_{n}=2\phi_2+\phi_0$, after which the $K$ matrix becomes (Appendix \ref{app:zero-GSD})
\begin{equation}
K'=\left(\begin{array}{ccc}
2&0&0\\
0&-2&0\\
0&0&1\\
\end{array}\right),\
\left(\begin{array}{c}
\phi_s\\
\phi_{\bar{s}}\\
\phi_n\\
\end{array}\right).
\end{equation}
Accordingly, the charge vector is transformed to $q'_I=(U^{-1}q)_I=(1,1,0)^\rT$, and the velocity matrix becomes $V'_{IJ}=(U^{-1T}VU^{-1})_{IJ}$. In this basis, the chiral bosons $\phi_s$ of level $2$ and $\phi_{\bar{s}}$ of level $-2$ have a correspondence with charge $e/2$ semions in the Pf bulk and charge $-e/2$ anti-semions in the APf bulk, respectively, while $\phi_n$ is a charge neutral level $1$ chiral boson identical to a complex fermion \cite{lee2007,levin2007}. The charged sector of the domain wall is therefore a nonchiral double-semion theory, while the neutral sector is fully chiral.

The nonchiral charged sector of the domain wall can be properly gapped out by interactions \cite{haldane1995,1212.4863WW,levin2013}. Generically, the allowed interaction terms on the domain wall must be bosonic (statistically trivial), non-fractionalized, nonchiral and charge conserving (Appendix \ref{app:Lstability}), which constrains the most relevant interaction term to be (Appendix \ref{app:zero-GSD})
\begin{equation}\label{Interaction}
S_{\text{int}}=\int dtdx \left[g\ e^{2i(\phi_s+\phi_{\bar{s}})}+h.c.\right]\ ,
\end{equation}
where $g$ is the coupling constant. Without loss of generality, hereafter we shall assume $g>0$ is real. The scaling dimension of the vertex operator $e^{2i(\phi_s+\phi_{\bar{s}})}=e^{2i(\phi_1+\phi_2+\phi_0)}$, and thus the scaling dimension of $g$, depend on the velocity matrix $V_{IJ}$. For instance, if the original velocity matrix $V_{IJ}$ is symmetric under exchange of $\phi_1$ and $\phi_2$, namely $V_{11}=V_{22}$ and $V_{01}=V_{02}$, the dimension of $g$ can be obtained as
$d_g=2-2\sqrt{\frac{2V_{00}+V_{11}+V_{12}-4V_{01}}{2V_{00}+V_{11}+V_{12}+4V_{01}}}.$
In particular, when $V_{01}=V_{02}>0$, we have $d_g>0$, and the interaction (\ref{Interaction}) is relevant and will gap out $\phi_s$ and $\phi_{\bar{s}}$. This is likely to be the case, since $V_{IJ}$ mainly comes from Coulomb repulsions between charges $\partial_x\phi_I$ and $\partial_x\phi_J$. However, we note that even if interaction (\ref{Interaction}) is perturbatively irrelevant, it can still gap out $\phi_s$ and $\phi_{\bar{s}}$ in the strongly interacting nonperturbative regime \cite{1212.4863WW}.
In either case, $\phi_s$ and $\phi_{\bar{s}}$ are gapped out by condensation of semion anti-semion pairs $s\bar{s}$ on the domain wall, which yields a nonzero expectation value
\begin{equation}\label{condense}
\langle e^{2i(\phi_s+\phi_{\bar{s}})}\rangle=1\ ,
\end{equation}
and minimizes the interaction energy in Eq.~(\ref{Interaction}). As we shall show in
Sec.~\ref{Sec5}, such a condensation also induces additional zero modes associated with the domain walls.

Given $\phi_s$ and $\phi_{\bar{s}}$ gapped out, the remaining edge theory only contains a level $1$ left-moving neutral chiral boson $\phi_n$ with a renormalized velocity $V'_{nn}$, and two left-moving chiral Majorana fermions $\chi_1$ and $\chi_2$ with velocities $v_1$ and $v_2$. As we argued in Sec.~\ref{Sec3}, the domain wall should already exhibit a U(1)$\times$U(1) emergent symmetry before we further add any disorder terms $S_d$.
We expect this to result from a combined approximate discrete $\Z_2$-symmetry $R_2\mathcal{C}$,
which is respected by the neighborhood of the domain wall, where
$R_2$ is the $\pi$ rotation (2-fold rotation) about a point on the domain wall, and $\mathcal{C}$ is the particle-hole transformation in the first Landau level \cite{girvin1984a}. Such a symmetry transformation exchanges $\phi_1$ with $\phi_2$ and $\chi_1$ with $\chi_2$, therefore ensures $v_1=v_2$.
The domain wall then has two independent U(1) symmetries, which are the U(1) phase rotation of $e^{i\phi_n}$ and
the SO(2) flavor rotation symmetry of Majorana modes $(\chi_1,\chi_2)^\rT$, respectively.

One may concern that our definition of the neutral mode $\phi_n=2\phi_2+\phi_0$ is asymmetric under exchange of $\phi_1$ and $\phi_2$, and does not respect the $R_2\mathcal{C}$ symmetry. In fact, due to the condensation in Eq.~(\ref{condense}), we have $2(\phi_s+\phi_{\bar{s}})=2(\phi_1+\phi_2+\phi_0)=0\ (\text{mod}\ 2\pi)$, or $2\phi_2+\phi_0=-2\phi_1-\phi_0\ (\text{mod}\ 2\pi)$. Therefore, $\phi_n$ becomes $-\phi_n$ up to multiples of $2\pi$ under $R_2\mathcal{C}$, and the $R_2\mathcal{C}$ symmetry is still respected.

A small violation of $R_2\mathcal{C}$ symmetry due to small distortions of the domain wall will not affect the U(1)$\times$U(1) emergent symmetry. This is because such perturbations only generates a random velocity difference $\delta v(x)=v_1(x)-v_2(x)$ between $v_1$ and $v_2$, which satisfies $\langle \delta v(x)\delta v(x')\rangle\propto\delta(x-x')$. By an RG analysis similar to that done in Sec.~\ref{subsec32}, one would find $\delta v(x)$ is an irrelevant perturbation.

The U(1)$\times$U(1) emergent symmetry will, however, be lost when the domain wall is severely bent or distorted, namely, when the size of a single domain $\ell_0$ is comparable to $\ell_B$. This leads to a breakdown energy scale of the emergent U(1)$\times$U(1) symmetry roughly round $\Lambda_2\lesssim e^2/\epsilon\ell_B$, which is in agreement with our estimation in Sec.~\ref{Sec3}.

\subsection{Emergent SO(4) symmetry at energy scales $\Lambda<\Lambda_1$}

In the presence of disorders, a disorder term $S_d$ arises from random backscattering of electrons on the domain wall. Such disorder may come directly from chemical potential fluctuations on the domain wall, or indirectly from charge $\pm e/4$ quasiparticles in Pf and APf domains forced to arise by the local filling fraction $\nu(\mathbf{r})$. There are three electron creation operators one could write down on the domain wall between Pf and APf: $\chi_1e^{2i\phi_1}$, $\chi_2e^{2i\phi_2}$ and $e^{-i\phi_0}$ \cite{lee2007,levin2007}, each of which has charge $e$ and fermionic statistics. The most relevant backscattering action can then be expressed as
\begin{equation}
S_d=\int dtdx\left[\xi_1(x)\chi_1e^{2i\phi_1+i\phi_0}+\xi_2(x)\chi_2e^{2i\phi_2+i\phi_0}+h.c.\right],
\end{equation}
where $\xi_I(x)$ are random functions satisfying $\overline{\xi_I^*(x)\xi_J(x)}=W_I\delta_{IJ}\delta(x-x')$ ($I=1,2$) and have zero means. Since $\phi_s$ and $\phi_{\bar{s}}$ are gapped out, one can then rewrite the vertex operators in the new basis as $e^{2i\phi_1+i\phi_0}=e^{2i(\phi_s+\phi_{\bar{s}})}e^{-i\phi_n}$ and $e^{2i\phi_2+i\phi_0}=e^{i\phi_n}$, and then replace $e^{2i(\phi_s+\phi_{\bar{s}})}$ by its expectation value $1$. The disorder term then becomes
\begin{equation}
S_d=\int dtdx\left[\xi_1(x)\chi_1e^{-i\phi_n}+\xi_2(x)\chi_2e^{i\phi_n}+h.c.\right].
\end{equation}
The scaling dimension of the disorder strengths $W_1$ and $W_2$ can be easily found to be $d_1=d_2=1/2$, therefore the disorder term $S_d$ is a relevant perturbation. The problem can be solved by redefining the chiral boson mode $\phi_n$ as two chiral Majorana fermions $e^{i\phi_n}=\chi_3+i\chi_4$. The disorder term $S_d$ can then be rewritten as
\[S_d=\int dtdx\ \omega_a(x)\chi^\rT T^a\chi\ ,\]
where $\chi=(\chi_1,\chi_2,\chi_3,\chi_4)^\rT$ are the Majorana fields, $T^a$ ($1\le a\le6$) are the SO(4) group generators, and we have defined $\omega_a(x)=(0,\mbox{Re} \xi_1(x),\mbox{Im}\xi_1(x),\mbox{Re} \xi_2(x),\mbox{Im}\xi_2(x),0)$. By doing a unitary transformation $\chi(x)=O(x)\chi'(x)$ where $O(x)=P \exp\left(-\frac{i}{\bar{v}}\int_{-\infty}^{x}dx'\omega_a(x')T^a\right)$, one can eliminate $S_d$ in the action \cite{kane1994,lee2007,levin2007}, with the rest action takes the form
\begin{equation}
S=\int dtdx\left[\chi'^\rT(i\partial_t-i\bar{v}\partial_x)\chi'-i\chi'^\rT\delta v'(x)\partial_x\chi' \right]\ ,
\end{equation}
where $\bar{v}=(v_1+v_2+2V'_{nn})/4$ is the mean velocity, while $\delta v'(x)=O(x)^\rT(v-\bar{v})O(x)$ is the randomly rotated velocity anisotropy, with the original velocity matrix $v=\text{diag}(v_1,v_2,V'_{nn},V'_{nn})$. The correlation of $\delta v'(x)$ is short-ranged, which can be estimated to be $\text{tr}\overline{\delta v'(x)\delta v'(x')}\sim e^{-(W_1^*+W_2^*)|x-x'|/\bar{v}^2}\sim W_v\delta(x-x')$, where $W_I^*$ is the UV value of the disorder strengths $W_I$ ($I=1,2$). It is then straightforward to see the scaling dimension of $W_v$ is $d_v=-1<0$, so the $\delta v'(x)$ term is irrelevant. As a result, the action $S$ gains an SO(4) emergent symmetry of rotation of four chiral Majorana fermions $\chi'_I$ at low energies. The breakdown energy scale of the SO(4) emergent symmetry can be estimated as $\Lambda_1\sim (W_1^*+W_2^*)/\bar{v}$. We thus reach the same conclusion as that in Sec.~\ref{Sec3}.

\section{Non-Local Topological Degeneracy and Heat Capacity}\label{Sec5}

We have been focusing on the neutral chiral Majorana fermion sector of the domain walls between Pf and APf so far. In this section, we shall show the charged non-chiral double-semion sector ($\phi_s$ and $\phi_{\bar{s}}$), although gapped out, still contributes a ground state degeneracy (GSD), which scales up exponentially with respect to the number of 1+1D compact domain walls.\footnote{By a 1+1D compact domain wall (or interface), we mean that
a 1D spatial closed domain wall, homotopic equivalent to a spatial $S^1$ circle, without 0+1D ends.} 
Thus, these ground states participate in the low energy physics such as the specific heat.

\begin{figure}[t!] 
  \centering
  \includegraphics[width=4.4in]{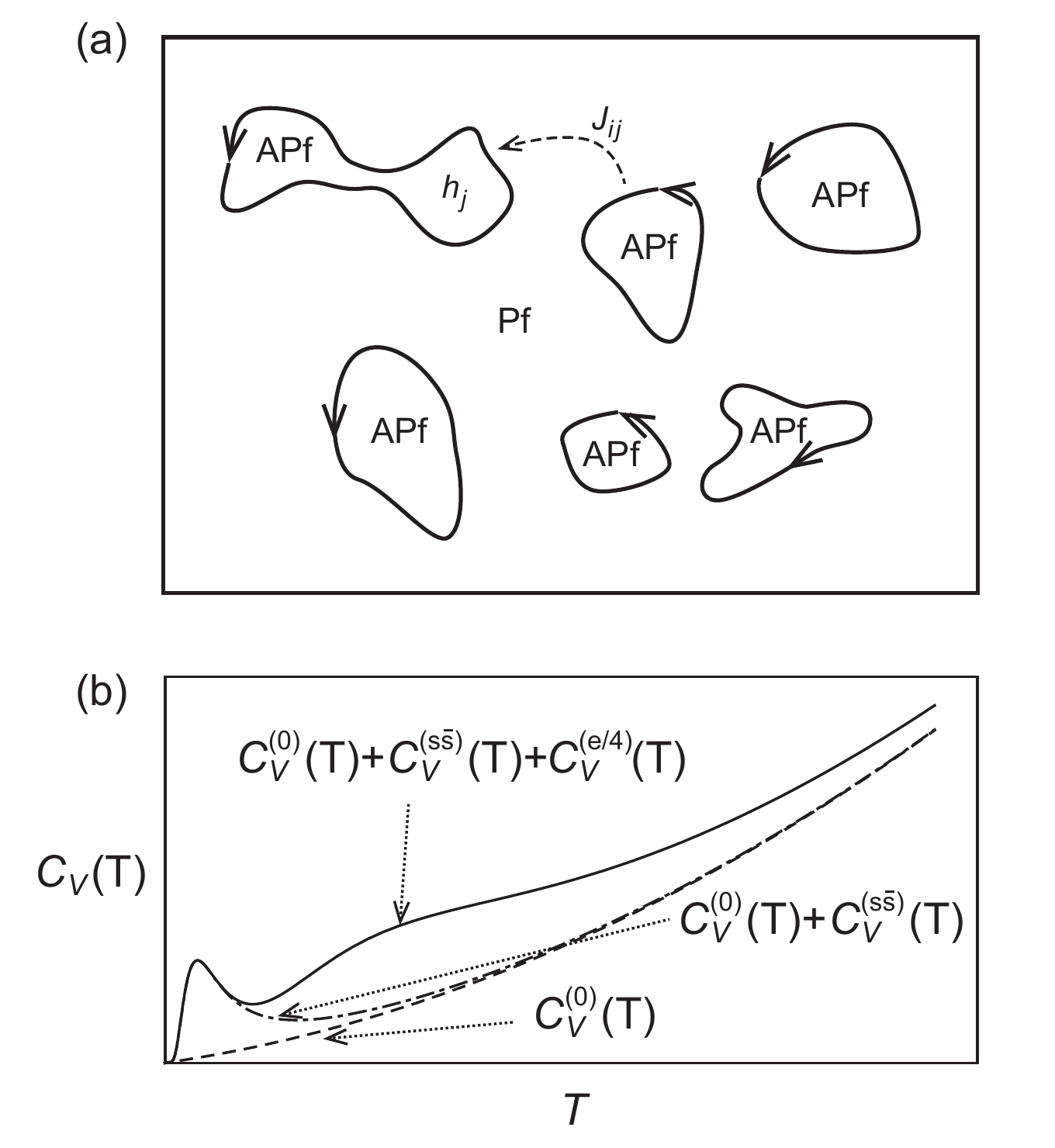}
  \caption{(a) The double-semion sector of $N_D$ domain walls between Pf and APf states contribute a GSD $2^{N_D-1}$. The GSD are lifted into low energy modes by the double semion hoppings $J_{ij}$ and local energy splittings $h_i$, which are nonzero when the sizes and distances of the domain walls are finite. (b) Heat capacity of the system, which exhibit two peaks (solid line) due to contributions of the double-semion GSD $\left(C_V^{(s\bar{s})}(T)\right)$ and charge $e/4$ non-Abelian quasiparticles $\left(C_V^{(e/4)}(T)\right)$.}\label{figGSD}
\end{figure}

The GSD due to gapped $\phi_s$ and $\phi_{\bar{s}}$ on compact domain walls between Pf and APf can be most easily seen from the condensation condition of
Eq.~(\ref{condense}), which is required by the minimization of interaction energy (\ref{Interaction}). On each compact domain wall,
Eq.~(\ref{condense}) is satisfied by two
saddle point solutions\footnote{More precisely, here we mean the zero modes $\phi_{0,s}$ and $\phi_{{0,\bar{s}}}$
can be pinned
down thus localized at strong coupling
due to non-perturbative effects.
Their conjugate momenta, winding modes
$P_{\phi,s}$ and $P_{\phi,\bar s}$,
 actually fluctuate and can hop in the quantized winding mode Hilbert space.
See Appendix \ref{app:zero-GSD} on the derivation of zero modes/winding modes counting and GSD calculations.}
\begin{equation}
\phi_s+\phi_{\bar{s}}=0\ \text{or}\ \pi\ (\text{mod}\ 2\pi)\ .
\end{equation}
One can understand $\phi_s+\phi_{\bar{s}}$ as the phase angle of the semion anti-semion pairing order parameter. When there is just one compact domain wall, $\phi_s+\phi_{\bar{s}}=0$ and $\pi$ describe the same physical ground state, since they only differ by a global redefinition of $\phi_s\rightarrow\phi_s+\pi$. When there are two compact domain walls far apart from each other, the angle $\phi_s+\phi_{\bar{s}}$ on each domain wall can be either $0$ or $\pi$. Again, a global $\pi$ shift of $\phi_s+\phi_{\bar{s}}$ on both domain walls does not change the physical state. The phase difference of $\phi_s+\phi_{\bar{s}}$ between the two domain walls, however, is physically meaningful, which can be either $0$ or $\pi$. This is analogous to the phase difference of two superconductors in a Josephson junction. Therefore, the system with two domain walls has ${\text{GSD}}=2$ degenerate ground states in the double-semion sector corresponding to $\phi_s+\phi_{\bar{s}}$ phase differences $0$ and $\pi$, respectively.

In general, when there are $N_D$ compact domain walls far away from each other (Fig.~\ref{figGSD}(a)), each domain wall could have $\phi_s+\phi_{\bar{s}}=0$ or $\pi$, and a global $\pi$ shift on all domain walls does not alter the ground state. Therefore, there are ${\text{GSD}}=2^{N_D-1}$ degenerate ground states in the double-semion sector. A more detailed derivation of the GSD can be found in Appendix \ref{app:zero-GSD} and Ref. \cite{1212.4863WW}.

By viewing $\tau_z^{j}=e^{i(\phi_s+\phi_{\bar{s}})}|_{j}=\pm1$ on the $j$-th domain wall as an Ising spin, the double-semion sector GSD can
be approximately
modeled as the degrees of freedom of $n$ Ising spins up to a global flip of all spins \cite{iadecola2014}. When quantum fluctuations of $\phi_s+\phi_{\bar{s}}$ are considered, the Ising spins acquire a local energy splitting $h_j\propto e^{-L_j/\xi_s}$ and a spin interaction between domain walls $J_{ij}\propto e^{-L_{ij}/\xi_s}$, where $L_j$ is the size of the $j$-th domain wall, and $L_{ij}$ is the distance between the $i$-th and $j$-th domain walls. Physically, $h_j$ is the tunneling between potential wells at $\phi_s+\phi_{\bar{s}}=0$ and $\pi$, while $J_{ij}$ is the hopping of semion anti-semion pairs $s\bar{s}$ between domain walls $i$ and $j$. The $N_D$ Ising spins then form a $2+1$D transverse field Ising model with an effective Hamiltonian
\begin{equation}
H_{s\bar{s}}=\sum_{j=1}^{N_D} h_j\tau_x^{j}+\sum_{1=i<j}^{N_D} J_{ij}\tau_z^{i}\tau_z^{j}\ ,
\end{equation}
subject to a restriction $\prod_{j=1}^{N_D}\tau_x^j=1$ which eliminates the global spin flip redundancy. These charge neutral low energy modes together with chiral Majorana fermions on the domain walls constitute the effective theory of the system under disorders.

One effect of $H_{s\bar{s}}$ is its contribution to heat capacity at low temperatures, as noted by \cite{iadecola2014}.
For instance, in the limit $J_{ij}\rightarrow0$, the heat capacity at temperature $T=1/k_B\beta$ due to $H_{s\bar{s}}$ is
\begin{equation}
C_v^{(s\bar{s})}=k_B\beta^2\frac{\partial^2\ln\left(\text{tr}e^{-\beta H_{s\bar{s}}}\right)}{\partial\beta^2}=\sum_{j=1}^{N_D}\frac{h_j^2}{k_BT^2}\ \mbox{sech}^2\left(\frac{h_j}{k_BT}\right).
\end{equation}
The total heat capacity is then given by $C_v(T)=C_v^{(0)}(T)+C_v^{(s\bar{s})}(T)$, where $C_v^{(0)}(T)\approx\gamma_1T+\gamma_{2}T^2$ is the contribution from chiral Majorana fermions and phonons. In particular, the GSD Hamiltonian $H_{s\bar{s}}$ contributes a peak at $k_BT$ around the order of magnitude of $h_j$ and $J_{ij}$, as shown in Fig.~\ref{figGSD}(b), and the peak height is proportional to the number of domain walls $N_D$. Since both $h_j$ and $J_{ij}$ are exponentially small, the temperature of the peak is expected to be low.
Furthermore, when the system is $n$ electrons away from filling fraction $\nu=5/2$,
there will be $N_{e/4} \approx 4n$ non-Abelian quasiparticles of charge $e/4$, which contribute another factor ${2}^{(N_{e/4}/2)-1}$ to the GSD. When the hopping and mobility among these quasiparticles are taken into account, they will also contribute a heat capacity $C_v^{(e/4)}(T)$ peaked at $k_BT$ around the energy scale of their hopping. Since the $e/4$ quasiparticles are point-like and free to move, we expect their hopping to be much larger than that between two domain walls. As a result, the total heat capacity $C_v(T)=C_v^{(0)}(T)+C_v^{(s\bar{s})}(T)+C_v^{(e/4)}(T)$ will exhibit a lower peak due to double-semion GSD and a higher peak due to charge $e/4$ non-Abelian quasiparticles, as illustrated by the solid line in Fig.~\ref{figGSD}(b).
This provides a way to tell whether the $\nu=5/2$ FQH system in the experiment is formed by a large number of random Pf and APf domains.

In addition, these almost zero energy modes will also contribute to the longitudinal thermal conductance $\kappa_{xx}$. They will however not affect $\kappa_{xy}$, since they have a nonchiral nature and are decoupled from chiral Majorana fermions on the domain walls. Therefore, all our earlier conclusions about $\kappa_{xy}$ remain unchanged.

Lastly, we remark that when there are closed Pf$\mid$APf$\mid$Pf domain walls, namely, narrow Corbino rings of APf in the background of Pf, such domain walls may contain gapped non-Abelian sector that contributes to the GSD differently (see Appendix \ref{app:Bootstrap}, and Ref.~\cite{1408.6514LWW}).
In practice, such domain walls may rarely occur. If they are present for some reason, the above physical picture still holds, except that the scaling behavior of the GSD with respect to the number of domain walls may be different.

\section{Conclusion and Discussion}

In conclusion, under the assumption that the disordered $\nu=5/2$ FQH system is formed by random domains of Pf and APf states, we studied the low energy theory and emergent symmetries of the domain walls separating Pf and APf, and proposed a possible phase diagram for the system with respect to disorders.

The study is aimed at explaining the thermal Hall conductance $\kappa_{xy}=5/2$ in the $\nu=5/2$ FQH state as revealed by the recent experiment \cite{banerjee2017}, while a clean Pf (APf) state has $\kappa_{xy}=7/2$ ($\kappa_{xy}=3/2$). When random domains of Pf and APf are formed in the system, each domain wall contains four chiral Majorana fermion modes, and it has been suggested that the percolation of these chiral Majorana fermions on the domain walls in the bulk may stabilize a $\kappa_{xy}=5/2$ phase under disorders \cite{wangc2017,mross2017}. In this work, we studied in details the domain wall edge theory between Pf and APf. In particular, we show the theory has an SO(4) emergent symmetry at low energy scales $\Lambda<\Lambda_1$, which is lowered to U(1)$\times$U(1) emergent symmetry at intermediate energy scales $\Lambda_1<\Lambda<\Lambda_2$, and is then lowered to the $\mathbb{Z}_2^F$ fermion parity symmetry at high energies $\Lambda>\Lambda_2$.
The energy scale dependent emergent symmetries enable us to propose a more specific phase diagram Fig. \ref{PD} regarding the mean filling fraction $\nu$ and the disorder energy scale $\Lambda=\bar{v}/\ell_0$, where $\bar{v}$ is the mean velocity of chiral Majorana fermions on the domain walls, and $\ell_0$ is the rough size of a single Pf or APf domain. When $\Lambda<\Lambda_1$, the system involves only a single phase transition from Pf state to APf state with respect to $\nu$, as forced by the emergent SO(4) symmetry. When $\Lambda_1<\Lambda<\Lambda_2$, the U(1)$\times$U(1) emergent symmetry allows two phase transitions with respect to $\nu$, and an intermediate $\kappa_{xy}=5/2$ phase arises between Pf and APf phases. Under strong disorders $\Lambda>\Lambda_2$, the system may undergo four phase transitions with $\nu$, each of which changes $\kappa_{xy}$ by $1/2$, or the system may become a gapless thermal metal with non-quantized $\kappa_{xy}$.

We also show the gapped double semion sector of the domain wall between Pf and APf contributes a GSD $2^{N_D-1}$ when there are $N_D$ compact domain walls. When the $N_D$ domain walls are small and spatially close, these GSD may be lifted to neutral low energy modes, which contributes a peak to the heat capacity $C_v$ and increases the longitudinal thermal conductance $\kappa_{xx}$. This may be used to detect whether the system consists of large number of domain walls in experiments.

However, we emphasize that the explanation for $\kappa_{xy}=5/2$ is still unsettled. The domain wall energy (e.g., positive or negative) between Pf and APf may significantly affect whether random Pf and APf domains are favored by the system, and such energy calculations has not been done yet. It is also important to have the tunability of disorders in the experiment to verify such a random Pf and APf domain percolation theory. Besides, whether the disordered $\kappa_{xy}=3$, $5/2$ and $2$ phases can be identified with the known $K=8$, PH-Pf and $113$ FQH states in the clean limit should be further studied.

Below we comment final theoretical remarks and future directions:
\begin{enumerate}

\item \emph{Emergent symmetry v.s. spontaneous symmetry breaking}: We remark that in our study, as the disorder energy scale decreases/increases, the emergent symmetry is enhanced/reduced.
This phenomenon is \emph{opposite} to tuning energy scales in spontaneous symmetry breaking (SSB) systems.
In the more familiar SSB systems, such as ferromagnets and superfluids,
when the temperature scale decreases/increases, the symmetry is broken/restored.
Another SSB example is quantum chromodynamics,
when the temperature scale decreases/increases, the chiral symmetry and time reversal symmetry can be broken/restored.

\item \emph{Field theory and topological order data}:
We refine the information in literatures in order to identify the fermionic spin topological quantum field theories (fTQFTs)
for Pf/PH-Pf/APf states as well as 331, $K = 8$, 113 FQH states,
and their bosonic TQFT (bTQFT) sectors. These data, including TQFTs and edge theories,
are gathered together in Appendix \ref{app:data}.\footnote{
\cblue{By spin TQFT (fTQFT), we mean that defining such TQFT requires a manifold with spin structures.}
}
We find that their spin fTQFTs can be obtained as
U(1)$_{\pm 8}$ Chern-Simons theory with additional $\mathcal{\nu} \in \Z_8$-class TQFT, subject to a $\Z_2$-quotient constraint,
which we may denote as $\frac{\mathrm{U(1)}_{\pm 8}\times (\mathcal{\nu} \in \Z_8 \text{TQFT})}{\Z_2}$,
with other details in Appendix \ref{app:data}.
Our TQFT notations follow \cite{1602.04251SW, 1612.09298PWY}.
We also list down their modular SL$(2,\Z)$ data $\cS, \cT$ matrices and chiral central charges $c_-$ for bosonic sectors,
which are modular theories in terms of category theory.
The fTQFT however is pre-modular,
one can extend our analysis to full fTQFTs based on the methods of \cite{1602.04251SW, 1612.09298PWY}.
Since global symmetries play important roles for these topological orders (TQFTs),
in modern terminology, we provide the essential data to
examine their symmetry enriched topologically ordered (SET) states.
It will be worthwhile to revisit related Pf/PH-Pf/APf systems, considering the interplay of symmetries,
in our framework, for a larger families of related SET states.

\item \emph{Non-perturbative gapping criteria  analysis}:
Based on the methods of \cite{1212.4863WW, 1408.6514LWW},
we provide non-perturbative gapping criteria on the domain walls of related quantum Hall systems.
Even though certain interaction terms may be irrelevant in a perturbative RG sense, they can still affect domain wall properties at non-perturbative strong couplings.
We are aware that the precise identifications of Pf/PH-Pf/APf and their domain walls require spin fTQFTs.
We note that our methods in Appendix \ref{app:Lstability} and \ref{app:zero-GSD} are applicable to Abelian bosonic/fermionic TQFTs.
For the convenience of bootstrapping the topological domain wall based on modular data, in
Appendix  \ref{app:Bootstrap},
we focus on the \emph{non-Abelian} bTQFT sectors.
At this moment, we believe that domain walls of bTQFT capture the major interesting effects.
The further investigation of the domain wall theories of full fTQFTs is left for the future work.

\item \emph{Abelian topological domain walls v.s. non-Abelian anyons}:
In the main text, we mainly focus on the gapped double-semion domain wall sectors, which cause additional highly degenerate ground states
in the Pf$\mid$APf percolation picture (Fig.~\ref{figGSD} (a)).
This is an astonishing long-range entanglement phenomenon.
In contrast, the usual symmetry-breaking degeneracy scales as a product of individual degeneracy on each symmetry-breaking domain wall.
As stressed in  \cite{1212.4863WW}, the domain wall topological degeneracy contains non-local
long-range entangled information between distinct domain walls (Appendix \ref{app:zero-GSD}/\ref{app:Bootstrap}).

The double-semion sectors are purely Abelian topological orders.
Even though many Abelian anyons (trapped by vortices or defect punctures) on a flat substrate has only a unique ground state (GSD=1),
the GSD of many topological domain walls of Abelian TQFT can still grow \emph{exponentially}, respect to the number of anyon insertions, even on a flat substrate.
The later GSD phenomenon from Abelian TQFT's topological domain walls is more similar to that of \emph{non-Abelian} anyons instead.

\item
\emph{Exotic non-Abelian topological domain walls}:
In Appendix \ref{app:nAbDW}
and \ref{app:nAb-GSD}, we include
additional solutions of further exotic non-Abelian domain walls obtained from Pf$\mid$APf$\mid$Pf interface.
The bosonic non-Abelian sector part of interface can be described by tunneling data between two sets of Ising $\times$ SU(2)$_{2}$ TQFTs.
Ising and SU(2)$_{2}$ TQFTs consist of a set of anyon $\{1, \sigma_ \mathcal{V}, \psi_ \mathcal{V}\}$,
labeled by $\mathcal{V}= 1, 3 \in \Z_8$ respectively in terms of 2+1D $\Z_8$-class TQFTs.
For example, we find one intriguing set of interesting tunneling data of Pf$\mid$APf$\mid$Pf as
$1 \oplus \psi \psi_3 \leftrightarrow 1 \oplus \psi \psi_3$,
  $\psi \oplus \psi_3 \leftrightarrow  \psi \oplus \psi_3$ and $2 \sigma \sigma_3 \leftrightarrow  2 \sigma \sigma_3$.
It is tempting to speculate their implications of junction-like device.
The tunneling data may be relevant to testable experimental settings.

\end{enumerate}

\section{Acknowledgments}
We thank Jie Wang and Michael Zaletel for helpful information and discussions.
JW thanks Nathan Seiberg and Edward Witten for conversations,
and thanks collaborators for a previous collaboration on \cite{1612.09298PWY}.
B.L. is supported by Princeton Center for Theoretical Science at Princeton University.
J.W. gratefully acknowledges the support from Corning Glass Works Foundation Fellowship, NSF Grant PHY-1314311
and Institute for Advanced Study.

\appendix

\newpage
\begin{center}
{\bf\LARGE{Appendix}}
\end{center}
\section{Data of QFT/Topological Order: Pf/APf/PH-Pfaffian states}
\label{app:data}

\begin{table}[!h]	
\footnotesize
    \hspace{-.0em}\begin{tabular}{ | c | c | c | c | c | c | c | }
    \hline
    State &
    \begin{minipage}[c]{2.3in}
    2+1D TQFT description;\\
    1+1D Edge modes
    \end{minipage}
    & $\begin{array}{c}(c_L,-c_R); c_-\\ (q_L,q_R);  (\rt_L,\rt_R); \\ \sigma_{xy} \text{ in $(\frac{e^2}{h})$};\\
    \kappa_{xy} \text{ in $(\frac{\pi^2 k_B^2}{3 h}T)$.}  \end{array}$  & 
    $\GSD_{T^2}$
     &
         \begin{minipage}[c]{1.1in}
     Description: \\
     BCS pairing \\
     of fermions
         \end{minipage}
     \\ \hline
\begin{minipage}[c]{.45in}
(I) \\
Pfaff\\[2mm]
(Moore-\\
Read)\\[2mm]
$\kappa_{xy}=\frac{7}{2}$
\end{minipage}
&
\begin{minipage}[c]{2.45in}
fTQFT (spin): \cblue{$\rU(2)_{2,-8} \times \rU(1)_1$ CS}\\
$\simeq  (\rSU(2)_2 \times \rU(1)_{-8})/ \Z_2 \times \rU(1)_1$ CS\\
$\simeq$ $(\rU(1)_{8} \times {\text{Ising}})/\Z_2$ CS\\
$\simeq$ Moore-Read state. \\[2mm]
bTQFT: Semion $\times$ Ising +
$\frac{q}{2 \pi} \int A da$.\\
Anyons: $\{1, s\} \otimes \{1, \sigma, \psi\} $.\\[2mm]
$S_{\text{edge, Pf}}=$
$\int_L dt dx \frac{2}{4 \pi} \partial_{x} \phi_s (\partial_{t} \phi_s
- v_{1}  \partial_{x} \phi_s )$\\
$+\chi_L (i \partial_t - i v_L \partial_x) \chi_L
+ \frac{q}{2 \pi} \epsilon^{\mu \nu} A_\mu \partial_{\nu} \phi_s \mid_{q=e 1}$.\\[0mm]
\end{minipage}
 &
 $\begin{array}{c}
 (\frac{3}{2},0); \frac{3}{2}\\ (1+0,0)_q; \\(\frac{e}{2} \cdot 1+0,0)_{\rt}; \\
\sigma_{xy}=\frac{1}{2};\\
\kappa_{xy}=c_-+2=\frac{7}{2}
 \end{array}$
 &
 \begin{minipage}[c]{.55in}
6 (b+f)
\end{minipage}
&
  \begin{minipage}[c]{1.15in}
  Cooper pairs:\\
  $p+ip$-wave of CF\\
 ($d+id$-wave of CDF)
 \end{minipage}
\\ \hline
\begin{minipage}[c]{.45in}
(II)\\
PH-Pfaff
\\[2mm]
$\kappa_{xy}=\frac{5}{2}$
\end{minipage}
&
\begin{minipage}[c]{2.45in}
fTQFT (spin): $(\rU(1)_{8} \times \overline{\text{Ising}})/\Z_2$ CS.\\[2mm]
bTQFT:  Semion $\times$ $\overline{\text{Ising}}$+
$\frac{q}{2 \pi} \int A da$.  \\
Anyons: $\{1, s\} \otimes \{1, \bar \sigma, \bar \psi\} $,\\[2mm]
$S_{\text{edge, PH-Pf}}=$ $\int_L dt dx \frac{2}{4 \pi} \partial_{x} \phi_s (\partial_{t} \phi_s
- v_{1}  \partial_{x} \phi_1 )$\\
$+\int_R dt dx \chi_R (i \partial_t + i v_R \partial_x) \chi_R$\\
$+\int dt dx \frac{q}{2 \pi} \epsilon^{\mu \nu} A_\mu \partial_{\nu} \phi_s \mid_{q=e 1}$.\\
\end{minipage}
 &
 $\begin{array}{c}(1,\frac{-1}{2}); \frac{1}{2}\\ ( 1,0 )_q; \\
 (\frac{e}{2} \cdot 1,0 )_{\rt};\\
 \sigma_{xy}=\frac{1}{2};\\
 \kappa_{xy}=c_-+2=\frac{5}{2}.
 \end{array}$
&
 \begin{minipage}[c]{.55in}
6 (b+f)
\end{minipage}
&
  \begin{minipage}[c]{1.in}
  Cooper pairs:\\
  $p-ip$-wave of CF\\
 ($s$-wave of CDF)
 \end{minipage}
\\ \hline
\begin{minipage}[c]{.45in}
(III) \\
A-Pfaff\\[2mm]
$\kappa_{xy}=\frac{3}{2}$
\end{minipage}
&
\begin{minipage}[c]{2.45in}
fTQFT (spin): \cblue{$\rU(2)_{-2,8} $ CS}\\
$\simeq  (\rU(1)_{8} \times \rSU(2)_{-2} )/ \Z_2$ CS.\\[2mm]
bTQFT: Semion $\times$  $\rSU(2)_{-2}$ CS +
$\frac{q}{2 \pi} \int A da$. \\
Anyons: $\{1, s\} \otimes \{1,  \bar \sigma_3, \bar \psi_3\} $,\\[2mm]
$S_{\text{edge, APf (i)}}=$\\
$\int dt dx \frac{1}{4 \pi}
(
\left(
\begin{array}{cc}
 1 & 0 \\
0 & -2 \\
\end{array}
\right)_{IJ} \partial_{x} \phi_I \partial_{t} \phi_J$\\
$- V_{IJ} \partial_{x} \phi_I  \partial_{x} \phi_J )$\\
$+\int dt dx \frac{q_I}{2 \pi} \epsilon^{\mu \nu} A_\mu \partial_{\nu} \phi_I \mid_{q=e (\begin{smallmatrix} 1 \\1 \end{smallmatrix})}$\\
$+\int_R dt dx \chi_R (i \partial_t +i v_R \partial_x) \chi_R$, \\
\text{ here $\phi_I \equiv (\phi_0, \phi_2)$.}\\
$\to$
$S_{\text{edge, APf (ii)}}=$\\
$\int dt dx \frac{1}{4 \pi}
(
\left(
\begin{array}{cc}
 2 & 0 \\
0 & -1 \\
\end{array}
\right)_{IJ} \partial_{x} \phi_I \partial_{t} \phi_J$\\
$- V_{IJ} \partial_{x} \phi_I  \partial_{x} \phi_J )$\\
$+\int dt dx \frac{q_I}{2 \pi} \epsilon^{\mu \nu} A_\mu \partial_{\nu} \phi_I \mid_{q=e (\begin{smallmatrix} 1 \\0 \end{smallmatrix})}$\\
$+\int_R dt dx \chi_R (i \partial_t +i v_R \partial_x) \chi_R$, \\
\text{ here $\phi_I \equiv (\phi_s, \phi_n)$.}\\
$\to$
$S_{\text{edge, APf (iii)}}=$\\
$\int_L dt dx \frac{2}{4 \pi} \partial_{x} \phi_s (\partial_{t} \phi_s - v_{s}  \partial_{x} \phi_s )$\\
$+\int dt dx \frac{q}{2 \pi} \epsilon^{\mu \nu} A_\mu \partial_{\nu} \phi_s \mid_{q= e 1}$\\
$+\int_R dt dx \sum_{j=1}^3 \chi_{Rj} (i \partial_t +i v_R \partial_x) \chi_{Rj}$.\\[0mm]
\end{minipage}
 &
 $\begin{array}{c}(1,\frac{-3}{2});\frac{-1}{2}\\
( 1,0 )_q; \\
{(e,\frac{e}{2} \cdot 1 +0)_{\rt}}\\
 \to{(\frac{e}{2},  0)_{\rt}};\\
 \sigma_{xy}=\frac{1}{2};\\
 \kappa_{xy}=c_-+2=\frac{3}{2}.
\end{array}$
&
 \begin{minipage}[c]{.55in}
6 (b+f)
\end{minipage}
&
  \begin{minipage}[c]{1.15in}
  Cooper pairs:\\
  $f-if$-wave of CF\\
 ($d-id$-wave of CDF)
 \end{minipage}
\\ \hline
    \end{tabular}
    \caption{Data of Pf/PH-Pf/APf states. (Non-)Abelian Chern-Simons theories (CS) are identified.
    We provide both the fermionic spin TQFT (fTQFT) and the simplified bosonic TQFT (bTQFT) versions.
  Semion theory has a bulk action $\rU(1)_2$ CS as $\frac{2}{4 \pi} \int a da$.
   {The $\phi_s/\phi_{\bar s}$ stands for the chiral boson of semion/anti-semion (with a level $k=2$).}
   {The $\phi_0$ is the neutral chiral boson of a level $k=1$.}
   The $c_L/c_R$ and
   $q_L/q_R$ are chiral central charges and charge couplings,
   $\rt_{L/R}= K^{-1} q_{L/R}$ where $K$ is given by a $K$-matrix CS.
   The $\sigma_{xy}, \kappa_{xy}$ are quantum/thermal Hall conductances.
    }
    \label{table:Sbulkedge}
\end{table}

\begin{table}[!h]	
\footnotesize
    \hspace{-2.8em}\begin{tabular}{ | c | c | c | c | c | c | c | }
    \hline
    State &
    \begin{minipage}[c]{2.3in}
    2+1D TQFT description;\\
    1+1D Edge modes
    \end{minipage}
    & $\begin{array}{c}(c_L,-c_R); c_-\\ q_I; \quad \rt_I; \\ \sigma_{xy} \text{ in $(\frac{e^2}{h})$};\\
    \kappa_{xy} \text{ in $(\frac{\pi^2 k_B^2}{3 h}T)$.}  \end{array}$  & 
    $\GSD_{T^2}$
     &
         \begin{minipage}[c]{1.1in}
     Description: \\
     BCS pairing \\
     of fermions
         \end{minipage}
     \\ \hline
\begin{minipage}[c]{.45in}
 $K=8$ \\[2mm]
$\kappa_{xy}=3$
         \end{minipage}
&
\begin{minipage}[c]{2.9in}
$K=8$ as a spin U(1)$_8$-Abelian CS
+ $\frac{q}{2 \pi} \int A da \mid_{q=2 e}$;\\
fTQFT may be denoted as U(1)$_8/\Z_2$-CS.\\
$K$-matrix multiplet chiral bosons \eqn{eq:Sedge}.
 \end{minipage}
&
 $\begin{array}{c}(1,0);0\\
(2)_q;  {(\frac{1}{4})_{\rt}}\\
 \sigma_{xy}=\frac{1}{2};\\
 \kappa_{xy}=c_-+2=3.
\end{array}$
 & 8 &
  \begin{minipage}[c]{1.15in}
  Cooper pairs:\\
  $s$-wave of CF\\
 ($p+ip$-wave of CDF)
          \end{minipage}
\\
\hline
%
\begin{minipage}[c]{.52in}
 113-state\\[2mm]
$\kappa_{xy}=2$
 \end{minipage}
&
\begin{minipage}[c]{2.8in}
$K=
\left(
\begin{array}{cc}
1 & 3  \\
3 & 1
\end{array}
\right)$-Abelian CS+ $\frac{q}{2 \pi} \int A da \mid_{q=e (\begin{smallmatrix} 1 \\1 \end{smallmatrix})}$\\
$\simeq
\left(
\begin{array}{cc}
1 & 0  \\
0 & -8
\end{array}
\right)$-Abelian CS+ $\frac{q}{2 \pi} \int A da \mid_{q=e (\begin{smallmatrix} 1 \\2 \end{smallmatrix})}$;\\
$K$-matrix multiplet chiral bosons \eqn{eq:Sedge}.
         \end{minipage}
&
 $\begin{array}{c}(1,-1);0\\
(1,1)_q;  {(\frac{1}{4}, \frac{1}{4})_{\rt}};\\
 \sigma_{xy}=\frac{1}{2};\\
 \kappa_{xy}=c_-+2=2.
\end{array}$
& 8 &
  \begin{minipage}[c]{1.15in}
  Cooper pairs:\\
  $d-id$-wave of CF\\
 ($p-ip$-wave of CDF)
          \end{minipage}
\\
\hline
\hline
\multicolumn{5}{c}{ Other Root States (Not directly related to the $\nu=5/2$-experiment and our phase diagram Fig.~\ref{PD})}\\[2mm]
\hline
\begin{minipage}[c]{.52in}
 331-state\\[2mm]
$\kappa_{xy}=4$
         \end{minipage}
&
\begin{minipage}[c]{2.8in}
$K=
\left(
\begin{array}{cc}
3 & 1  \\
1 & 3
\end{array}
\right)$-Abelian CS+ $\frac{q}{2 \pi} \int A da \mid_{q=e (\begin{smallmatrix} 1 \\1 \end{smallmatrix})}$;\\
$K$-matrix multiplet chiral bosons \eqn{eq:Sedge}.
         \end{minipage}
&
 $\begin{array}{c}(2,0);0\\
(1,1)_q;  {(\frac{1}{4}, \frac{1}{4})_{\rt}}\\
 \sigma_{xy}=\frac{1}{2};\\
 \kappa_{xy}=c_-+2=4.
\end{array}$
& 8 &
  \begin{minipage}[c]{1.15in}
  Cooper pairs:\\
  $d+id$-wave of CF\\
 ($f+if$-wave of CDF)
          \end{minipage} \\
\hline
\hline
 $\mathcal{V}=1$ & $\begin{array}{c}
 \text{Ising TQFT}\, \cong \,   \text{gauge $\Z_2^f$ of } \,{\text{spin-Ising TQFT}} \\
  \cong \text{gauge $\Z_2^f$ of } {p_x+ip_y} \text{ superconductor} \cong  \\
\cblue{\rU(2)_{2,-4} \;{\text{CS}}}     \;
\cong (\rSU(2)_2 \times \rU(1)_{-4})/ \Z_2 \;{\text{CS}}
       \end{array}$
&
$\begin{array}{c}(\frac{1}{2},0); 0\\ (0,0); (0,0) \end{array}$
&
 3
  &
\\ \hline
 $\mathcal{V}=3$ & $\rSU(2)_{2}$ CS
 &
 $\begin{array}{c}(\frac{3}{2},0); 0\\ (0,0); (0,0)  \end{array}$
&
 3
  &
\\ \hline
    \end{tabular}
    \caption{
     The set-up follows the same as Table \ref{table:Sbulkedge}.
    Data of 331/$K=8$/113 quantum Hall states.
   Here $\mathcal{V}=1,3 \in \Z_8$ state means some root of TQFT, where one can find other related bosonic/fermionic (spin-)TQFT data
     in \cite{1612.09298PWY}.
    }
        \label{table:Sbulkedge-2}
\end{table}

\begin{table}[!t] 
\footnotesize
    \hspace{-3.05em}\begin{tabular}{ | c | c | c | c |  c|}
    \hline
    State & 2+1D TQFT description  & $\cS^{xy}$ & $\cT^{xy}$ & $c_-$  \\ \hline
 $\mathcal{V}=1$ & $\begin{array}{c}
 \text{Ising TQFT}\cong \\
\rU(2)_{2,-4} \;{\text{CS }} \cong
\\
 (\rSU(2)_2 \times \rU(1)_{-4})/ \Z_2
  \end{array}$
 &
 $\frac{1}{2}
 \left(\begin{array}{cccc}
1  & {\sqrt{2}} & 1 \\
 {\sqrt{2}}  & {0} & -{\sqrt{2}}  \\
{1} & -{\sqrt{2}} & 1  \\
\end{array}
\right)
$ & $\left(
\begin{array}{cccc}
 1 & 0 & 0  \\
 0 & e^{\frac{\pi i}{8}} & 0  \\
 0 & 0 & -1
\end{array}
\right)$ & $\frac{1}{2}$
\\ \hline
 $\mathcal{V}=3$ & $\rSU(2)_{2}$ CS
&
$\frac{1}{2}
 \left(\begin{array}{cccc}
1  & {\sqrt{2}} & 1 \\
 {\sqrt{2}}  & {0} & -{\sqrt{2}}  \\
{1} & -{\sqrt{2}} & 1  \\
\end{array}
\right)
$  &  $\left(
\begin{array}{cccc}
 1 & 0 & 0  \\
 0 & e^{\frac{3\pi i}{8}}  & 0  \\
 0 & 0 & -1
\end{array}
\right)$  & $\frac{3}{2}$
\\ \hline
Semion & $\rU(1)_2$ CS: $\frac{2}{4 \pi} \int a da$  &
$\frac{1}{\sqrt 2}
  \begin{pmatrix}
    1&1\\
    1&-1
  \end{pmatrix}$
& $\diag(1,\ii)$ & 1 \\ \hline
Pfaff &
\begin{minipage}[c]{2.15in}
fTQFT $\to$ simplified bTQFT:\\
Semion $\times$ Ising,\\
Anyons: $\{1, s\} \otimes \{1, \sigma, \psi\} $\\
\end{minipage}
&
$
  \begin{pmatrix}
    \frac{1}{{\sqrt 2}} &\frac{1}{{\sqrt 2}}\\
    \frac{1}{{\sqrt 2}}&-\frac{1}{{\sqrt 2}}
  \end{pmatrix}
  \otimes
\left(\begin{array}{cccc}
\frac{1}{2}  & \frac{1}{\sqrt{2}} & \frac{1}{2} \\
 \frac{1}{\sqrt{2}}  & {0} & -\frac{1}{\sqrt{2}}  \\
\frac{1}{2} & -\frac{1}{\sqrt{2}} & \frac{1}{2}  \\
\end{array}
\right)
  $
&
\begin{minipage}[c]{1.15in}
$\diag(1,\ii)$ \;$\otimes$\\
$\left(
\begin{array}{cccc}
 1 & 0 & 0  \\
 0 & e^{\frac{\pi i}{8}}  & 0  \\
 0 & 0 & -1
\end{array}
\right)$
\end{minipage}  & $\frac{1}{2}$
\\ \hline
PH-Pfaff &
\begin{minipage}[c]{2.15in}
fTQFT $\to$ simplified bTQFT:\\
Semion $\times$ $\overline{\text{Ising}}$, \\
Anyons: $\{1, s\} \otimes \{1, \bar \sigma, \bar\psi\} $
\end{minipage}
&
$
  \begin{pmatrix}
    \frac{1}{{\sqrt 2}} &\frac{1}{{\sqrt 2}}\\
    \frac{1}{{\sqrt 2}}&-\frac{1}{{\sqrt 2}}
  \end{pmatrix}
  \otimes
\left(\begin{array}{cccc}
\frac{1}{2}  & \frac{1}{\sqrt{2}} & \frac{1}{2} \\
 \frac{1}{\sqrt{2}}  & {0} & -\frac{1}{\sqrt{2}}  \\
\frac{1}{2} & -\frac{1}{\sqrt{2}} & \frac{1}{2}  \\
\end{array}
\right)
  $
&
\begin{minipage}[c]{1.15in}
$\diag(1,\ii)$ \;$\otimes$\\
$\left(
\begin{array}{cccc}
 1 & 0 & 0  \\
 0 & e^{-\frac{\pi i}{8}}  & 0  \\
 0 & 0 & -1
\end{array}
\right)$
\end{minipage}
 & $\frac{-1}{2}$
\\ \hline
A-Pfaff &
\begin{minipage}[c]{2.15in}
fTQFT $\to$ simplified bTQFT:\\
Semion $\times$  $\rSU(2)_{-2}$ CS, \\
Anyons: $\{1, s\} \otimes \{1, \bar \sigma_3, \bar\psi_3\} $
\end{minipage}
&
$
  \begin{pmatrix}
    \frac{1}{{\sqrt 2}} &\frac{1}{{\sqrt 2}}\\
    \frac{1}{{\sqrt 2}}&-\frac{1}{{\sqrt 2}}
  \end{pmatrix}
  \otimes
\left(\begin{array}{cccc}
\frac{1}{2}  & \frac{1}{\sqrt{2}} & \frac{1}{2} \\
 \frac{1}{\sqrt{2}}  & {0} & -\frac{1}{\sqrt{2}}  \\
\frac{1}{2} & -\frac{1}{\sqrt{2}} & \frac{1}{2}  \\
\end{array}
\right)
  $
&
\begin{minipage}[c]{1.15in}
$\diag(1,\ii)$ \;$\otimes$\\
$\left(
\begin{array}{cccc}
 1 & 0 & 0  \\
 0 & e^{-\frac{3 \pi i}{8}}  & 0  \\
 0 & 0 & -1
\end{array}
\right)$
\end{minipage}
 & $\frac{-3}{2}$
\\ \hline
    \end{tabular}
    \caption{
  Hereby the semion theory, we mean the action $S_{\text{semion}}=\frac{2}{4 \pi} \int a da$.
  The $\cS^{xy}$ and $\cT^{xy}$ are written in the bases of $\{1, \sigma_\mathcal{V}, \psi_\mathcal{V}\}$, here $\mathcal{V}=1,3 \in \Z_8$-class TQFTs.
    One can find other related $\mathcal{V} \in \Z_8$ fermionic (spin-)TQFT data 
  in \cite{1612.09298PWY}.
  The chiral central charge $c_-\equiv c_L-c_R$ mod 8 is determined by the relation $(\cS^{xy} \cT^{xy})^3=e^{2i \pi c_-/8}(\cS^{xy})^2$.
    }
        \label{table:TO-ST}
\end{table}

Here we first gather the data of topological orders and their topological quantum field theories (TQFTs) in Table \ref{table:Sbulkedge}, \ref{table:Sbulkedge-2} and \ref{table:TO-ST}.
Later we can use these data to determine all possible interfaces/domain walls.

In Table \ref{table:Sbulkedge} and  \ref{table:Sbulkedge-2}'s first column, we list down the underlying
 topological orders (Pfaffian/Anti-Pfaffian/PH-Pfaffian and also
 331/$K=8$/113 quantum Hall states) relevant for our work.
Table \ref{table:Sbulkedge} and  \ref{table:Sbulkedge-2}'s second column,
we identify their bulk TQFTs and their gapless edge theories.
By a symmetric bilinear
$K$-matrix Abelian Chern-Simons theory (CS), we mean the action of (see more discussions in Appendix \ref{app:bulk-bdry-ab}):
\bea
S_{\text{bulk}}[a,A] &=&
\frac{K_{IJ}}{4\pi} \int a_I d a_J + \frac{q_I}{2 \pi} \int A da_I.
\eea
We denote $q$ as the charge vector coupling to U(1)-background (electromagnetic) field $A$,
and its transpose $q^\rT$, introduced later in \eqn{eq:bulk-charge}.
The $K$-matrix CS with any diagonal entry of $K_{II}$ as an odd integer is a fermionic TQFT 
(here as spin TQFT or simply fTQFT\footnote{By spin TQFT or fTQFT, we mean the TQFT defined on \emph{spin} manifolds.
The underlying fTQFT can be viewed as the low energy physics theory emergent from a lattice model with fermionic degrees of freedom at high energy cutoff scale.}).
By
a non-Abelian U$(2)_{k_2,k_1}\equiv (\rSU(2)_{k_2} \times \rU(1)_{k_1})/ \Z_2$ Chern-Simons theory, we mean the action of:
\bea
S_{\text{bulk}} &=&  \int \frac{k_2}{4 \pi}  \Tr[bdb+ \frac{2}{3} b^3]+ \frac{k_1}{4 \pi}  a da ,
\eea
where $b$ is an SU(2) gauge field and $a$ is a U(1)  gauge field,
with a $\Z_2$-quotient constraint, following the notations of recent work \cite{1602.04251SW,1612.09298PWY}.
The $\Z_2$-quotient is effectively obtained by gauging its $\Z_2$ 1-form global symmetry, which
results in the modification of line operators (anyons) spectrum.
In our case, this modification yields spin fTQFTs.
The path integral is written as
  $$
  Z[A] = \int [Da] [Db] \exp(i S[a,b, A]),
  $$
 where $a,b$ are dynamical internal gauge fields that are summed over, while the $A$ is the probed background field.
 For the identification of Pfaffian Moore-Read state as a fermionic spin fTQFT, we follow a remarkable Ref.~\cite{1602.04251SW}.

For the gapless edge theories,
we study the Majorana-Weyl modes and additional $K$-matrix multiplet chiral boson theories (see Appendix \ref{app:quantize-cboson}).
Indeed, we can rigorously show that, from Table \ref{table:Sbulkedge},
$S_{\text{edge, A-Pfaff (i)}}$ can be transformed to $S_{\text{edge, A-Pfaff (ii)}}$ by the standard GL$(N,\Z)$ field redefinition.
Then we can rewrite the neutral $\phi_n$ chiral boson
by fermionization to a complex fermion, then to two chiral Majorana-Weyl modes $\chi_{R,2}$ and $\chi_{R,3}$ (see Appendix \ref{app:PfAPf-KDS}).

Table \ref{table:Sbulkedge} and  \ref{table:Sbulkedge-2}'s  third column,
the $c_L$/$c_R$ are left/right central charges.
The $q$ and $\rt$ are the charge vectors (see Sec.~\ref{eq:bulk-charge}).
The $q_L/q_R$ are the charge coupling between internal gauge field $a$ and external background electromagnetic $A$ field.
The $\sigma_{xy}$ and $\kappa_{xy}$ are quantum Hall (here only the Abelian $K$-matrix part is charged) and thermal Hall conductances:
\bea \label{eq:Hall-der}
\sigma_{xy} = q^\rT K^{-1} q(\frac{e^2}{h}),\;\;\;\;\;\; \kappa_{xy}=(c_L-c_R)\frac{\pi^2 k_B^2}{3 h}T.
\eea

In Table \ref{table:Sbulkedge} and  \ref{table:Sbulkedge-2}'s fourth column, we show ground state degeneracy (GSD) on a spatial torus $T^2$, where $T^2$
can have spin structures as ${T^2_\text{e}}$ for the even and ${T^2_\text{o}}$ for the odd \cite{1612.09298PWY, 1801.05416WOP}.
The even spin structure ${T^2_\text{e}}$ means that both $S^1$ 1-cycle on $T^2$ has at least one in anti-periodic condition.
The odd spin structure ${T^2_\text{o}}$ means that both $S^1$ 1-cycle on $T^2$ are both in periodic conditions.
Follow \cite{1612.09298PWY, 1801.05416WOP},
the $\GSD_{T^2}=\GSD_{T^2_\text{e}|T^2_\text{o}}$ can be computed as the partition function $Z({T^2_\text{e}}\times  S^1_{\text{anti-periodic}})$ and
$Z({T^2_\text{o}}\times S^1_{\text{anti-periodic}} )$ respectively.
The GSD can have ``b'' for bosonic and ``f'' for fermionic sectors.

In Table \ref{table:Sbulkedge} and  \ref{table:Sbulkedge-2}'s fifth column, we summarize the pairing symmetry of composite fermions (CF) or
composite Dirac fermions (CDF), 
in terms of Bardeen-Cooper-Schrieffer (BCS) mechanism,
following a description in footnote \ref{foot:composite-pair} in the main text.

In Table \ref{table:TO-ST}, we show the modular SL(2,$\Z$) representation data, $\cS^{xy}$ and $\cT^{xy}$ matrices, in the good anyon quasiparticle basis
(namely, in a good Wilson line basis), and again their chiral central charges $c_-$.
Here we present their bosonic TQFT sectors. For the full spin fermionic TQFT, one can include the fermionic ($f$) sector $\{1, f\}$ with additional constraints.
We can consider the generators of mapping classes groups of $T^2$, the SL$(2, \Z)$
modular data, permuting the spin structures on $T^2$, see Section 8 of \cite{1612.09298PWY}.

In 2+1D, bosonic TQFT theory is modular in terms of category theory.
Physically speaking, modular theory means that all nontrivial anyons have nontrivial (mutual) braiding statistics at least
with respect to one other particle.
Fermionic TQFT theory however is pre-modular.
Fermionic theory has a fermion (electron) that has trivial  (mutual) braiding statistics with respect to all other particles.

We write the rank-2 $\cS^{xy}$ and $\cT^{xy}$ matrices of semion theory in the $\{1, s\}$ basis (where $s$ is the semion anyon).
We  write the rank-3 $\cS^{xy}$ and $\cT^{xy}$ matrices $ \mathcal{V} \in \Z_8$-theory (here $ \mathcal{V}=\pm 1, -3$)  in $\{1, \sigma_ \mathcal{V}, \psi_ \mathcal{V}\}$ basis.
The $\sigma_\mathcal{V}$ is a non-Abelian Majorana zero mode also called $\sigma$-anyon, that is trapped at the core of $\pi$-vortex
($\Z_2$-gauge flux line operator, with the vison at ends).
The $\psi$ is a Bogoliubov fermion solved from the Bogoliubov-de Gennes (BdG) equation.
The fusion rules (denoted as a ``$\times$'' operation) of anyons follow:
 \bea
&& 1 \times s = s, \quad  s \times s = 1.\\
&& \sigma_\mathcal{V} \times \sigma_\mathcal{V} = 1 \oplus \psi_\mathcal{V}, \quad
\sigma_\mathcal{V}  \times \psi_\mathcal{V} = \sigma_\mathcal{V}, \quad \psi_\mathcal{V}  \times \psi_\mathcal{V} =1.
\eea
See the renown work \cite{0506438Kitaev} for an introduction to category theory and fusion/braiding properties of 2+1D TQFT.
The classic developments root in \cite{1989MS}.


\section{Canonical Quantization of $K$-matrix Chiral Boson Theory/Luttinger Liquids}

\label{app:quantize-cboson}

In Appendix \ref{app:data}, we had describe the TQFT description of various Pf/PH-Pf/A-Pf states.
When we only focus on the Abelian sectors, we can study them by simply using $K$-matrix Chern-Simons (CS) theory.
In this section, we will focus on the Abelian CS and their edge $K$-matrix chiral boson theory/Luttinger liquids.
Then we will use this info to demonstrate stability of gapless modes in the next Appendix \ref{app:Lstability}.

\subsection{Bulk and boundary actions}

\label{app:bulk-bdry-ab}

The bulk action $S_{\text{bulk}}$ of Abelian fractional quantum Hall state (described by an Abelian $K$-matrix CS theory)
in the bulk $\mathcal{M}=\mathcal{M}^3$,
and the boundary action $S_{\partial}$ of its edge states (described by $K$-matrix chiral bosons/Luttinger liquids) on the boundary
$\partial \mathcal{M}=\partial (\mathcal{M}^3) \equiv \Sigma^2$
are:
\bea \label{eq:Sbulk}
S_{\text{bulk}} &=&
\frac{K_{IJ}}{4\pi}\int_\mathcal{M}  dt\; d^2x \; \epsilon^{\mu\nu\rho} a_{I,\mu} \partial_\nu a_{J,\rho}, \\
 \label{eq:Sedge}
S_{\partial}&=& \frac{1}{4\pi} \int_{\partial \mathcal{M}} dt \; dx \; ( K_{IJ} \partial_t \phi_{I} \partial_x \phi_{J} -V_{IJ}\partial_x \phi_{I}   \partial_x \phi_{J} ).
\eea
Here $K_{IJ}$ and $V_{IJ}$ are symmetric integer bilinear $N \times N$ matrices,
$a_{I,\mu}$ is the 1-form emergent gauge field's $I$-th component in the multiplet.
The $a$ gauge fields are emergent degree of freedom after integrating out the bulk gapped matter fields.
The above theories are RG fixed point TQFT in the bulk, and the gapless theory on the edge.
The $V_{IJ}$ is positive definite for the \emph{potential energy} like term to be bounded from below (see. \eqn{eq:H-ChB}).
Each mode has individual chiral central charge $|c_L-c_R|={1}$.
The number of left moving modes subtracts the number of right moving modes, say the total chiral central charge
 $$c_- \equiv c_L-c_R= \text{signature}(K),$$
is the signature of $K$ matrix (Numbers of positive eigenvalues subtracts negative eigenvalues).
We will allow add interactions later in Appendix \ref{app:Lstability} to explore the gap edge/domain wall:
\bea  \label{eq:Sedge-int}
S_{\partial,\text{interaction}}&=&
 \int_{\partial \mathcal{M}} dt \; dx\;  \sum_{\ra} g_{\ra}  \cos(\ell_{\ra,I}^{} \cdot\phi_{I}),
\eea
where $\ra$ are different components.
However, below we like to first canonical quantize the gapless edge theory \eqn{eq:Sedge} in Sec.~\ref{subsec:canon-q}.

\subsection{Canonical Quantization of multiplet Chiral Boson Field Theory}
\label{subsec:canon-q}

In the literature, there are vast but none generic treatments on canonical quantization of multiplets-chiral boson theory.
So we aim to be self-contained,  we start from scratch. Below we will be pedagogical for our convention and definition.

The time-independent mode expansion of $\phi_I(x)$ on a \emph{compact} circle $x \in [0, L)$ that we construct is:
\bea \label{eq:phi-mode}
\phi_I(x)
&\equiv& {\phi_{0}}_{I}+K^{-1}_{IJ} P_{\phi_J} \frac{2\pi}{L}x+i \sum_{n\neq 0} \frac{1}{n} \alpha_{I,n} e^{-in x \frac{2\pi}{L}}
\eea
We write zero mode part as $ {\phi_{0}}_{I}$. The conjugate winding momentum is $P_{\phi_I}$. All of
${\phi_{0}}_{I}$, $P_{\phi_J}$,
and non-zero mode part, Fourier modes $\alpha_{I,n}$, should regard as operators (instead of complex numbers).
Our canonical quantization is performed by imposing following commutation relations.
The canonical conjugation relation of zero mode and winding momentum is
\be
[{\phi_{0}}_{I},  P_{\phi_J}]=i\delta_{IJ}.
\ee
The non-zero Fourier modes part satisfy a generalized Kac-Moody algebra:
\be
[\alpha_{I,n} , \alpha_{J,m} ]= n K^{-1}_{IJ}\delta_{n,-m}.
\ee
The conjugate momentum field $\Pi_{I}(x)$ of $\phi_I(x)$ is:
\be
\Pi_{I}(x) = \frac{\delta {L}}{\delta (\partial_t \phi_{I} )} =\frac{1}{4\pi} K_{IJ} \partial_x \phi_{J} = \frac{1}{4\pi}\frac{2\pi}{L}( P_{\phi_I}  + \sum_{n \neq 0} K_{I J} \alpha_{J,n} e^{-in x \frac{2\pi}{L}}).
\ee
We can check consistently the canonical conjugation relation of operators $\phi_I(x_1)$ and $\Pi_{J}(x_2)$:
\bea \label{eq:canon-commutator}
[\phi_I(x_1),\Pi_{J}(x_2)]&=& \frac{1}{4\pi} \frac{2\pi}{L} i \delta_{IJ} (1+ \sum_{n \neq 0}   e^{-in (x_1-x_2) \frac{2\pi}{L}} )\\
&=& \frac{1}{4\pi} \frac{2\pi}{L} i \delta_{IJ} \sum_{n \in \mathbb{Z}}   e^{-in (x_1-x_2) \frac{2\pi}{L} } \nonumber\\
&=& \frac{1}{2} i  \delta_{IJ} \delta(x_1-x_2), \nonumber
\eea
where we derive and apply some Fourier transformation formulas.\footnotemark
\footnotetext{
We derive Fourier transformation formulas:
\bea
\frac{1}{2\pi} \int^{2\pi}_{0} e^{i{(m-n)\varphi} } d\varphi  &=&\delta_{m,n},\\
\frac{1}{2\pi} \int^{L}_{0} e^{i \frac{2\pi}{L} n x} dx &=& \frac{L}{2\pi} \delta_{n,0},\\
\frac{1}{2\pi} \int^{\infty}_{-\infty} e^{ikx} dx &=& \delta(k),\\
\sum_{n \in \mathbb{Z}}   e^{-in (x_1-x_2) \frac{2\pi}{L} } &=& L \delta(x_1-x_2).
\eea
The first line is proved by complex analysis:
$$\delta_{m,n}=\frac{1}{2\pi i} \oint_{|z|=1} z^{m-n-1} dz=\frac{1}{2\pi}  \int^{2\pi}_{0} e^{i{(m-n)\varphi} } d\varphi. $$
The second line is proved by replacing the first line to:
$$
\delta_{m,n}=\frac{1}{2\pi}  \int^{2\pi}_{0} e^{i{(m-n)\varphi} } d\varphi =\frac{1}{L} \int^{x=L}_{x=0} e^{i \frac{2\pi(m-n)}{L}x} dx.
$$ with $x\equiv \frac{L}{2\pi} \varphi$, $k \equiv \frac{2\pi(m-n)}{L}$, so
$$
\delta_{m,n}=\frac{1}{2\pi}  \int^{\pi}_{-\pi} e^{i{(m-n)\varphi} } d\varphi =\frac{1}{L} \int^{x=\pi \frac{L}{2\pi}}_{x=-\pi \frac{L}{2\pi}} e^{i \frac{2\pi(m-n)}{L}x} dx.
$$ The third line is proved by taking $L\to \infty$:
$$
\frac{1}{2\pi} \int^{\infty}_{-\infty} e^{i k x} dx=\lim_{L\to\infty}\frac{L}{2\pi} \delta_{m,n} =\delta(\frac{2\pi}{L}(m-n))= \delta(k).
$$
The fourth line is shown by relating a sum to a continuous limit (with $k_n=\frac{2\pi n}{L}$):
$$
\sum_{n \in \mathbb{Z}}   e^{-in (x_1-x_2) \frac{2\pi}{L} } =\sum_{n \in \mathbb{Z}} e^{-i k_n (x_1-x_2) } (k_{n+1}-k_n)\frac{L}{2\pi}  \to \int e^{-ik (x_1-x_2)} dk \frac{L}{2\pi} =L \delta(x_1-x_2).
$$
} We comment that the factor is $\frac{1}{2}$ instead of $1$ in $\frac{1}{2} i  \delta_{IJ} \delta(x_1-x_2)$, due to that we have each mode as chiral modes.
If we have two modes together with the combined left and right non-chiral modes, the sum of two modes give $i  \delta_{IJ} \delta(x_1-x_2)$.\\

\subsubsection{Equation of Motion}
The Equation of Motion (E.O.M) of \eqn{eq:Sedge} is
\bea
\frac{\partial {L}}{\partial (\phi_{I} )} -\partial_\mu(\frac{\partial {L} }{\partial (\partial_\mu \phi_{I} )})=\frac{1}{4\pi}(-2)\partial_x (K_{IJ} \partial_t \phi_{J}- V_{IJ}\partial_x \phi_{J})=0 \nonumber \\
\Rightarrow K_{IJ} \partial_t \phi_{J}- V_{IJ}\partial_x \phi_{J} = f(t), \text{ and }
 (K_{IJ} \partial_t \rho_{J}- V_{IJ}\partial_x \rho_{J})=0.
\eea
Here $\rho_J \equiv \frac{1}{2\pi} \partial_x \phi_{J}(x)$ can be regarded as the density field. 
The $V_{IJ}$ is positive definite matrix, so the sign of eigenvalues of $K_{IJ}$ determines the left or right moving modes.
Positive eigenvalues are left (L) moving, and negative eigenvalues are right (R) moving. One simple trick to see this L/R moving is based on
the E.O.M. of the forms: $(\partial_t-v\partial_x) \phi_L(vt+x)$ and $(\partial_t+v\partial_x) \phi_R(vt-x)$, then simply drawing their wave packets in the $(t,x)$ plane.

\subsubsection{Hamiltonian}
The time-independent Hamiltonian is:
\bea \label{eq:H-ChB}
H&=&\int^{L}_{0} dx \;[\Pi^{I} \frac{\partial {L}}{\partial (\partial_t \phi_{I} )} -L]=\int^{L}_{0} dx \;[ V_{IJ} \partial_x \phi_{I} \partial_x \phi_{J}] \nonumber\\
&=& \frac{1}{4\pi} \frac{(2\pi)^2}{L} [ V_{IJ} K^{-1}_{I l1} K^{-1}_{J l2} P_{\phi_{l1}} P_{\phi_{l2}}+\sum_{n\neq0} V_{IJ} \alpha_{I,n} \alpha_{J,-n}],
\eea
Notice that the Hamiltonian only depends on the winding mode parts ($P_{\phi}$), and the positive-definite velocity matrix ($V_{IJ}$)
gives rise to a potential like term for Fourier modes
($\alpha_{I,n}$).

\subsubsection{Time-in/dependent (Schr\"odinger/Heisenberg picture) mode expansion}

The mode expansion we use is a time-independent operator (Schr\"odinger picture) $\phi_I(x)$ in \eqn{eq:phi-mode}, we now check whether the time-dependent operator (Heisenberg picture) $\phi_I(x,t)$ satisfies the E.O.M., and whether the time-dependent part show the left, right moving modes explicitly.

One quick method to calculate $\phi_I(x,t)$ is going reversely to find the constraints from E.O.M. For example, writing:
\be \label{eq:phi-mode-t}
\phi_I(x,t)={\phi_{0}}_{I}+K^{-1}_{IJ} P_{\phi_J} \frac{2\pi}{L}x+
\tilde V_{I l} P_{\phi_l} \frac{2\pi}{L}t+i \sum_{n\neq 0} \frac{1}{n}( \alpha_{l',n} e^{-in t \frac{2\pi}{L} {M}_{Il'} } e^{-in x \frac{2\pi}{L}} ).
\ee
Check that $K_{IJ} \partial_t \phi_{J}- V_{IJ}\partial_x \phi_{J} =0$ imposes:
$\tilde V_{I l} =K^{-1}_{I I'} V_{I'J'} K^{-1}_{J' l} $ and $ {M}_{Il'}=K^{-1}_{Il}V_{ll'}$.

The alternative standard method, we can calculate
$
\phi_I(x,t)=e^{iHt} \phi_I(x)e^{-iHt}.
$
We find:
\bea
[H,{\phi_{0}}_{I}]&=&\frac{1}{4\pi} 2 \frac{(2\pi)^2}{L} (-i) K^{-1}_{I I'} V_{I'J'}  K^{-1}_{J' l} P_{\phi_l}, \\ \;
[H,  P_{\phi_I}]&=&0, \\ \;
[H, \alpha_{I,n}]&=&\frac{1}{4\pi} 2 \frac{(2\pi)^2}{L} (-n) K^{-1}_{J'I} V_{I'J'} \alpha_{I',n} =\frac{1}{4\pi} 2 \frac{(2\pi)^2}{L} (-n) K^{-1}_{I J} V_{J I'}  \alpha_{I',n}.
\eea
Plug in we derive the exactly consistent result above:
\be
\phi_I(x,t)=e^{iHt} \phi_I(x)e^{-iHt}= \text{\eqn{eq:phi-mode-t}}.
\ee
This concludes our canonical quantization of the gapless multiplet chiral boson theory.
\subsection{Global symmetry and charge vector}

Additionally we can consider a charge vector $q_I$ coupling to an external (e.g. electromagnetic) field $A_\mu$ of U(1) global symmetry,
by adding a coupling term $A_\mu q_I J^\mu_I$ with a symmetry current
${{q}_I} J^{\mu}_I= \frac{{q}_I}{2\pi} \; \epsilon^{\mu\nu\rho} \partial_\nu a_{I,\rho}$
as
\bea \label{eq:bulk-charge}
\int dt d^2x \; \frac{{q}_I}{2\pi} \; \epsilon^{\mu\nu\rho} A_{\mu} \partial_\nu a_{I,\rho}
\eea
into the ${S}_{\text{bulk}}$ of \eqn{eq:Sbulk}.
From reading the E.O.M. of ``\eqn{eq:Sbulk} plus the external probe \eqn{eq:bulk-charge},''
\bea
e\, q_J \, J_J^\mu=-q_I \frac{e^2}{h} K^{-1}_{IJ} q_J \epsilon^{\mu \nu \rho} \partial_\nu A_\rho,
\eea
we can derive that the Hall conductance $\sigma_{xy} = q^\rT K^{-1} q(\frac{e^2}{h})$ as \eqn{eq:Hall-der} from $J_x \propto \sigma_{xy}  E_y$.
Meanwhile on the boundary,
we can add the following coupling term
\bea
\int dt dx \; \frac{{q}_I}{2\pi} \; \epsilon^{\mu\nu} A_{\mu} \partial_\nu  \phi_{I},
\eea
with
a boundary symmetry current ${{q}_I} j_I^{ \mu}=  \frac{{q}_I}{2\pi}   \; \epsilon^{\mu\nu} \partial_\nu \phi_{I}$
into the ${S}_{\partial}$ of  \eqn{eq:Sedge}.
The symmetry operator on boundary is generated by the symmetry current, thus
$\text{U}_{\text{sym}}=\exp(\ti \,\theta\, \frac{{q}_I}{2\pi}   \int  \partial_x \phi_{I})$,
with any U(1) angle $\theta$.
The induced symmetry transformation acts on $\phi_I $ as:
\bea \label{chiralboson-sym}
&&(\text{U}_{\text{sym}}) \phi_I (\text{U}_{\text{sym}})^{-1}=\phi_I- \ti \theta\int dx \frac{{q}_l}{2\pi}[\phi_I,\partial_x \phi_l]
=\phi_I + \theta  (K^{-1})_{I l}  {{q}_l}  \equiv \phi_I +\theta {\rt}_I,
\eea
here we use the canonical commutation relation \eqn{eq:canon-commutator} as $[\phi_I,\partial_x \phi_l]=2\pi \ti \,(K^{-1})_{I l}  $.
A different \emph{charge vector} $\rt_I$ is commonly defined by a $K$-inverse with the original charge vector $q$-vector:
\bea
\rt_I \equiv (K^{-1})_{I l}   {{q}_l}.
\eea
Given any internal line operator $\exp(i \ell_{I} \int a_I)$ in the bulk, we can compute
its associated charge,
\bea \label{eq:bulk-charge}
q_I K^{-1}_{IJ} {\ell}^{}_{J} = \rt_J  {\ell}^{}_{J},
\eea
viewed as the charge of anyon (living on the end of an open line operator),
generated by U(1) symmetry operator $\text{U}_{\text{sym}}$.
 Then there is also a charge
\bea \label{eq:bdry-charge}
q_I Q_{I}= q_I \frac{1}{2\pi}\int^L_0 \partial_x \Phi_I dx=q_I K^{-1}_{IJ} P_{\phi_{J}}
\eea associated to each edge mode $\phi_I$ on the boundary.

\subsection{GL$(N,\Z)$ or SL$(N,\Z)$ field redefinition}
We can implement a field redefinition under $U \in \GL(N,\Z)$ (or $U \in \SL(N,\Z)$) such that the original and new quantities are related
\bea  \label{eq:relabel-1}
\vec{\tilde \phi} = U \vec{ \phi},  \;\;\;  \tilde K={U^\rT}^{-1} K U^{-1},   \;\;\;\tilde q^\rT =q^\rT U^{-1},
\eea
with the vector $ \vec{ \phi}$ abbreviates ${ \phi}_I$.
Several quantities in the action are invariant, e.g.
\bea \label{eq:relabel-2}
\begin{array}{c}
\phi_I^\rT K_{IJ}\phi_J = ({ \phi^\rT} U^\rT) ({U^\rT}^{-1} K U^{-1}) (U \phi) =\tilde \phi_I^\rT \tilde K_{IJ}\tilde \phi_J.\\
 q^\rT \phi= ( q^\rT  U^{-1}) (U  \phi) = \tilde q^\rT\tilde \phi.
\end{array}
\eea
This method will be used in Appendix \ref{app:zero-GSD}.
%
\section{Gapped Boundary, Topological Stability and Lagrangian Subgroup}

\label{app:Lstability}

Now we like to  include \eqn{eq:Sedge-int}'s interaction, the sine-Gordon cosine term $S_{\partial,\text{interaction}}$ $=$
 $\int_{\partial \mathcal{M}} dt \; dx$ $\sum_{\ra} g_{\ra}  \cos(\ell_{\ra,I}^{} \cdot\phi_{I})$
into the free-quadratic action of gapless theory
\eqn{eq:Sedge}.
One well-known issue is under what criteria that the gapless edge modes can be gapped under $S_{\partial,\text{interaction}}$.
Here we will focus on the \emph{non-perturbative} analysis,
consider the full interacting action:
\bea \label{eq:S-edge-all}
S_{\text{edge}} &=&S_{\partial}+ S_{\partial,\text{interaction}} \nonumber \\
&=& \frac{1}{4\pi} \int_{\partial \mathcal{M}} dt \; dx \; ( K_{IJ} \partial_t \phi_{I} \partial_x \phi_{J} -V_{IJ}\partial_x \phi_{I}   \partial_x \phi_{J} )
+\int_{\partial \mathcal{M}} dt \; dx \sum_{\ra} g_{\ra}  \cos(\ell_{\ra,I}^{} \cdot\phi_{I}).\;\;\;\;
\eea
By \emph{non-perturbative}, we mean that without limiting to \emph{relevant} operators in the RG sense,
the \emph{irrelevant} term at a strong coupling $g$ at the high-energy ultraviolet (UV) lattice scale,
can still \emph{gap} the gapless modes.
The classic analysis is firstly done by \cite{haldane1995}, hence the name of Haldane stability.
Our analysis follows more modern view derived in \cite{1212.4863WW} (See also a closely related work \cite{levin2013}).
In a more mathematical term, to obtain the topological gapped boundary, one needs to implement a
\emph{Lagrangian subgroup} structure \cite{1008.0654KS} at the field theory.

Below we like to derive the stability of topological boundary condition in a self-contained consistent modern view, using both
field theory and condensed matter intuition.
To this end,
first we recall that the Abelian mutual/self statistics of Abelian anyons, of the internal line operator $\exp(i \ell_{\ra,I} \int a_I)$
of the bulk action \eqn{eq:Sbulk}, is given by
\begin{equation}
\begin{array}{rcl}
\exp[i \theta_{\text{mutual}}]  &=&\exp[i \theta_{\ra\rb}]= \exp[ i \, 2\pi \, \ell_{\ra,I}^{} K^{-1}_{IJ} \ell_{\rb,J}^{}],\;\;\\
\exp[i \theta_{\text{self}}]   &=&\exp[i \frac{\theta_{\ra\ra}}{2}]= \exp[ i \pi \, \ell_{\ra,I}^{} K^{-1}_{IJ} \ell_{\ra,J}^{}].\;\;
\end{array}
\end{equation}
This can be easily derived as the path integral of Hopf link of two line operators labeled by $ \ell_{\ra}$/$\ell_{\rb}$  with proper normalization,
see Sec.~III of \cite{1612.09298PWY} for a derivation.
In terms of a recent description in \cite{1801.05416WOP}, we can view that the topological boundary condition on the edge ${\partial \mathcal{M}}$
sets the boundary gauge degrees of freedom vanishes,
\bea \label{eq:bdry-cond-a}
\ell_{\ra,I}  a_I \bigg\rvert_{\partial \mathcal{M}} =0.
\eea
The boson modes $\phi_{I}$, originally related by the gauge transformation $a_I \to a_I + d \lambda_I$ and $\phi_{I} \to \phi_{I}- \lambda_I$,
now {may be} able to condense on the boundary,
\bea
\langle \exp[i  (\ell_{I}^{} \cdot\phi_{I})] \rangle \bigg\rvert_{\partial \mathcal{M}} \neq 0, \;\;\;\; \text{more precisely,  indeed }
\langle \exp[i  (\frac{\ell_{I}}{|\gcd({\ell_{I}}) |} \cdot\phi_{I})]  \rangle \bigg\rvert_{\partial \mathcal{M}} \neq 0,
\eea
where $\gcd({\ell_{I}}) \equiv \gcd(\ell_1, \ell_2, \dots, \ell_N)$ is the greatest common divisor (gcd) of the all components of $\ell$.
This condensation is precisely triggered by the cosine term at strong coupling
\bea
g \int_{\partial \mathcal{M}} dt \; dx   \cos(\ell_{I}^{} \cdot\phi_{I}).
\eea
Moreover, the set of \emph{condensed anyons} should be generated by a subset (not necessarily the full set) of
\bea \label{eq:anyon-line}
\ell_{I}' = n \frac{\ell_{I}}{|\gcd({\ell_{I}}) |}, \quad n \in \Z_{|\gcd({\ell_{I}}) |}.
\eea
We say that  \emph{anyons} labeled by $\ell_{I}'$ corresponds to the internal line operator $\exp(i \ell_{\ra,I}' \int a_I)$ in the bulk;
the anyons living on the open ends of this line operator that can have open ends on the boundary as \eqn{eq:bdry-cond-a}.
With the above information, below we summarize the criteria in \cite{1212.4863WW}
answering ``\emph{which set of interaction terms and consequential condensations can obtain a stable gapped boundary?}''
\begin{enumerate}
\item
Trivial self statistics:
${\ell'}_{\ra,I}^{} K^{-1}_{IJ} {\ell'}_{\ra,J}^{} \in 2 \mathbb{Z}$ even integers for bosonic systems, or in $\mathbb{Z}$ odd integers for fermionic systems.
This means that the self-statistics of ${\ell'}_{a}$ line operator is bosonic/fermionic, with $\theta_{\text{self}}$ a multiple $2\pi$ or $\pi$ phase.

\item Trivial mutual statistics:
${\ell'}_{\ra,I}^{} K^{-1}_{IJ} {\ell'}_{\rb,J}^{} \in \mathbb{Z}$ integers.
Anyons labeled by ${\ell'}_{\ra}$ braid around ${\ell'}_{\rb}$ must yield a trivial mutual bosonic statistical phase.
In a spacetime picture,
the line operator labeled by ${\ell'}_{\ra}$ linked with ${\ell'}_{\rb}$ yields a path integral without any complex phase.

\item Non-fractionalized interaction terms:
The ${\ell}_{}^{}$ in $\cos(\ell_{I}^{} \cdot\phi_{I})$ term must be excitations of
non-fractionalized degrees of freedom (e.g. electrons or local bosons).
In terms of Chern-Simons $K$ matrix, the column/row vector of $K$ represents non-fractionalized operator
(e.g. $\exp(i \sum_J \rc_J K_{IJ} \int a_I)$ is non-fractionalized line operator),
thus
\be
\ell_{I}^{}= \sum_J \rc_J K_{IJ}, \quad {\rc_J}    \in \mathbb{Z}.
\ee
This criterion imposes integral charges for the bulk object's charge \eqn{eq:bulk-charge}
as well as boundary charge \eqn{eq:bdry-charge} on $\ell_I \phi_I$ edge mode.

In contrast, the anyonic operator ${\ell'}$ in \eqn{eq:anyon-line} takes integer values.
The ${\ell'}$ is \emph{not} a linear combination of column vectors of $K$ matrix,
thus fractional with respect to $K$.

\item Completeness: We find all the condensed anyon $\{\ell_\ra',\ell_\rb', \dots\}$ as a complete set, by including
every possible term $\ell_\rc'$ that has trivial self braiding statistics (criterion 1)
and trivial mutually braiding statistics (criterion 2) respect to all the elements $\{\ell_\ra',\ell_\rb', \dots\}$.

\item Non-chiral: To fully gap out the gapless modes, we need that the chiral central charge $c_L-c_R=0$,
thus \eqn{eq:Hall-der}'s thermal Hall conductance requires $\kappa_{xy}=0$.

\end{enumerate}

The detailed derivation of above criteria can be found in \cite{1212.4863WW}. Here we only like to remark some physics intuitions behind.

\noindent
$\bullet$ First, from the edge theory,
in order to pin down the zero mode at strong coupling, we require
$$
g_\ra \int_0^{L} dx\;
\cos(\ell_{\ra,I}^{} \cdot\phi_{I}) \to
\int_0^{L}  dx\;
\cos(\ell_{\ra,I}^{} \cdot ({\phi_{0}}_{I}+K^{-1}_{IJ} P_{\phi_J} \frac{2\pi}{L}x) )
=g_{\ra} L \; \cos(\ell_{\ra,I}^{} \cdot {\phi_{0}}_{I})  \,\delta_{(\ell_{\ra,I}^{} \cdot K^{-1}_{IJ} P_{\phi_J} ,0)}.
$$
The approximation is firstly due to a strong coupling thus focusing on zero/winding modes to determine the lowest energy spectrum.
The second equality holds, if $\ell_{\ra,I}$ has trivial statistics (criteria 1 and 2, but obviously to see
if it satisfies the \emph{null} condition), see \cite{1212.4863WW},
we can pin down zero modes by strong coupling $g_{\ra} \cos(\ell_{\ra,I}^{} \cdot {\phi_{0}}_{I})$
provided $\ell_{\ra,I}^{} \cdot K^{-1}_{IJ} P_{\phi_J} =0$.

\noindent
$\bullet$
Second, from the bulk theory, the trivial statistics $\exp[i \theta_{\text{mutual}}]=\exp[i \theta_{\text{self}}] =1$ helps to stabilize the path integral $Z$, thus
helping to achieving the quantum stability for gapped phases without unwanted fluctuations causing gapless modes.

\noindent
$\bullet$ Third, the non-fractionalized interaction term for ${\ell}_{}^{}$ (criterion 3) is important to
calculate the precise bulk and boundary/domain wall GSD \cite{1212.4863WW}, later shown in Appendix \ref{app:zero-GSD}.

We comment that the similar/\emph{simplified} criteria are derived but formulated in terms of Lagrangian subgroup in \cite{levin2013}.
These results \cite{1212.4863WW,{levin2013}} only work for Abelian topological order/TQFT.
Later we will implement the generalized criteria for non-Abelian case
\cite{1408.6514LWW} in Appendix \ref{app:Bootstrap}.

In addition, if we like to the interaction terms to preserve a symmetry (e.g. U(1) charge conservation in the main text, in Sec.~\ref{Sec4}),
say the boson/fermion charge symmetry as $\phi_I \to  \phi_I + K^{-1}_{IJ} q_J  \theta = \phi_I + t_I \theta$
and $\psi_I \to \psi_I e^{i K^{-1}_{IJ} q_J \theta} = \psi_I e^{i t_I \theta}$, then we require that
$\cos(\ell_{I}^{} \cdot\phi_{I}) = \cos(\ell_{I}^{} \cdot (\phi_{I} + K^{-1}_{IJ} q_J \theta))$.
For a U(1)-charge symmetry, this requires that $\ell_{I} K^{-1}_{IJ} q_J =0$.

\section{Derivations: Zero Modes and GSD Counting with Gapped Domain Walls}

\label{app:zero-GSD}

\subsection{Gapped sector of Pf$\mid$APf domain wall}

\label{app:PfAPf-KDS}

We like to implement the gapping criteria Appendix \ref{app:Lstability} to our domain wall theory on Pf$\mid$APf with data summarized in Appendix \ref{app:data}.
The only sectors of edge modes are Abelian are the chiral boson (e.g. semion) sectors.
Here we follow the treatment of gapping interaction terms of \cite{1212.4863WW}.

Based on Table \ref{table:Sbulkedge}, we can study the domain wall of Pf$\mid$APf,
by considering the actions $S_{\text{edge, Pf}}+ \bar S_{\text{edge, APf}}$.
Here we write $\bar S$ means that reversing the chirality of edge modes in $S$ due to the folding/orientation at the interface.
Starting from Table \ref{table:Sbulkedge}'s $S_{\text{edge, Pf}}+ \bar S_{\text{edge, APf (i)}}$, denoting
its $K$-matrix data as $K_{\text{(i)}}$, we can find an SL$(3,\Z)$ matrix $U_{\text{(i $\leftrightarrow$ ii)}}$ transforming
the theory to Table \ref{table:Sbulkedge}'s $S_{\text{edge, Pf}}+ \bar S_{\text{edge, APf (ii)}}$, based on \eqn{eq:relabel-1}/(\ref{eq:relabel-2}):
\bea
&&K_{\text{(i)}}=
\left(
\begin{array}{ccc}
 2 & 0 & 0 \\
0 & -1 & 0 \\
0 & 0 & 2
\end{array}
\right),
\left(
\begin{array}{c}
\phi_{1}   \\
\phi_{0} \\
\phi_{2} \\
\end{array}
\right),
\;
q = \Big(\begin{smallmatrix} 1\\1 \\1 \end{smallmatrix}\Big)
\overset{U^{}_{\text{(i $\leftrightarrow$ ii)}}}{
\Longleftrightarrow
}
K_{\text{(ii)}}=
\left(
\begin{array}{ccc}
 2 & 0 & 0 \\
0 & -2 & 0 \\
0 & 0 & 1
\end{array}
\right),
\left(
\begin{array}{c}
\phi_s   \\
\phi_{\bar s} \\
\phi_{n} \\
\end{array}
\right),
\;
q = \Big(\begin{smallmatrix} 1\\1 \\0 \end{smallmatrix}\Big),
\nonumber
\\
&&U_{\text{(i $\leftrightarrow$ ii)}}=\left(
\begin{array}{ccc}
 1 & 0 & 0 \\
 0 & 1 & 1 \\
 0 & 1 & 2 \\
\end{array}
\right),
\;\;\;
{U^{-1}_{\text{(i $\leftrightarrow$ ii)}}=
\left(
\begin{array}{ccc}
 1 & 0 & 0 \\
 0 & 2 & -1 \\
 0 & -1 & 1 \\
\end{array}
\right)}.
\eea
After rewriting the neutral chiral boson $\phi_n$ to two chiral Majorana-Weyl modes, we obtain
Table \ref{table:Sbulkedge}'s
$S_{\text{edge, Pf}}+ \bar S_{\text{edge, APf (iii)}}$. Since the net chiral central charge has $c_L-c_R=2$ (e.g. 4 chiral Majorana-Weyl modes),
they cannot be fully gapped, due to violating the criterion 5 in Appendix \ref{app:Lstability}.
However the double semion (DS) sectors
within $K_{\text{(ii)}}$, with
$K_{\text{DS}}=$
$
\left(
\begin{array}{ccc}
 2 & 0  \\
0 & -2
\end{array}
\right)$
and boson modes
$
\left(
\begin{array}{c}
\phi_s   \\
\phi_{\bar s}
\end{array}
\right)
$ can be fully gapped.\footnote{Indeed, this is the only Abelian sector that can be fully gapped.
However, using the gapping criteria for more exotic non-Abelian in Appendix \ref{app:Bootstrap}, we can explore other types of domain wall.}
The criterion 5 holds because $c_L-c_R=1-1=0$ for the double semion's $K_{\text{DS}}$.
The criteria 1, 2, and 4 says that we should condense a set of anyons labeled by $\ell'$ as
\bea
\{\ell'\}=\{(1,1), (1,-1), (2,2), (2,-2), (2,0), (0,2), \dots\} =\{(1,1)\zeta, (1,-1)\zeta' \; \vert \; \zeta, \zeta' \in \Z  \dots\}.
\eea
We say $\ell'=(1,1)$ implies the condensation of double-semions $s \bar s$ on the domain wall, while other terms are related by only non-fractionalized objects.
(Namely, it has a $\Z_2$-fusion rule, thus it has mod 2 invariant.)
The criterion 3 says that $S_{\partial,\text{interaction}}$ in \eqn{eq:S-edge-all} requires:
\bea \label{S-int-s-s}
g  \int_{\text{domain}} dt \; dx   \cos( 2(\phi_{s} + \phi_{\bar s} )) +\dots
\eea
In principle, naively the $\dots$ terms can include
$g'  \int_{\text{domain}} dt \; dx   \cos( 2(\phi_{s} - \phi_{\bar s} ))$
$+g_s  \int dt \; dx   \cos( 2 \phi_{s})$ $+$ $g_{\bar s}  \int dt \; dx   \cos( 2 \phi_{\bar s})$.\footnote{
However, there is a subtlety that a stronger criterion named the \emph{Haldane null condition} can be used \cite{haldane1995}.
The condition requires the statistical phase to be strictly zero, ${\ell}_{\ra,I}^{} K^{-1}_{IJ} {\ell}_{\ra,J}^{} =0$, instead of just being trivial.
In the strict null condition case, we find that $\ell=(2,2)$ is indeed \emph{incompatible} with the coexistence of $\ell=(2,0),(0,2),(2,2)$, see \cite{1212.4863WW}.
The later $\ell$-vectors have trivial statistics but not null statistics with respect to $\ell=(2,2)$.
In the realistic implementation, the stability depends on the relative strength of $g$ couplings.}
We remark that, in the context of Sec.~\ref{Sec4}, under the U(1)-charge conservation constraint,
only this \eqn{Interaction}, the $\cos( 2(\phi_{s} + \phi_{\bar s} ))$ term is allowed.
But we will see that, these additional terms, on one hand, do not affect the counting of GSD contributed from zero/winding modes (see Fig.~\ref{fig:zero-mode-DS-1}),
and on the other hand, they are \emph{not required} to fully gap out the edge modes as long as  we have \eqn{S-int-s-s}.
As noted in \cite{1212.4863WW},  we only require a half of the rank of $K$, say a $\text{rank(K)}/2$ (here 2/2=1) number of interaction term(s)
to fully gap the non-chiral (Abelian) edge modes. Eqn.~(\ref{S-int-s-s}) is the single required term,
satisfying all gapping criteria in Appendix \ref{app:Lstability}.

\subsection{{Zero modes and GSD counting} for percolating Pf$\mid$APf domain walls: Double-Semion sector}

Now we count the ground state degeneracy (GSD) for a number of $n$ Pf$\mid$APf domain walls in the percolation picture.
Since the Pfaffian-Anti-Pfaffian domain wall has the edge states of $\Z_2$-double-semion (twisted  $\Z_2$-gauge theory), below we focus on demonstrate
counting the GSD and zero modes for such a gapped theory.
The zero modes are topological robust and are non-local long-ranged entangled (LRE) phenomenon.
On the space without nontrivial-cycle (here the trivial homology class $H_1(\CM,\Z_2)$), such as a spatial sphere $S^2$ or a flat substrate experimental sample $\R^2$, we like to show that
$n$-double-semion domain walls contribute additional $2^{n-1}$ ground states.
We can say that there is
\bea
\GSD_{\CM_{\text{flat space}}^2} 
=\dim \cH_{\CM_{\text{flat space}}^2}=|Z_{{\CM_{\text{flat space}}^2}\times
S^1_{\text{time}}}|=2^{n-1},
\eea
simply on the flat space $\CM_{\text{flat space}}^2$.
Here $Z$ is the path integral of the whole system (including the bulk and domain walls/boundaries),
while the $\CM_{\text{flat space}}^2$ is the space with domain walls/boundaries.
Of course, in the Pfaffian-Anti-Pfaffian domain wall in a percolation picture, there are \emph{additional four gapless neutral chiral Majorana-Weyl modes} ($|c_L-c_R|=2$)
on each domain wall.
However, due to finite size effect of volume $V$,
 the energy split $\Delta_E$ of topological zero modes are exponentially small and close ($\Delta_E \simeq e^{-\# V} $),
while the energy split for ``gapless modes'' is slightly larger ($\Delta_E \simeq V^{-\#} $)  above topological GSD.
Theoretically we can  isolate and focus on the effect topologically robust zero modes.
Phenomenologically, the degenerate zero modes can contribute low $T$ heat capacity $C_V$.

Our analysis, on GSD with gapped domain walls, closely follows \cite{1212.4863WW} (See also a related work \cite{1306.4254K}).
The periodicity for ${\phi_{0}} \sim {\phi_{0}} + 2 \pi $ imposes the quantization of its conjugate variable $P_{\phi} \in \mathbb{Z}$.
Given $[\hat{\phi_{0}}, \hat{P_{\phi}}]=i$, the operator relations follow
$$
e ^{i \hat{{P_{\phi}}} \rm} | {\phi_{0}} \rangle = \; | {\phi_{0}} - \rm \rangle, \;\;\;
e^{i \rn \hat{{\phi_{0}}}} | {P_{\phi}} \rangle = | {P_{\phi}}+\rn\rangle, \;\;\;
e^{-i \rn \hat{{\phi_{0}}}} \hat{{P_{\phi}}} e^{i \rn \hat{\phi}} = \hat{{P_{\phi}}} +\rn.
$$
The $g \cos(\ell_{I}^{} \cdot {\phi_{0}}_{I})$ plays two rules: One is the
potential pinning down the zero mode ${\phi_{0}}$ to its minimum,
and the other is the hopping term, hopping between the winding mode ${P_{\phi}}$-lattice.

\begin{figure}[h!]
  \centering
    \includegraphics[width=5.in]{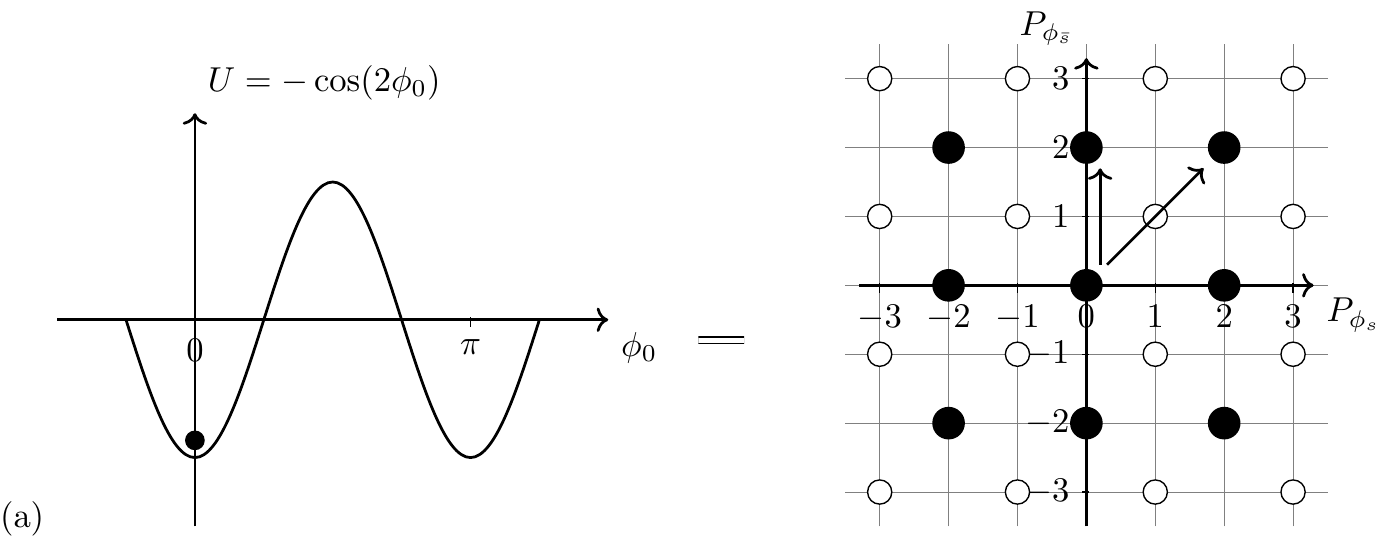}
  \includegraphics[width=5.in]{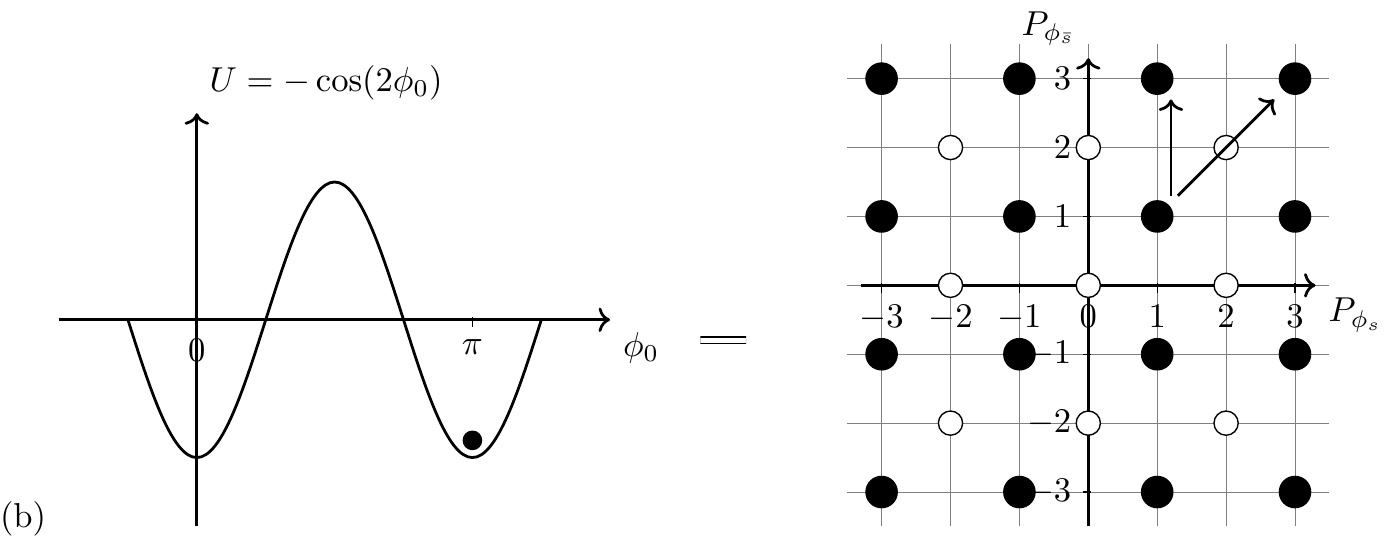}
  \caption{
The left-hand side shows the
zero mode ${\phi_{0}}$-Hilbert space: The  ${\phi_{0}}$ has a $2\pi$-periodicity.
At the strong coupling cosine term localizes
${\phi_{0}}$ to 0 and $\pi$ $\in [0, 2 \pi)$.
Note here
$-\cos(2{\phi_{0}})$ and ${\phi_{0}}$ include a set of allowed potentials
$-\cos(2{\phi_{0,s}})$,$-\cos(2{\phi_{0,\bar s}})$ and $-\cos(2({\phi_{0,s}}\pm{\phi_{0,\bar s}}))$, say in \eqn{S-int-s-s}.
The right-hand side shows the winding mode
$P_{\phi_s}$
and $P_{\phi_{\bar s}}$-Hilbert space which forms an integral lattice.
However, a ground state of localized zero mode
$| {\phi_{0}}_s, {\phi_{0}}_{\bar s}  \rangle$ projected to $P_{\phi}$ only forms a sublattice (of unit 2) of the full Hilbert space.
The arrows show the hopping from the cosine term $\cos(\ell_{I}^{} \cdot {\phi_{0}}_{I})$, shown along the $\ell_{I}=(2,2)$ and $(2,0)$ directions.
In general, the hoppings are allowed in $\ell_{I}=(2 \rn, 2 \rm)$ for $\forall$ $\rn, \rm \in \Z$.
Thus, the black larger disk shows a ground state occupying a sublattice, while the white dot shows another sublattice that
 can be occupied by another representation of a ground state, in  the same $P_{\phi_{}}$-Hilbert space.
}
\label{fig:zero-mode-DS-1}
\end{figure}

In the zero mode ${\phi_{0}}$-Hilbert space $\mathcal{H}$,
we see that $\mathcal{H}=\{ | {\phi_{0}}_s, {\phi_{0}}_{\bar s}  \rangle \}$
where $({\phi_{0}}_s, {\phi_{0}}_{\bar s})$ $=$ $(0,0),$ $(\pi,0),$ $(0,\pi),$ $(\pi,\pi)$  $\mod (2 \pi, 2 \pi)$,
due to the possible presence of all these terms:
$\cos( 2(\phi_{s} \pm \phi_{\bar s} ))$,
$\cos( 2 \phi_{s})$ and $ \cos( 2 \phi_{\bar s})$ in \eqn{S-int-s-s}.
This is shown in the left hand side of Fig.~\ref{fig:zero-mode-DS-1} (a)/(b).

In the winding mode ${P_{\phi}}$-Hilbert space $\mathcal{H}$,
we see that $\mathcal{H}=\{ | {P_{\phi_s}} , {P_{\phi_{\bar s}}}   \rangle \}$ with both ${P_{\phi_I}} \in \Z$, thus it forms an integral (2-dimensional) lattice,
shown in the right hand side of Fig.~\ref{fig:zero-mode-DS-1} (a)/(b).

How do we compute topological GSD contributed from zero/winding modes?
Naively, the minimum of all $({\phi_{0}}_s, {\phi_{0}}_{\bar s})$ contribute different GSD.
However, the $\cos(\ell_{I}^{} \cdot {\phi_{0}}_{I})$ may be shifted $\cos(\ell_{I}^{} \cdot {\phi_{0}}_{I}+\delta)$ by some $\delta$,
thus the minimum of $({\phi_{0}}_s, {\phi_{0}}_{\bar s})$ could contain \emph{accidental symmetry breaking} GSD, not
 the \emph{topological} GSD.
 Moreover, we should be able to switch topological sector to different ground states,
 if we transport the condensed anyons from one edge/domain to the other edge/domain, similar to Laughlin's thought experiment.
 For example, derived in \cite{1212.4863WW}, by threading background flux $\Delta \Phi_B$ by a unit,
 the winding mode also jumps a unit, as $q_I \Delta \Phi_B/(\frac{h}{e})=\Delta P_{\phi,I}$, up to the charge coupling.
If we consider a single edge/domain setting, those minimums of zero modes cannot be transported to different edge,
but they contribute GSD naively, thus some of them (in $(0,0),$ $(\pi,0),$ $(0,\pi)$) are accidental or redundant GSD.

\begin{figure}[t!]
  \centering
      \includegraphics[width=5.2in]{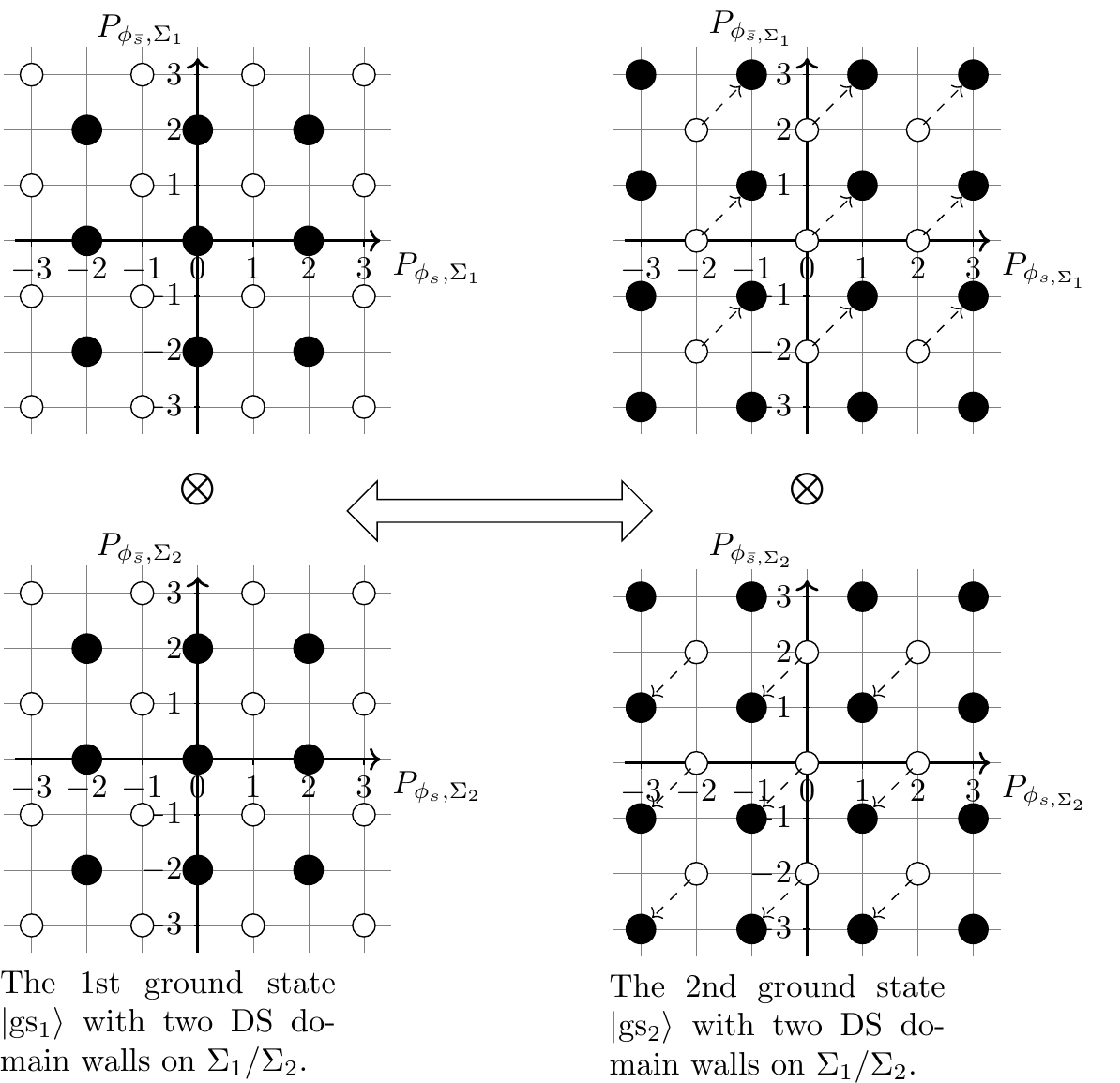} 
\caption{We illustrate the two ground states, shown in the $P_{\phi_{}}$-Hilbert space,
when the quantum state is on a real space with no nontrivial 1-cycle (e.g. flat sample $R^2$, or $S^2$) with two gapped double semion (DS) condensed domain walls ($\Sigma_1$/$\Sigma_2$).
The 1st ground state $|${gs}${}_1\rangle$, on the left side, can be understood as a tensor product ($\otimes$) of two copies of Fig.~\ref{fig:zero-mode-DS-1} (a) on two edges/domains $\Sigma_1$ and $\Sigma_2$.
The 2nd ground state $|${gs}${}_2\rangle$, on the right side, can be understood as a tensor product ($\otimes$) of
two copies of Fig.~\ref{fig:zero-mode-DS-1} (b) on two edges/domains $\Sigma_1$ and $\Sigma_2$.
The $|${gs}${}_1\rangle$ and $|${gs}${}_2\rangle$ are related by transporting a pair of double semion $s \bar s$ from one edge to another edge,
causing $(\Delta P_{\phi,s}, \Delta P_{\phi,\bar s})=\ell'=\pm (1,1)$ respectively.
This process is shown in terms of dashed arrows.
}
\label{fig:zero-mode-DS-2-dw}
\end{figure}

To correctly capture the topological GSD, we could better use the winding mode ${P_{\phi}}$-Hilbert space $\mathcal{H}$.
We could project any zero mode basis to winding mode via
$\sum_{ \underset{n_\ra\in \mathbb{Z}, \; \forall \ra}{{P_{\phi_J}} =n_\ra \ell_{\ra,J},}} | {P_{\phi_J}} \rangle  \cdot \langle{P_{\phi_J}} | {\phi_{0}}_{I} \rangle,$
and see that the ${P_{\phi_J}}$ forms a sublattice of integral lattice, see the right hand side of Fig.~\ref{fig:zero-mode-DS-1} (a)/(b).
It is a sublattice with a minimal distance of 2 due to the hopping term from the $\cos(\ell_{I}^{} \cdot {\phi_{0}}_{I})$
$=\frac{(e^{i\ell_{I}^{} \cdot {{\phi_{0}}}_{I}}+e^{-i\ell_{I}^{} \cdot {{\phi_{0}}}_{I}})}{2}$
 with a multiple of 2 for $\ell_{I}$ in \eqn{S-int-s-s}.
Again, on a \emph{single} domain, there is no way to transport the condensed double-semions ($s \bar s$) to a different domain,
the two ways of ${P_{\phi_J}}$-Hilbert space projection in Fig.~\ref{fig:zero-mode-DS-1} (a)/(b) do not imply topological GSD but only a redundancy.
However, if we have two edges/domains (say, at $\Sigma_1, \Sigma_2$), there are two robust topological GSD, shown in Fig.~\ref{fig:zero-mode-DS-2-dw}.
%
%
There
we draw on the left hand side one ground state in
$| {P_{\phi_s}, \Sigma_1} , {P_{\phi_{\bar s}},\Sigma_1}   \rangle$ $\otimes$
$| {P_{\phi_s},\Sigma_2} , {P_{\phi_{\bar s}},\Sigma_2}   \rangle$ basis,
and the right hand side another ground state in the same basis.
The left hand side ground state (up to a choice of projection for a shifting redundancy),
however, now can be transported to a different ground state, if we (say, adiabatically)
drag one double-semion pair ($s \bar s$, that is $\ell'=(1,1)$) from one edge
to another edge then condense this $s \bar s$.
%
In Fig.~\ref{fig:zero-mode-DS-2-dw}, this ground state changing from $|${gs}${}_1\rangle$ to $|${gs}${}_2\rangle$  is demonstrated in terms of
the dashed arrows.
In this case,
$q_I \Delta \Phi_B/(\frac{h}{e})=(\Delta P_{\phi,s}, \Delta P_{\phi,\bar s})=(1,1) $ say on one edge $\Sigma_1$'s Hilbert space,
but $(\Delta P_{\phi,s}, \Delta P_{\phi,\bar s})=(-1,-1)$ on the other edge  $\Sigma_2$'s Hilbert space.
The $|${gs}${}_1\rangle$ and $|${gs}${}_2\rangle$ occupy distinct topological GSD in the Hilbert space.
In a field theory view \cite{1801.05416WOP},
we can say that the topological vacua are tunneled into each other via an extended operator
(here the semionic line operator $\int (a_s + a_{\bar s})$ in
\eqn{eq:Sbulk})
with open ends, ending on two sides of different domain walls. Each end has a pair of double-semions $s \bar s$, and there are two pairs of such anyons
on two ends.
However, when $(\Delta P_{\phi,s}, \Delta P_{\phi,\bar s})=(0,0) \mod 2$, we go back to the same ground state.
Namely, transporting even units of ($s \bar s$) from one to another domain wall has \emph{no} effect on changing ground states.
More generally, we can deduce that for $n$ domain walls, there are additional topological GSD=$2^{n-1}$ contributed from these zero/winding mode sectors.

\section{Bootstrap (non-)Abelian Topological Domain Walls}
\label{app:Bootstrap}

Here we like to report some analysis of a more general domain wall theory suitable for generic non-Abelian topological orders/TQFT in 2+1D, developed in
\cite{1408.6514LWW}.
This can be viewed as a generalization of Lagrangian subgroup algebra, from Abelian to non-Abelian cases.
We like to implement the approach \cite{1408.6514LWW} to study the domain walls of non-Abelian Pf$\mid$APf quantum Hall states.
This will \emph{either} confirm our previous Abelian analysis in Appendix \ref{app:Lstability}, or \emph{reveal}
something new (which is missed by Abelian Lagrangian subgroup) intrinsically for non-Abelian topological orders.

First, we require the data of topological orders/TQFT, the SL(2,$\Z$) modular representation of
$\cS^{xy} \equiv \cS$ and $\cT^{xy} \equiv \cT$ matrices, given in Table \ref{table:TO-ST}.
The $\cS$ and $\cT$ capture the mutual braiding statistics and self exchange statistics (the later is equivalent to spin statistics),
when written in the canonical quasi-particle (anyon) basis.
In the canonical basis, we have a diagonal $\cT$ matrix such that each entry $\cT_{IJ}=\exp(i 2 \pi s_I) \delta_{IJ}$ tells the spin statistics of anyon labeled by the $\int a_I$ line operator.

\subsection{Non-perturbative bootstrap on topological 2-surface defects}

\label{app:bootstrap-1}

Below we implement a non-perturbative bootstrap on
 topological 1+1D domain walls/boundaries of topological orders,
 based on the method introduced in Ref.\cite{1408.6514LWW}.
This means that we can bootstrap 2-dimensional topological surface defects (2-surface defects),
given a 2+1D TQFT, for both Abelian and non-Abelian TQFTs.
Mathematically, classifying topological 2-surface defects
implies a classification of bimodule categories between modular tensor categories (for bosonic TQFTs).

Ref.\cite{1408.6514LWW} labels the
topological 2-surface defect (1+1D topological domain wall)
as a \emph{tunneling matrix} $\cW$ that we reviewed below.

Given two generic (non-)Abelian TQFTs, say topological orders of $A$ and $B$, with the data of modular matrices and chiral central charges
$\cS^A, \cT^A,c^A_-$ and $\cS^B,\cT^B,c^B_-$.
Say we have $M$ and $N$ types of line operators (anyons) for TQFT$_A$ and TQFT$_B$,
then respectively the rank of modular matrices are $M$ and $N$.
In our treatment, we can first isolate the gapless sector (those chiral sectors cannot be gapped out)
away from the possible gappable sectors.
If $A$ and $B$ are connected by a gapped 2-surface defect, their central
charges require to be equal $c^A_-=c^B_-$, at least required for the gappable sector.
Ref.\cite{1408.6514LWW} introduces the domain wall defect labeled by
a $N \times M$ \emph{tunneling matrix} $\cW$. Each entry $\cW_{Ia}$ represents \emph{fusion-space dimensions} within
 non-negative integers $\mathbb{Z}_{\geq 0}$:
\begin{align}
\cW_{Ia}\in\mathbb{Z}_{\geq 0},
 \label{Winteger}
\end{align}
satisfying
a \emph{commuting criterion} \eqref{commute}:
\begin{align}
  \cW \cS^A = \cS^B \cW,\quad  \cW \cT^A = \cT^B \cW,
  \label{commute}
\end{align}
(similar to Lagrangian subgroup,  \eqn{commute} imposes the consistency of anyon statistics to condense on a gapped domain wall),
and a \emph{stable criterion} \eqref{stable}:
\begin{align}
  \cW_{ia}\cW_{jb}\leq\sum_{lc} (\cN^B)_{ij}^l \cW_{lc} (\cN^A)_{ab}^c\,.
  \label{stable}
\end{align}
Here
$a,b,c,\dots$/$i,j,k,\dots$ are anyon (line operator) indices for TQFT$_A$/TQFT$_B$.
Given modular $\cS^A/\cS^B$, the fusion rules
$(\cN^A)_{ab}^c$ and $(\cN^B)_{ij}^k$, for TQFT$_A$/TQFT$_B$,
are easily determined by Verlinde formula \cite{Verlinde:1988sn},
$\cN_{ab}^c=\sum_{\al} \frac{\cS_{a \al} \cS_{b \al}{\cS_{c \al}^*}}{\cS_{1 \al}}\in \mathbb{Z}_{\geq 0}$.
The criteria whether there exists topological 2-surface defect/domain wall, is equivalent to,
whether there exists a non-zero solution $\cW$ under \eqref{Winteger},\eqref{commute} and \eqref{stable}.
(Although additional subtleties can happen, see Reference in \cite{1408.6514LWW}.)
We can bootstrap topological 2-surface defect/domain wall
between two TQFTs by analytically exhausting all solutions of $\cW$.

A tunneling matrix entry $\cW_{ia}$ means that the anyon $a$ in
TQFT$_A$ has a number of
$\cW_{ia}$-splitting channels from $a$ to $i$ after going through domain wall
 to TQFT$_B$.
Moreover, it is well-known that we can use the \emph{folding trick} to relate a gapped domain wall to a gapped boundary.
Thus we can bootstrap topological 2-surface defect both in the bulk domain wall and on the boundary.

\subsection{Gapped Pf$\mid$APf$\mid$Pf domain walls: Curious non-Abelian examples} 

\label{app:Pf/APf/Pf}

\subsubsection{Pf$\mid$APf domain wall} 

Now we like to bootstrap various types of topological domain walls at the interface of Pf$\mid$APf
discussed in the Figures.~\ref{RD}, ~\ref{Modes} and \ref{figGSD}.
First we remind the readers that we will focus on the bosonic sectors of Pf/APf quantum Hall states.
For bosonic sectors,
we recall Table \ref{table:Sbulkedge} that Pf quantum Hall state is a Semion $\otimes$ Ising topological order (which is a $\rU(1)_2 \times$ Ising TQFT)
and APf quantum Hall state is a Semion $\otimes$ $\rSU(2)_{-2}$ topological order (which is a $\rU(1)_2 \times \rSU(2)_{-2}$ TQFT).
The Pf$\mid$APf interface can be viewed as a domain wall between
$\rU(1)_2 \times$ Ising and $\rU(1)_2 \times \rSU(2)_{-2}$ TQFTs.
By folding trick,
we can view it as a boundary of $\rU(1)_2 \times \rU(1)_{-2}$ CS $\times$ Ising $\times$ SU(2)$_{2}$ TQFT to the trivial vacuum,
i.e. a boundary of double-semions $\times$ Ising $\times$ SU(2)$_2$ TQFT to a vacuum.
Since the $c_-=2$ has 4 chiral Majorana-Weyl modes on the 1+1D interface,
we find that the bootstrap method in Appendix \ref{app:bootstrap-1} for this non-Abelian TQFT
actually gives the same result as the Abelian version of Lagrangian subgroup in \ref{app:Lstability}.
Namely, the sector that can be gapped out is the double-semion theory with $K_{\text{DS}}$ in \ref{app:PfAPf-KDS}.

\subsubsection{Gapped Pf$\mid$APf$\mid$Pf domain walls}

\label{app:nAbDW}

Next we consider Pf$\mid$APf$\mid$Pf interface, shown at the corner within the dashed circle of Fig.~\ref{RD}, if the
Pf$\mid$APf$\mid$Pf  interface is very closely
joined together like a junction.\footnote{
However, we stress that there is a \emph{limitation} of applying the bootstrap method to this example of domain walls in Fig.~\ref{RD}.
Ideally, we require that the Pf$\mid$APf$\mid$Pf interface to be perfectly joined together in a compact circle in a spatial region.
But this is not precisely the case in Fig.~\ref{RD}. The applicability of new Pf$\mid$APf$\mid$Pf domain wall studied in this section depends on
the shared length and the size of Pf$\mid$APf$\mid$Pf region, in experimental set-ups.
}
Again by folding trick, we can study the problem in terms of
the boundary of
 $(\rU(1)_2 \times \rU(1)_{-2})^2$ $\times$ Ising $\times$ $\overline{\text{Ising}}$
 $\times$ SU(2)$_{2}$ $\times$ SU(2)$_{-2}$ TQFT to the trivial vacuum,
i.e. a boundary of (double-semions)$^2$ $\times$ double-Ising $\times$ double-SU(2)$_2$ TQFT to a vacuum.
The rank of modular matrices $\cS$ and $\cT$ for such theory is much higher, as $2^4 \cdot 3^4=1296$.
It means that there are 1296 distinct anyons sectors/line operators, and the GSD on $T^2$-spatial torus is 1296.
For simplicity, we can consider that, the Abelian sector from  $(\rU(1)_2 \times \rU(1)_{-2})^2$ CS theory (with 16 anyons), and
non-Abelian sector from Ising $\times$ $\overline{\text{Ising}}$
 $\times$ SU(2)$_{2}$ $\times$ SU(2)$_{-2}$ TQFT (with 81 anyons), separately.

Again, the Abelian sector from  $(\rU(1)_2 \times \rU(1)_{-2})^2$ CS theory can be tackled by the simpler method in Appendix \ref{app:Lstability},
gapping by two sets of cosine terms. Moreover, in terms of  Appendix \ref{app:bootstrap-1}'s bootstrap method, we can obtain the tunneling data between
a double-semion theory and a trivial vacuum as a $4\times 1$ tunneling matrix  \cite{1408.6514LWW}:
\bea
\cW =
\left(
\begin{array}{ccccc}
1 &  s & \bar s   & s  \bar s  & \\
\hline
1 & 0 & 0 & 1 \;\vline & 1
\end{array}
\right),
\eea
In this language, we can view a double semion condensation to $s \bar s$ to a trivial vacuum after crossing the boundary,
\bea
1\oplus  s \bar s \leftrightarrow  1.
\eea
Alternatively, we can also view as the domain wall between two semion theories with this (relatively trivial) tunneling data
\bea
1\leftrightarrow 1, \quad s   \leftrightarrow  s.
\eea

The non-Abelian sector from Ising $\times$ $\overline{\text{Ising}}$
 $\times$ SU(2)$_{2}$ $\times$ SU(2)$_{-2}$ TQFT, requires the Appendix \ref{app:bootstrap-1}'s bootstrap method.
 We can use the folding trick again to consider this problem equivalently as
interfaces between Ising $\times$ SU(2)$_{2}$ TQFT
and  Ising $\times$ SU(2)$_{2}$ TQFT
(each of them has a set of 9 anyons $\{1, \sigma, \psi\} \otimes \{1,\sigma_3, \psi_3\}$).
 In particular, we find two types of domain wall data, $\cW_1$ and $\cW_2$:
\bea
\cW_1&=&\left(
\begin{array}{cccccccccc}
1 &  \sigma_3 & \psi_3 & \sigma & \sigma  \sigma_3 & \sigma \psi_3 & \psi & \psi  \sigma_3 & \psi  \psi_3   & \\
\hline
 1 & 0 & 0 & 0 & 0 & 0 & 0 & 0 & 0 \;\vline & 1 \\
 0 & 1 & 0 & 0 & 0 & 0 & 0 & 0 & 0  \;\vline & \sigma_3\\
 0 & 0 & 1 & 0 & 0 & 0 & 0 & 0 & 0 \;\vline &\psi_3\\
 0 & 0 & 0 & 1 & 0 & 0 & 0 & 0 & 0 \;\vline &\sigma\\
 0 & 0 & 0 & 0 & 1 & 0 & 0 & 0 & 0 \;\vline &\sigma  \sigma_3\\
 0 & 0 & 0 & 0 & 0 & 1 & 0 & 0 & 0 \;\vline  &\sigma \psi_3\\
 0 & 0 & 0 & 0 & 0 & 0 & 1 & 0 & 0 \;\vline &\psi\\
 0 & 0 & 0 & 0 & 0 & 0 & 0 & 1 & 0 \;\vline &\psi  \sigma_3\\
 0 & 0 & 0 & 0 & 0 & 0 & 0 & 0 & 1  \;\vline &\psi  \psi_3\\
\end{array}
\right),\\
\cW_2&=&\left(
\begin{array}{cccccccccc}
1 &  \sigma_3 & \psi_3 & \sigma & \sigma  \sigma_3 & \sigma \psi_3 & \psi & \psi  \sigma_3 & \psi  \psi_3   & \\
\hline
 1 & 0 & 0 & 0 & 0 & 0 & 0 & 0 & 1 \;\vline & 1 \\
 0 & 0 & 0 & 0 & 0 & 0 & 0 & 0 & 0  \;\vline & \sigma_3\\
 0 & 0 & 1 & 0 & 0 & 0 & 1 & 0 & 0  \;\vline &\psi_3\\
 0 & 0 & 0 & 0 & 0 & 0 & 0 & 0 & 0  \;\vline &\sigma\\
 0 & 0 & 0 & 0 & 2 & 0 & 0 & 0 & 0  \;\vline &\sigma  \sigma_3\\
 0 & 0 & 0 & 0 & 0 & 0 & 0 & 0 & 0  \;\vline  &\sigma \psi_3\\
 0 & 0 & 1 & 0 & 0 & 0 & 1 & 0 & 0  \;\vline &\psi\\
 0 & 0 & 0 & 0 & 0 & 0 & 0 & 0 & 0  \;\vline &\psi  \sigma_3\\
 1 & 0 & 0 & 0 & 0 & 0 & 0 & 0 & 1  \;\vline &\psi  \psi_3\\
\end{array}
\right)
\eea
The $\cW_1$  in terms of a $9 \times 9$ matrix reveals the tunneling data between two Ising $\times$ SU(2)$_{2}$ TQFTs:
\bea
\begin{array}{c}
1\leftrightarrow 1, \sigma_3 \leftrightarrow \sigma_3, \psi_3 \leftrightarrow \psi_3,\\
\sigma \leftrightarrow \sigma , \sigma  \sigma_3 \leftrightarrow  \sigma  \sigma_3 ,
\sigma \psi_3 \leftrightarrow \sigma \psi_3 , \\
\psi \leftrightarrow \psi,
\psi  \sigma_3 \leftrightarrow \psi  \sigma_3,
\psi  \psi_3 \leftrightarrow \psi  \psi_3.
\end{array}
\eea
Equivalently, if we view $\cW_1$ as the boundary to the trivial vacuum,
by the folding trick,  we can rewrite it as a $81 \times 1$ matrix whose tunneling data is:
\bea
  1\oplus  \sigma \bar  \sigma \oplus  \psi  \bar \psi \oplus
   \sigma_3 \bar  \sigma_3 \oplus (\sigma_3 \bar  \sigma_3)(\sigma \bar  \sigma)
   \oplus    (\sigma_3 \bar  \sigma_3) ( \psi  \bar \psi)
   \oplus ( \psi_3  \bar \psi_3)
      \oplus ( \psi_3  \bar \psi_3)(\sigma \bar  \sigma)
            \oplus ( \psi_3  \bar \psi_3)( \psi  \bar \psi)
   \leftrightarrow 1.
\eea
The $\cW_1$ is rather an obvious domain wall in terms of an identity (tunneling) map.
However, the $\cW_2$ reveals a different but more curious tunneling data between two Ising $\times$ SU(2)$_{2}$ TQFTs:
\bea
\begin{array}{rcl}
1 \oplus \psi \psi_3 &\leftrightarrow& 1 \oplus \psi \psi_3,\\
  \psi \oplus \psi_3 &\leftrightarrow&  \psi \oplus \psi_3,\\
    2 \sigma \sigma_3 &\leftrightarrow&  2 \sigma \sigma_3.\\
\end{array}
\eea
In some sense, the domain wall  $\cW_2$ is more \emph{non-Abelian} than the identity domain wall $\cW_1$.
We like to capture/contrast their physical properties by computing their GSDs in Appendix \ref{app:nAb-GSD}.

\subsection{Zero Modes and GSD from (non-)Abelian domain walls}

\label{app:nAb-GSD}

We can compute the topological GSD, here focusing on a flat substrate or a sphere $S^2$, in the presence of 1+1D topological domain walls (2-surface defects,
studied earlier in Appendix \ref{app:zero-GSD}/\ref{app:Pf/APf/Pf})
for 2+1D TQFTs:
\begin{itemize}
\item For $n$ double-semion ($\rU(1)_2 \times \rU(1)_{-2}$ Chern-Simons theory) domain walls, the topological GSD grows as $\GSD=2^{n-1}$.
\item For $n$ double-Ising (Ising $\times$ $\overline{\text{Ising}}$ TQFT) domain walls, the topological GSD grows as $\GSD=1,3, 10, 36, 136, \dots$
for $n=1,2,3,4,5, \dots$, which is much faster than $3^{n-1}$.

\item For $n$ Pf$\mid$APf$\mid$Pf domain walls of $\cW_1$ type,
the topological GSD grows as $\GSD=1,9$, $100, 1296$, $\dots$
for $n=1,2,3,4, \dots$, which is much faster than $8^{n-1}$.
(Note that the gapped double-semions$^2$ sectors contribute additional GSD.)

\item For $n$ Pf$\mid$APf$\mid$Pf domain walls of $\cW_2$ type,
the topological GSD grows as $\GSD=1, 12$, $160, 2304$, $\dots$
for $n=1,2,3,4, \dots$, which is much faster than $12^{n-1}$.
(Note that the gapped double-semions$^2$ sectors contribute additional GSD.)

\end{itemize}

Here are some remarks:

\noindent
1. We can derive the general GSD with domain walls, simply given the domain wall tunneling matrix $\cW$ and the fusion rule $\cN_{ab}^c$, based on
generalizing a formula in Ref.\cite{1408.6514LWW}.

\noindent
2. The topological GSD of a given TQFT on a flat substrate or a sphere $S^2$, in the presence of 0+1D anyons, can also be computed.
In TQFT, this means the path integral $Z$ with $n$ insertions ($n$ punctures) on a sphere,
say $\GSD=Z(S^2 \times S^1; \sigma_1,\sigma_2,\sigma_3, \dots)$.
This data is fully determined by the fusion rule $\cN_{ab}^c$ alone.
For instance:
\begin{itemize}
\item For $n$ semions $s$ (or $n$ double-semions) insertions in the semion (or double-semion) TQFT,
the $\GSD=0$ or $1$ (i.e. 0 means the configuration is not allowed).
This is a signature of an Abelian TQFT.
\item For $n$ anyons of $(\sigma \bar  \sigma)$ insertions in double-Ising TQFT,
the $\GSD=0, 1, 0, 4, 0, 16,0, 64, \dots$
for $n=1,2,3,4, 5,6, 7,8, \dots$.
The GSD goes as either $0$ (not allowed configurations) or $2^{n-2}$, where $2$ is the quantum dimension of $(\sigma \bar  \sigma)$ .
\item For any $n$ anyon $\alpha$ insertions in double-Ising $\times$ double-SU(2)$_2$ TQFT,
the GSD goes like $\GSD \simeq (d_\alpha)^n$ (or no allowed state GSD=0) for large $n$, bounded by
its anyon quantum dimension $d_\alpha$ to the $n$-th power. However,
$d_\alpha = 1, \sqrt{2}, 2, 2 \sqrt{2}, 4$ for this TQFT.
Thus its anyon-insertion $\GSD \leq  4^n  < 10^{n-1}  < 12^{n-1}$ grows again much slower than the domain wall  $\cW_1$ or $\cW_2$'s GSD.
\end{itemize}
In contrast, this GSD caused by anyon insertion grows much \emph{slower} than the \emph{domain wall GSD}.
The domain wall GSD for any (non-)Abelian topological orders/TQFT can still have an exponential growth for degenerate states
on a sphere $S^2$ (which is impossible for GSD caused by Abelian anyon insertions alone, by definition of the fusion rule).

\noindent
3. Since the domain wall GSD of $\cW_2$ grows more rapidly than that of $\cW_1$,
it suggests that the domain wall $\cW_2$ is more \emph{non-Abelian} in nature, in an intriguing way.
Detailed investigations on these domain walls are left for the future.

Finally, further more intricate domain walls from the joined Abelian and non-Abelian sectors
(with a $36 \times 36$ anyon tunneling matrix, or by folding trick
with a tunneling matrix of 1296 anyons to a trivial sector) for the bosonic TQFT sector of Pf$\mid$APf$\mid$Pf interface, and
also the full fermionic TQFT sector of Pf$\mid$APf$\mid$Pf interface (e.g. including additional fermionic sector and spin structures for each TQFT),
are planned to be studied in the future.

\bibliographystyle{arXiv}
\bibliography{TH-pf-Ref,juven-Ref}

\providecommand{\href}[2]{#2}\begingroup\raggedright\begin{thebibliography}{10}

\bibitem{willett1987}
R.~Willett, J.~P. Eisenstein, H.~L. St\"ormer, D.~C. Tsui, A.~C. Gossard and
  J.~H. English, \emph{Observation of an even-denominator quantum number in the
  fractional quantum hall effect},
  \href{http://dx.doi.org/10.1103/PhysRevLett.59.1776}{\emph{Phys. Rev. Lett.}
  {\bf 59} (Oct, 1987) 1776--1779}.

\bibitem{morf1998}
R.~H. Morf, \emph{Transition from quantum hall to compressible states in the
  second landau level: New light on the
  $\ensuremath{\nu}\phantom{\rule{0ex}{0ex}}=\phantom{\rule{0ex}{0ex}}5/2$
  enigma}, \href{http://dx.doi.org/10.1103/PhysRevLett.80.1505}{\emph{Phys.
  Rev. Lett.} {\bf 80} (Feb, 1998) 1505--1508}.

\bibitem{rezayi2000}
E.~H. Rezayi and F.~D.~M. Haldane, \emph{Incompressible paired hall state,
  stripe order, and the composite fermion liquid phase in half-filled landau
  levels}, \href{http://dx.doi.org/10.1103/PhysRevLett.84.4685}{\emph{Phys.
  Rev. Lett.} {\bf 84} (May, 2000) 4685--4688}.

\bibitem{peterson2008}
M.~R. Peterson, T.~Jolicoeur and S.~Das~Sarma, \emph{Finite-layer thickness
  stabilizes the pfaffian state for the 5/2 fractional quantum hall effect:
  Wave function overlap and topological degeneracy},
  \href{http://dx.doi.org/10.1103/PhysRevLett.101.016807}{\emph{Phys. Rev.
  Lett.} {\bf 101} (Jul, 2008) 016807}.

\bibitem{feiguin2009}
A.~E. Feiguin, E.~Rezayi, K.~Yang, C.~Nayak and S.~Das~Sarma, \emph{Spin
  polarization of the $\ensuremath{\nu}=5/2$ quantum hall state},
  \href{http://dx.doi.org/10.1103/PhysRevB.79.115322}{\emph{Phys. Rev. B} {\bf
  79} (Mar, 2009) 115322}.

\bibitem{wangh2009}
H.~Wang, D.~N. Sheng and F.~D.~M. Haldane, \emph{Particle-hole symmetry
  breaking and the $\ensuremath{\nu}=\frac{5}{2}$ fractional quantum hall
  effect}, \href{http://dx.doi.org/10.1103/PhysRevB.80.241311}{\emph{Phys. Rev.
  B} {\bf 80} (Dec, 2009) 241311}.

\bibitem{storni2010}
M.~Storni, R.~H. Morf and S.~Das~Sarma, \emph{Fractional quantum hall state at
  $\ensuremath{\nu}=\frac{5}{2}$ and the moore-read pfaffian},
  \href{http://dx.doi.org/10.1103/PhysRevLett.104.076803}{\emph{Phys. Rev.
  Lett.} {\bf 104} (Feb, 2010) 076803}.

\bibitem{rezayi2011}
E.~H. Rezayi and S.~H. Simon, \emph{Breaking of particle-hole symmetry by
  landau level mixing in the $\ensuremath{\nu}=5/2$ quantized hall state},
  \href{http://dx.doi.org/10.1103/PhysRevLett.106.116801}{\emph{Phys. Rev.
  Lett.} {\bf 106} (Mar, 2011) 116801}.

\bibitem{papic2012}
Z.~Papi\ifmmode~\acute{c}\else \'{c}\fi{}, F.~D.~M. Haldane and E.~H. Rezayi,
  \emph{Quantum phase transitions and the $\ensuremath{\nu}\mathbf{=}5/2$
  fractional hall state in wide quantum wells},
  \href{http://dx.doi.org/10.1103/PhysRevLett.109.266806}{\emph{Phys. Rev.
  Lett.} {\bf 109} (Dec, 2012) 266806}.

\bibitem{zaletel2015}
M.~P. Zaletel, R.~S.~K. Mong, F.~Pollmann and E.~H. Rezayi, \emph{Infinite
  density matrix renormalization group for multicomponent quantum hall
  systems}, \href{http://dx.doi.org/10.1103/PhysRevB.91.045115}{\emph{Phys.
  Rev. B} {\bf 91} (Jan, 2015) 045115}.

\bibitem{pakrouski2015}
K.~Pakrouski, M.~R. Peterson, T.~Jolicoeur, V.~W. Scarola, C.~Nayak and
  M.~Troyer, \emph{Phase diagram of the $\ensuremath{\nu}=5/2$ fractional
  quantum hall effect: Effects of landau-level mixing and nonzero width},
  \href{http://dx.doi.org/10.1103/PhysRevX.5.021004}{\emph{Phys. Rev. X} {\bf
  5} (Apr, 2015) 021004}.

\bibitem{moore1991}
G.~Moore and N.~Read, \emph{Nonabelions in the fractional quantum hall effect},
  \href{http://dx.doi.org/http://dx.doi.org/10.1016/0550-3213(91)90407-O}{\emph{Nucl.
  Phys. B} {\bf 360} (1991) 362 -- 396}.

\bibitem{read2000}
N.~Read and D.~Green, \emph{Paired states of fermions in two dimensions with
  breaking of parity and time-reversal symmetries and the fractional quantum
  hall effect}, \href{http://dx.doi.org/10.1103/PhysRevB.61.10267}{\emph{Phys.
  Rev. B} {\bf 61} (Apr, 2000) 10267--10297}.

\bibitem{levin2007}
M.~Levin, B.~I. Halperin and B.~Rosenow, \emph{Particle-hole symmetry and the
  pfaffian state},
  \href{http://dx.doi.org/10.1103/PhysRevLett.99.236806}{\emph{Phys. Rev.
  Lett.} {\bf 99} (Dec, 2007) 236806}.

\bibitem{lee2007}
S.-S. Lee, S.~Ryu, C.~Nayak and M.~P.~A. Fisher, \emph{Particle-hole symmetry
  and the $\ensuremath{\nu}=\frac{5}{2}$ quantum hall state},
  \href{http://dx.doi.org/10.1103/PhysRevLett.99.236807}{\emph{Phys. Rev.
  Lett.} {\bf 99} (Dec, 2007) 236807}.

\bibitem{1991Wen}
X.~G. Wen, \emph{Non-abelian statistics in the fractional quantum hall states},
  \href{http://dx.doi.org/10.1103/PhysRevLett.66.802}{\emph{Phys. Rev. Lett.}
  {\bf 66} (Feb, 1991) 802--805}.

\bibitem{kane1997}
C.~L. Kane and M.~P.~A. Fisher, \emph{Quantized thermal transport in the
  fractional quantum hall effect},
  \href{http://dx.doi.org/10.1103/PhysRevB.55.15832}{\emph{Phys. Rev. B} {\bf
  55} (Jun, 1997) 15832--15837}.

\bibitem{WenPhysRevLett.70.355}
X.-G. Wen, \emph{Topological order and edge structure of \ensuremath{\nu}=1/2
  quantum hall state},
  \href{http://dx.doi.org/10.1103/PhysRevLett.70.355}{\emph{Phys. Rev. Lett.}
  {\bf 70} (Jan, 1993) 355--358}.

\bibitem{banerjee2017}
M.~{Banerjee}, M.~{Heiblum}, V.~{Umansky}, D.~E. {Feldman}, Y.~{Oreg} and
  A.~{Stern}, \emph{{Observation of half-integer thermal Hall conductance}},
  {\emph{ArXiv e-prints} (Oct., 2017) },
  [\href{https://arxiv.org/abs/1710.00492}{{\tt arXiv:1710.00492}}].

\bibitem{zucker2016}
P.~T. Zucker and D.~E. Feldman, \emph{Stabilization of the particle-hole
  pfaffian order by landau-level mixing and impurities that break particle-hole
  symmetry},
  \href{http://dx.doi.org/10.1103/PhysRevLett.117.096802}{\emph{Phys. Rev.
  Lett.} {\bf 117} (Aug, 2016) 096802}.

\bibitem{son2015}
D.~T. Son, \emph{Is the composite fermion a dirac particle?},
  \href{http://dx.doi.org/10.1103/PhysRevX.5.031027}{\emph{Phys. Rev. X} {\bf
  5} (Sep, 2015) 031027}.

\bibitem{chen2014}
X.~Chen, L.~Fidkowski and A.~Vishwanath, \emph{Symmetry enforced non-abelian
  topological order at the surface of a topological insulator},
  \href{http://dx.doi.org/10.1103/PhysRevB.89.165132}{\emph{Phys. Rev. B} {\bf
  89} (Apr, 2014) 165132}.

\bibitem{wangc2017}
C.~{Wang}, A.~{Vishwanath} and B.~I. {Halperin}, \emph{{Topological Order from
  Disorder and the Quantized Hall Thermal Metal: Possible Applications to the
  $\nu = 5/2$ State}}, {\emph{ArXiv e-prints} (Nov., 2017) },
  [\href{https://arxiv.org/abs/1711.11557}{{\tt arXiv:1711.11557}}].

\bibitem{mross2017}
D.~F. {Mross}, Y.~{Oreg}, A.~{Stern}, G.~{Margalit} and M.~{Heiblum},
  \emph{{Theory of Disorder-Induced Half-Integer Thermal Hall Conductance}},
  {\emph{ArXiv e-prints} (Nov., 2017) },
  [\href{https://arxiv.org/abs/1711.06278}{{\tt arXiv:1711.06278}}].

\bibitem{senthil2000}
T.~Senthil and M.~P.~A. Fisher, \emph{Quasiparticle localization in
  superconductors with spin-orbit scattering},
  \href{http://dx.doi.org/10.1103/PhysRevB.61.9690}{\emph{Phys. Rev. B} {\bf
  61} (Apr, 2000) 9690--9698}.

\bibitem{chalker2001}
J.~T. Chalker, N.~Read, V.~Kagalovsky, B.~Horovitz, Y.~Avishai and A.~W.~W.
  Ludwig, \emph{Thermal metal in network models of a disordered two-dimensional
  superconductor},
  \href{http://dx.doi.org/10.1103/PhysRevB.65.012506}{\emph{Phys. Rev. B} {\bf
  65} (Dec, 2001) 012506}.

\bibitem{fulga2012}
I.~C. Fulga, A.~R. Akhmerov, J.~Tworzyd\l{}o, B.~B\'eri and C.~W.~J. Beenakker,
  \emph{Thermal metal-insulator transition in a helical topological
  superconductor},
  \href{http://dx.doi.org/10.1103/PhysRevB.86.054505}{\emph{Phys. Rev. B} {\bf
  86} (Aug, 2012) 054505}.

\bibitem{imry1975}
Y.~Imry and S.-k. Ma, \emph{Random-field instability of the ordered state of
  continuous symmetry},
  \href{http://dx.doi.org/10.1103/PhysRevLett.35.1399}{\emph{Phys. Rev. Lett.}
  {\bf 35} (Nov, 1975) 1399--1401}.

\bibitem{binder1983}
K.~{Binder}, \emph{{Random-field induced interface widths in Ising systems}},
  \href{http://dx.doi.org/10.1007/BF01470045}{\emph{Zeitschrift fur Physik B
  Condensed Matter} {\bf 50} (Dec., 1983) 343--352}.

\bibitem{1212.4863WW}
J.~C. {Wang} and X.-G. {Wen}, \emph{{Boundary degeneracy of topological
  order}}, \href{http://dx.doi.org/10.1103/PhysRevB.91.125124}{\emph{Phys. Rev.
  B} {\bf 91} (Mar., 2015) 125124},
  [\href{https://arxiv.org/abs/1212.4863}{{\tt arXiv:1212.4863}}].

\bibitem{levin2013}
M.~Levin, \emph{Protected edge modes without symmetry},
  \href{http://dx.doi.org/10.1103/PhysRevX.3.021009}{\emph{Phys. Rev. X} {\bf
  3} (May, 2013) 021009}.

\bibitem{read1996}
N.~Read and E.~Rezayi, \emph{Quasiholes and fermionic zero modes of paired
  fractional quantum hall states: The mechanism for non-abelian statistics},
  \href{http://dx.doi.org/10.1103/PhysRevB.54.16864}{\emph{Phys. Rev. B} {\bf
  54} (Dec, 1996) 16864--16887}.

\bibitem{milo1996}
M.~Milovanovi\ifmmode~\acute{c}\else \'{c}\fi{} and N.~Read, \emph{Edge
  excitations of paired fractional quantum hall states},
  \href{http://dx.doi.org/10.1103/PhysRevB.53.13559}{\emph{Phys. Rev. B} {\bf
  53} (May, 1996) 13559--13582}.

\bibitem{girvin1984a}
S.~M. Girvin, \emph{Particle-hole symmetry in the anomalous quantum hall
  effect}, \href{http://dx.doi.org/10.1103/PhysRevB.29.6012}{\emph{Phys. Rev.
  B} {\bf 29} (May, 1984) 6012--6014}.

\bibitem{halperin1993}
B.~I. Halperin, P.~A. Lee and N.~Read, \emph{Theory of the half-filled landau
  level}, \href{http://dx.doi.org/10.1103/PhysRevB.47.7312}{\emph{Phys. Rev. B}
  {\bf 47} (Mar, 1993) 7312--7343}.

\bibitem{jain1989}
J.~K. Jain, \emph{Composite-fermion approach for the fractional quantum hall
  effect}, \href{http://dx.doi.org/10.1103/PhysRevLett.63.199}{\emph{Phys. Rev.
  Lett.} {\bf 63} (Jul, 1989) 199--202}.

\bibitem{mross2015}
D.~F. Mross, A.~Essin and J.~Alicea, \emph{Composite dirac liquids: Parent
  states for symmetric surface topological order},
  \href{http://dx.doi.org/10.1103/PhysRevX.5.011011}{\emph{Phys. Rev. X} {\bf
  5} (Feb, 2015) 011011}.

\bibitem{geraedts2016}
S.~D. Geraedts, M.~P. Zaletel, R.~S.~K. Mong, M.~A. Metlitski, A.~Vishwanath
  and O.~I. Motrunich, \emph{The half-filled landau level: The case for dirac
  composite fermions},
  \href{http://dx.doi.org/10.1126/science.aad4302}{\emph{Science} {\bf 352}
  (2016) 197--201},
  [\href{https://arxiv.org/abs/http://science.sciencemag.org/content/352/6282/197.full.pdf}{{\tt
  arXiv:http://science.sciencemag.org/content/352/6282/197.full.pdf}}].

\bibitem{wangc2016}
C.~Wang and T.~Senthil, \emph{Composite fermi liquids in the lowest landau
  level}, \href{http://dx.doi.org/10.1103/PhysRevB.94.245107}{\emph{Phys. Rev.
  B} {\bf 94} (Dec, 2016) 245107}.

\bibitem{potter2016}
A.~C. Potter, M.~Serbyn and A.~Vishwanath, \emph{Thermoelectric transport
  signatures of dirac composite fermions in the half-filled landau level},
  \href{http://dx.doi.org/10.1103/PhysRevX.6.031026}{\emph{Phys. Rev. X} {\bf
  6} (Aug, 2016) 031026}.

\bibitem{geraedts2017}
S.~D. {Geraedts}, J.~{Wang}, E.~{Rezayi} and F.~D.~M. {Haldane}, \emph{{Berry
  phase and model wavefunction in the half-filled Landau Level}}, {\emph{ArXiv
  e-prints} (Nov., 2017) }, [\href{https://arxiv.org/abs/1711.07864}{{\tt
  arXiv:1711.07864}}].

\bibitem{chalker1988}
J.~T. Chalker and P.~D. Coddington, \emph{Percolation, quantum tunnelling and
  the integer hall effect}, {\emph{J. Phys. C} {\bf 21} (1988) 2665}.

\bibitem{kramer2005}
B.~Kramer, T.~Ohtsuki and S.~Kettemann, \emph{Random network models and quantum
  phase transitions in two dimensions},
  \href{http://dx.doi.org/10.1016/j.physrep.2005.07.001}{\emph{Phys. Rep.} {\bf
  417} (2005) 211--342}.

\bibitem{altland1997}
A.~Altland and M.~R. Zirnbauer, \emph{Nonstandard symmetry classes in
  mesoscopic normal-superconducting hybrid structures},
  \href{http://dx.doi.org/10.1103/PhysRevB.55.1142}{\emph{Phys. Rev. B} {\bf
  55} (Jan, 1997) 1142--1161}.

\bibitem{kane1994}
C.~L. Kane, M.~P.~A. Fisher and J.~Polchinski, \emph{Randomness at the edge:
  Theory of quantum hall transport at filling \ensuremath{\nu}=2/3},
  \href{http://dx.doi.org/10.1103/PhysRevLett.72.4129}{\emph{Phys. Rev. Lett.}
  {\bf 72} (Jun, 1994) 4129--4132}.

\bibitem{haldane1995}
F.~D.~M. Haldane, \emph{Stability of chiral luttinger liquids and abelian
  quantum hall states},
  \href{http://dx.doi.org/10.1103/PhysRevLett.74.2090}{\emph{Phys. Rev. Lett.}
  {\bf 74} (Mar, 1995) 2090--2093}.

\bibitem{wan2011}
X.~Wan, A.~M. Turner, A.~Vishwanath and S.~Y. Savrasov, \emph{Topological
  semimetal and fermi-arc surface states in the electronic structure of
  pyrochlore iridates},
  \href{http://dx.doi.org/10.1103/PhysRevB.83.205101}{\emph{Phys. Rev. B} {\bf
  83} (May, 2011) 205101}.

\bibitem{balents2011}
L.~Balents, \emph{Weyl electrons kiss},
  \href{http://dx.doi.org/10.1103/Physics.4.36}{\emph{Physics} {\bf 4} (May,
  2011) 36}.

\bibitem{pruisken1988}
A.~M.~M. Pruisken, \emph{Universal singularities in the integral quantum hall
  effect}, \href{http://dx.doi.org/10.1103/PhysRevLett.61.1297}{\emph{Phys.
  Rev. Lett.} {\bf 61} (Sep, 1988) 1297--1300}.

\bibitem{wangjing2014}
J.~Wang, B.~Lian and S.-C. Zhang, \emph{Universal scaling of the quantum
  anomalous hall plateau transition},
  \href{http://dx.doi.org/10.1103/PhysRevB.89.085106}{\emph{Phys. Rev. B} {\bf
  89} (Feb, 2014) 085106}.

\bibitem{he2017}
Q.~L. He, L.~Pan, A.~L. Stern, E.~C. Burks, X.~Che, G.~Yin et~al., \emph{Chiral
  majorana fermion modes in a quantum anomalous hall
  insulator{\textendash}superconductor structure},
  \href{http://dx.doi.org/10.1126/science.aag2792}{\emph{Science} {\bf 357}
  (2017) 294--299}.

\bibitem{lian2017}
B.~{Lian}, J.~{Wang}, X.-Q. {Sun}, A.~{Vaezi} and S.-C. {Zhang}, \emph{{Quantum
  phase transition of chiral Majorana fermion in the presence of disorder}},
  {\emph{ArXiv e-prints} (Sept., 2017) },
  [\href{https://arxiv.org/abs/1709.05558}{{\tt arXiv:1709.05558}}].

\bibitem{luttinger1960}
J.~M. Luttinger, \emph{Fermi surface and some simple equilibrium properties of
  a system of interacting fermions},
  \href{http://dx.doi.org/10.1103/PhysRev.119.1153}{\emph{Phys. Rev.} {\bf 119}
  (Aug, 1960) 1153--1163}.

\bibitem{balram2015}
A.~C. Balram, C.~T\ifmmode~\mbox{\H{o}}\else \H{o}\fi{}ke and J.~K. Jain,
  \emph{Luttinger theorem for the strongly correlated fermi liquid of composite
  fermions},
  \href{http://dx.doi.org/10.1103/PhysRevLett.115.186805}{\emph{Phys. Rev.
  Lett.} {\bf 115} (Oct, 2015) 186805}.

\bibitem{berezinski1971}
V.~L. {Berezinskii}, \emph{{Destruction of Long-range Order in One-dimensional
  and Two-dimensional Systems having a Continuous Symmetry Group I. Classical
  Systems}}, {\emph{Soviet Journal of Experimental and Theoretical Physics}
  {\bf 32} (1971) 493}.

\bibitem{kosterlitz1973}
J.~M. Kosterlitz and D.~J. Thouless, \emph{Ordering, metastability and phase
  transitions in two-dimensional systems}, {\emph{Journal of Physics C: Solid
  State Physics} {\bf 6} (1973) 1181}.

\bibitem{yang2013}
G.~Yang and D.~E. Feldman, \emph{Influence of device geometry on tunneling in
  the $\ensuremath{\nu}=\frac{5}{2}$ quantum hall liquid},
  \href{http://dx.doi.org/10.1103/PhysRevB.88.085317}{\emph{Phys. Rev. B} {\bf
  88} (Aug, 2013) 085317}.

\bibitem{barkeshli2015}
M.~Barkeshli, M.~Mulligan and M.~P.~A. Fisher, \emph{Particle-hole symmetry and
  the composite fermi liquid},
  \href{http://dx.doi.org/10.1103/PhysRevB.92.165125}{\emph{Phys. Rev. B} {\bf
  92} (Oct, 2015) 165125}.

\bibitem{wan2016}
X.~Wan and K.~Yang, \emph{Striped quantum hall state in a half-filled landau
  level}, \href{http://dx.doi.org/10.1103/PhysRevB.93.201303}{\emph{Phys. Rev.
  B} {\bf 93} (May, 2016) 201303}.

\bibitem{iadecola2014}
T.~Iadecola, T.~Neupert, C.~Chamon and C.~Mudry, \emph{Accessing topological
  order in fractionalized liquids with gapped edges},
  \href{http://dx.doi.org/10.1103/PhysRevB.90.205115}{\emph{Phys. Rev. B} {\bf
  90} (Nov, 2014) 205115}.

\bibitem{1408.6514LWW}
T.~{Lan}, J.~C. {Wang} and X.-G. {Wen}, \emph{{Gapped Domain Walls, Gapped
  Boundaries, and Topological Degeneracy}},
  \href{http://dx.doi.org/10.1103/PhysRevLett.114.076402}{\emph{Physical Review
  Letters} {\bf 114} (Feb., 2015) 076402},
  [\href{https://arxiv.org/abs/1408.6514}{{\tt arXiv:1408.6514}}].

\bibitem{1602.04251SW}
N.~{Seiberg} and E.~{Witten}, \emph{{Gapped boundary phases of topological
  insulators via weak coupling}},
  \href{http://dx.doi.org/10.1093/ptep/ptw083}{\emph{Progress of Theoretical
  and Experimental Physics} {\bf 2016} (Dec., 2016) 12C101},
  [\href{https://arxiv.org/abs/1602.04251}{{\tt arXiv:1602.04251}}].

\bibitem{1612.09298PWY}
P.~{Putrov}, J.~{Wang} and S.-T. {Yau}, \emph{{Braiding statistics and link
  invariants of bosonic/fermionic topological quantum matter in 2+1 and 3+1
  dimensions}}, \href{http://dx.doi.org/10.1016/j.aop.2017.06.019}{\emph{Annals
  of Physics} {\bf 384} (Sept., 2017) 254--287},
  [\href{https://arxiv.org/abs/1612.09298}{{\tt arXiv:1612.09298}}].

\bibitem{1801.05416WOP}
J.~{Wang}, K.~{Ohmori}, P.~{Putrov}, Y.~{Zheng}, Z.~{Wan}, M.~Y.~{Guo} et~al.,
  \emph{{Tunneling Topological Vacua via Extended Operators: (Spin-)TQFT
  Spectra and Boundary Deconfinement in Various Dimensions}}, {\emph{ArXiv
  e-prints} (Jan., 2018) }, [\href{https://arxiv.org/abs/1801.05416}{{\tt
  arXiv:1801.05416}}].

\bibitem{0506438Kitaev}
A.~{Kitaev}, \emph{{Anyons in an exactly solved model and beyond}},
  \href{http://dx.doi.org/10.1016/j.aop.2005.10.005}{\emph{Annals of Physics}
  {\bf 321} (Jan., 2006) 2--111},
  [\href{https://arxiv.org/abs/cond-mat/0506438}{{\tt
  arXiv:cond-mat/0506438}}].

\bibitem{1989MS}
G.~{Moore} and N.~{Seiberg}, \emph{{Classical and quantum conformal field
  theory}}, \href{http://dx.doi.org/10.1007/BF01238857}{\emph{Communications in
  Mathematical Physics} {\bf 123} (June, 1989) 177--254}.

\bibitem{1008.0654KS}
A.~{Kapustin} and N.~{Saulina}, \emph{{Topological boundary conditions in
  abelian Chern-Simons theory}},
  \href{http://dx.doi.org/10.1016/j.nuclphysb.2010.12.017}{\emph{Nuclear
  Physics B} {\bf 845} (Apr., 2011) 393--435},
  [\href{https://arxiv.org/abs/1008.0654}{{\tt arXiv:1008.0654}}].

\bibitem{1306.4254K}
A.~{Kapustin}, \emph{{Ground-state degeneracy for Abelian anyons in the
  presence of gapped boundaries}},
  \href{http://dx.doi.org/10.1103/PhysRevB.89.125307}{\emph{Phys. Rev. B} {\bf
  89} (Mar., 2014) 125307}, [\href{https://arxiv.org/abs/1306.4254}{{\tt
  arXiv:1306.4254}}].

\bibitem{Verlinde:1988sn}
E.~P. Verlinde, \emph{{Fusion Rules and Modular Transformations in 2D Conformal
  Field Theory}},
  \href{http://dx.doi.org/10.1016/0550-3213(88)90603-7}{\emph{Nucl. Phys.} {\bf
  B300} (1988) 360--376}.

\end{thebibliography}\endgroup

\end{document}